\RequirePackage{fix-cm}
\documentclass[epj]{svjour}              % onecolumn
\smartqed  % flush right qed marks, e.g. at end of proof
\usepackage{graphicx}
\usepackage{graphics}
\usepackage{epsfig}
\usepackage[toc,page]{appendix}
\usepackage{bm}
\usepackage{epic,eepic}
\usepackage{epsfig}
\usepackage{pifont}
\usepackage{subfigure}
\usepackage{caption}
\usepackage{amsmath,amssymb,amsfonts,mathrsfs,dsfont,mathtools}
\usepackage{url}

\newcommand{\Id}{\mathds{1}}            % D x D identity matrix
\newcommand{\be}{\begin{equation}}
\newcommand{\ee}{\end{equation}}

\newcommand{\tshear}{\tau_{\mbox{\tiny {shear}}}}
\newcommand{\tobs}{t_{\mbox {\tiny {obs}}}}
\newcommand{\Ca}{\mbox{Ca}}
\newcommand{\Cacr}{\mbox{Ca}_{\mbox{\tiny{cr}}}}
\newcommand{\Ren}{\mbox{Re}}
\newcommand{\De}{\mbox{De}}

\usepackage{float}

\usepackage{color}

\usepackage{mathptmx}
\usepackage{latexsym}


\begin{document}
\sloppy
\title{Effects of viscoelasticity on droplet dynamics and break-up in microfluidic T-Junctions: a lattice Boltzmann study}
%\subtitle{Do you have a subtitle?\\ If so, write it here}
\author{Anupam Gupta \& Mauro Sbragaglia% etc
% \thanks is optional - remove next line if not needed
%\thanks{\emph{Present address:} Insert the address here if needed}%
}                     % Do not remove
%
%\offprints{}          % Insert a name or remove this line
%
\institute{Department of Physics and INFN, University of ``Tor Vergata'', Via della Ricerca Scientifica 1, 00133 Rome, Italy}
\authorrunning{A. Gupta, M. Sbragaglia}
\titlerunning{A lattice Boltzmann study of the effects of viscoelasticity in microfluidic T-Junctions}
\date{Received: date / Revised version: date}
% The correct dates will be entered by Springer
%
\abstract{
The effects of viscoelasticity on the dynamics and break-up of fluid threads in microfluidic T-junctions are investigated using numerical simulations of dilute polymer solutions at changing the Capillary number ($\Ca$), i.e. at changing the balance between the viscous forces and the surface tension at the interface, up to $\Ca \approx 3 \times 10^{-2}$. A Navier-Stokes (NS) description of the solvent based on the lattice Boltzmann models (LBM) is here coupled to constitutive equations for finite extensible non-linear elastic dumbbells with the closure proposed by Peterlin (FENE-P model). We present the results of three-dimensional simulations in a range of $\Ca$ which is broad enough to characterize all the three characteristic mechanisms of breakup in the confined T-junction, i.e. {\it squeezing}, {\it dripping} and {\it jetting} regimes.  The various model parameters of the FENE-P constitutive equations, including the polymer relaxation time $\tau_P$ and the finite extensibility parameter $L^2$, are changed to provide quantitative details on how the dynamics and break-up properties are affected by viscoelasticity. We will analyze cases with {\it Droplet Viscoelasticity} (DV), where viscoelastic properties are confined in the dispersed (d) phase, as well as cases with {\it Matrix Viscoelasticity} (MV), where viscoelastic properties are confined in the continuous (c) phase. Moderate flow-rate ratios $Q \approx {\cal O}(1)$ of the two phases are considered in the present study. Overall, we find that the effects are more pronounced in the case with MV, as the flow driving the break-up process upstream of the emerging thread can be sensibly perturbed by the polymer stresses.
 % end of PACS codes
} %end of abstract
\maketitle

% \usepackage{mathptmx}      % use Times fonts if available on your TeX system
%
% insert here the call for the packages your document requires
%\usepackage{latexsym}
% etc.
%
% please place your own definitions here and don't use \def but
% \newcommand{}{}
%
% Insert the name of "your journal" with
% \journalname{myjournal}
%
\
\section{Introduction}
\label{intro}

Droplet-based microfluidic devices have gained a considerable deal of attention, due to their importance in studies that require control over droplet size~\cite{Christopher07,Seeman12,Christopher08,Teh08,Baroud10,Glawdeletal,Glawdeletalb,Glawdeletalbb}. Common droplet generator designs used in these devices are T-shaped~\cite{Demenech07,Demenech06} and flow-focusing~\cite{LiuZhang09,LiuZhang11,Garstecki13} geometries. In T-shaped geometries, a dispersed (d) phase is injected perpendicularly into the main channel containing a continuous (c) phase. Forces are created by the cross-flowing continuous phase which periodically produces break-up of droplets. The operational regime of these devices is primarily characterized by the Capillary number, which quantifies the importance of the viscous forces with respect to the surface tension forces at the non-ideal interface, and the droplet size and its distribution are dictated by the flow-rate ratio $Q=Q_d/Q_c$ of the two immiscible fluids. Distinct regimes of formation of droplets have been identified: {\it squeezing}, {\it dripping} and {\it jetting}, providing a unifying picture of emulsification processes typical of microfluidic systems~\cite{Demenech07,Demenech06,LiuZhang11,Garstecki06}. The squeezing mechanism of break-up is peculiar of all microfluidic systems, because of the physical confinement which naturally accompanies these geometries. In this regime, the break-up process is driven by the build-up of pressure upstream of the emerging thread. The dripping regime, while apparently homologous to the unbounded case, is also significantly influenced by the constrained geometry~\cite{Demenech07}, which modifies the scaling law for the size of the droplets derived from the balance of interfacial and viscous stresses. Finally, the jetting regime sets in only at very high flow rates, or with low interfacial tension, i.e. higher values of the Capillary number.\\
With few exceptions~\cite{Arratia08,Steinhaus,Husny06}, previous research has been mainly restricted to Newtonian fluids. However, the processing of biological fluids inevitably results in considering a non-Newtonian viscoelastic behaviour. Consistently, the study of viscoelastic liquids in flow-focusing geometries~\cite{Arratia08,Steinhaus,Arratia09} or T-junction geometries~\cite{Husny06} has gained some attention. The formation and the pinch-off mechanism of viscoelastic droplets in Newtonian continuous phases was investigated in various flow-focusing geometries by Steinhaus {\it et al.}~\cite{Steinhaus}, while the effect of polymer molecular weight on filament thinning was studied by Arratia {\it et al.}~\cite{Arratia08,Arratia09}. In a recent paper, Derzsi {\it et al.}~\cite{Garstecki13} presented an experimental study of the effects of viscoelasticity in microfluidic flow-focusing geometries. The authors find that the viscoelasticity of the focusing liquid stabilizes the jets facilitating formation of smaller droplets, and leads to transitions between various regimes at lower ratios of flow and at lower values of the Capillary numbers in comparison to the Newtonian focusing liquids. Complementing these results with systematic investigations by varying deformation rates and non-Newtonian constitutive parameters would be of extreme interest. This is witnessed by the various papers in the literature~\cite{Demenech07,Demenech06,LiuZhang09,LiuZhang11,Wang11,vandersman06,Legendre,Gupta09,Gupta10} addressing these kind of problems with the help of numerical simulations. \\
Here we present a three-dimensional numerical investigations of the interplay between viscoelasticity and geometry-mediated breaking in confined microfluidic T-junctions. Numerical simulations allow to address systematically the importance of the various free parameters in the viscoelastic model and visualize the distribution of the polymer feedback stresses, thus correlating the distribution of those stresses to the interface shape. Our numerical approach offers the possibility to tune the viscosity ratio of the two Newtonian phases, a fact that is instrumental to perform simulations with non-Newtonian phases and compare them with the results of fully Newtonian systems with the {\it same} viscosity ratio.\\ 
The paper is organized as follows: in Sec.~\ref{sec:model} we will present the necessary mathematical background for the problem studied, showing the relevant equations that we integrate in both the continuous and dispersed phases, and identifying the relevant dimensionless numbers useful for our investigation. Useful benchmarks for the shear rheology of the numerical model will be provided for the typical parameters used in our study. In section \ref{sec:results} we will present the numerical results and characterize the effects of viscoelasticity in the three distinct regimes of squeezing (subsection \ref{sec:squeezing}) and dripping/jetting (subsection \ref{sec:drippingjetting}). We will study both the droplet size soon after break-up as well as the characteristic time for break-up and compare them with the corresponding Newtonian cases. To explain the observed behaviour we will explore the distribution of feedback stresses in the non-Newtonian phases and correlate them with the characteristic mechanisms of break-up in the confined T-junction. Conclusions will follow in section \ref{sec:conclusions}.

\section{Theoretical Model}\label{sec:model}

\begin{figure*}[t!]
\subfigure[{\scriptsize Polymer shear stress}]
{
\includegraphics[width = 0.475\linewidth]{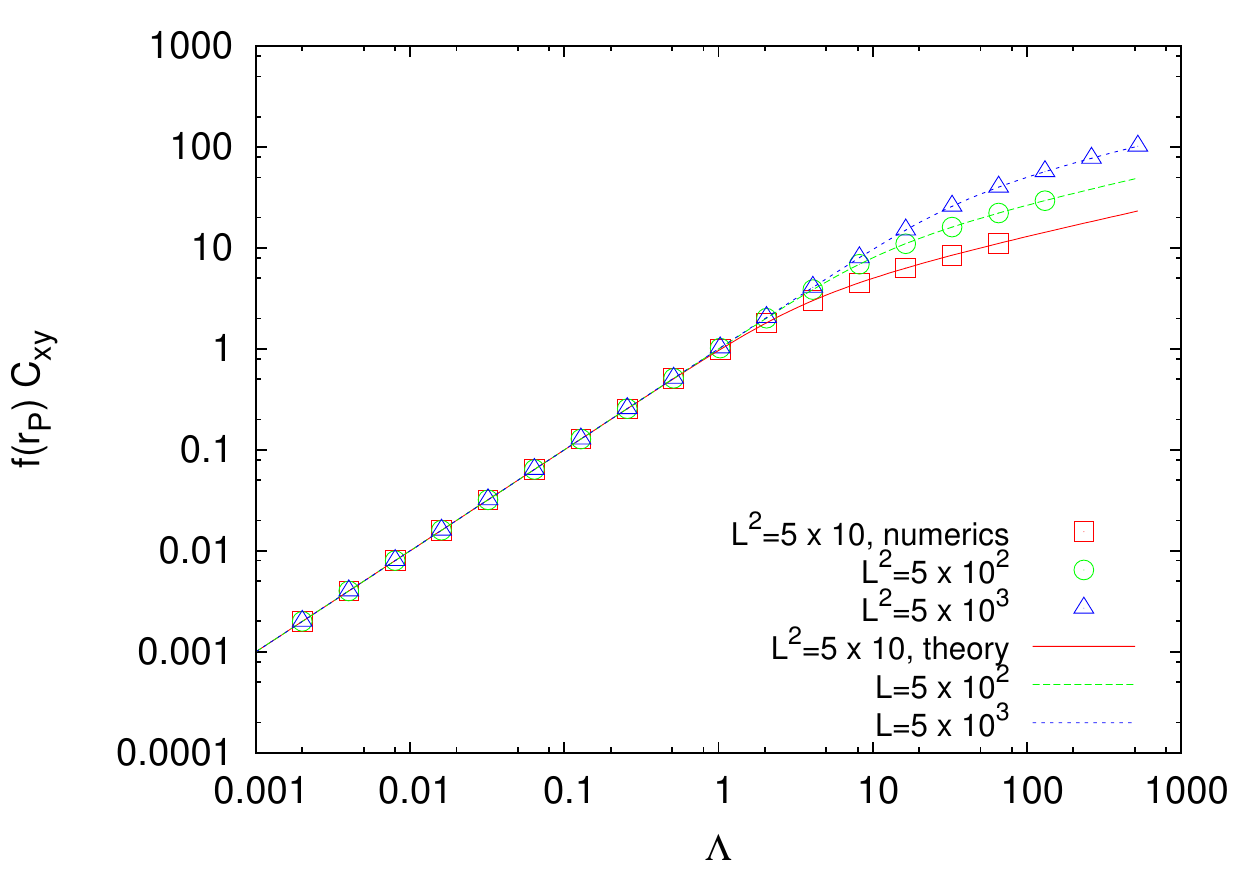}
}
\subfigure[{\scriptsize Polymer first normal stress difference}]
{
\includegraphics[width = 0.475\linewidth]{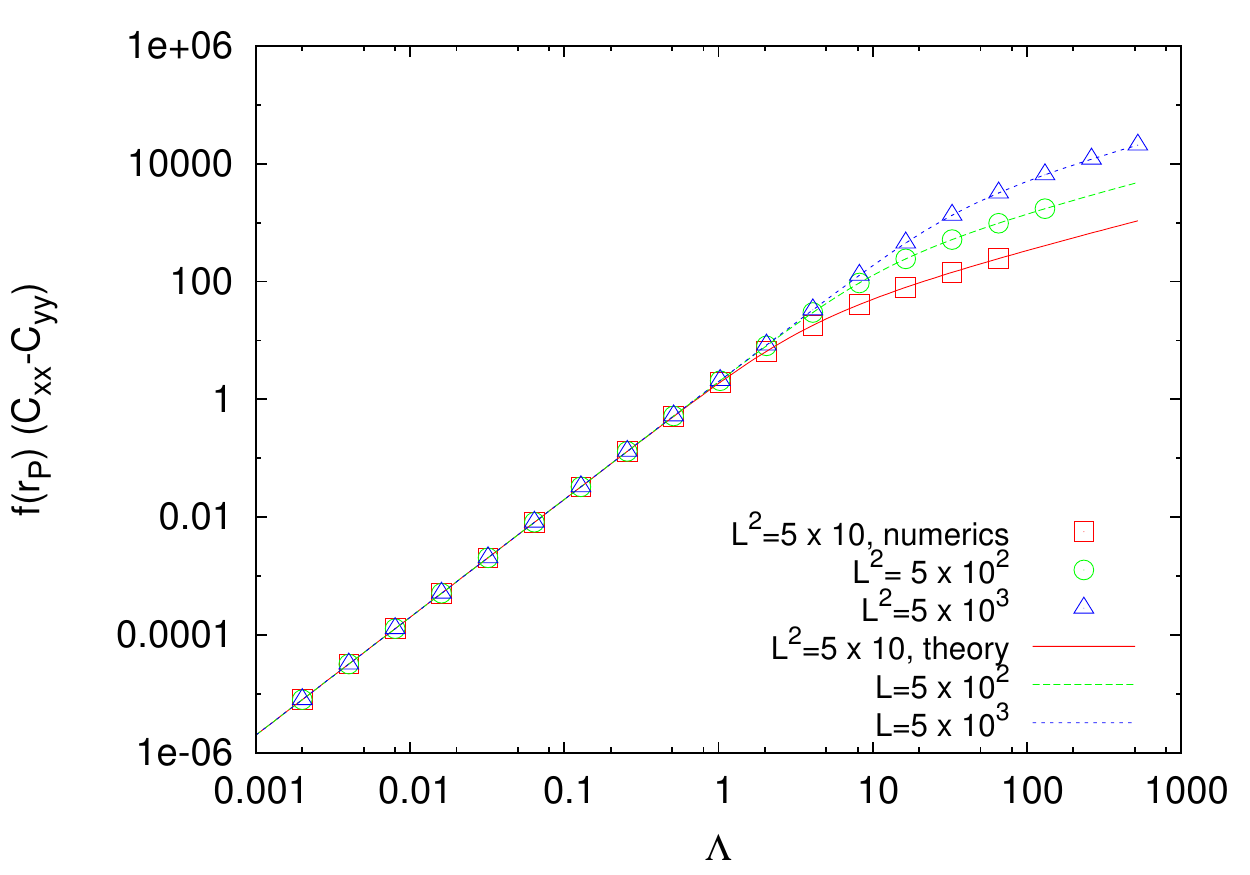}\\
}\\
\subfigure[{\scriptsize Polymer shear viscosity}]
{
\includegraphics[width = 0.475\linewidth]{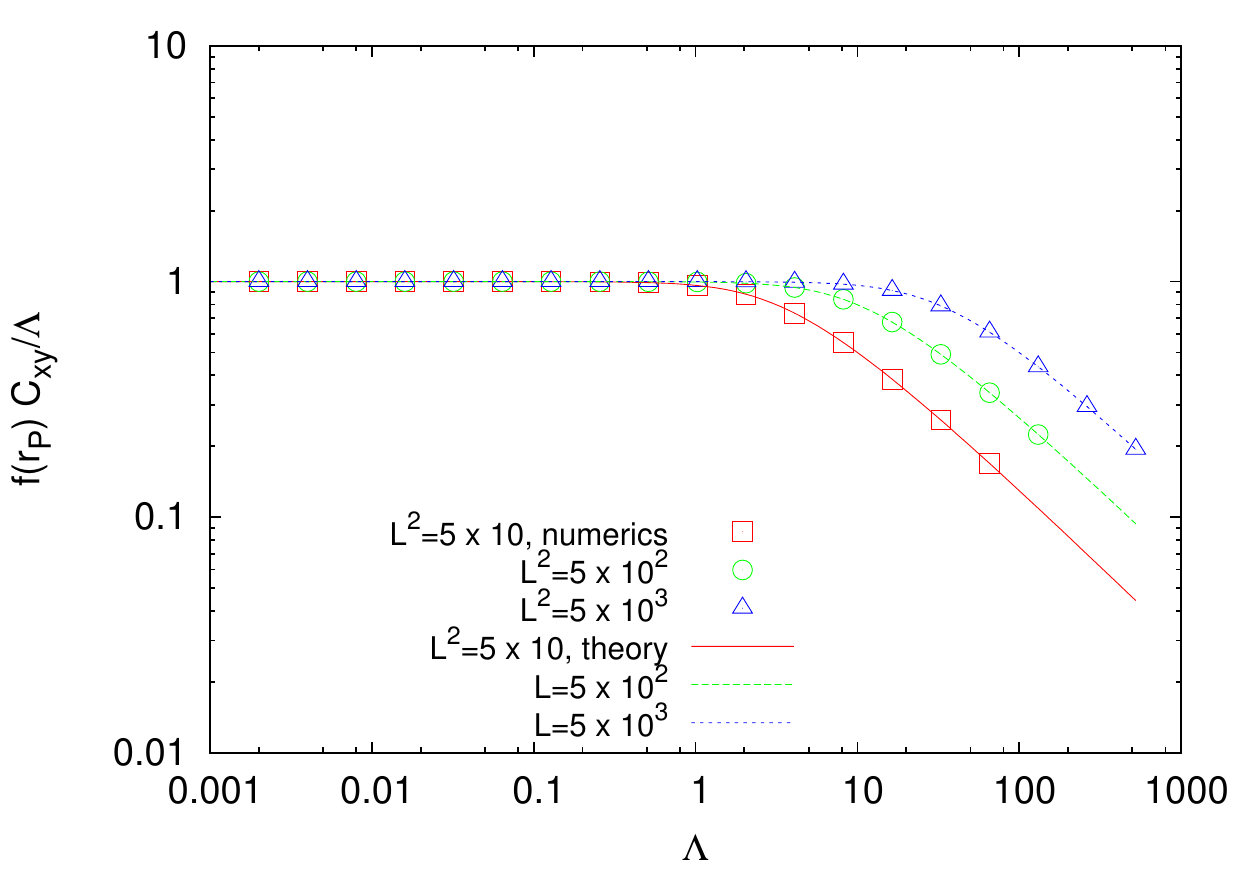}
}
\subfigure[{\scriptsize Polymer first normal stress coefficient}]
{
\includegraphics[width = 0.475\linewidth]{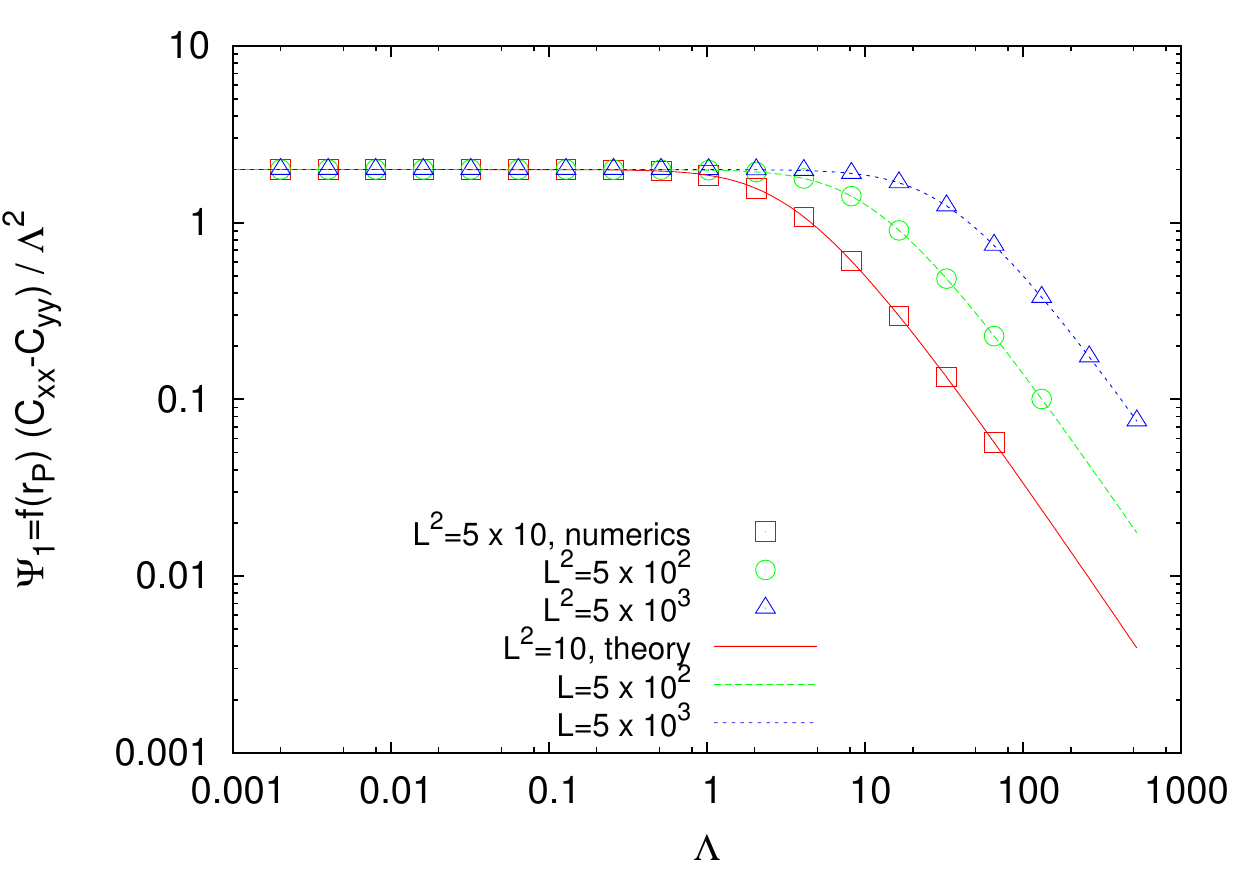}
}\\
\caption{Polymer shear rheology. Panels (a)-(b): we plot the polymer shear stress and the first normal stress difference (both scaled to the polymer viscosity $\eta_P$ and polymer relaxation time $\tau_P$, see \eqref{lindner}-\eqref{lindner2} and text for details) as a function of the dimensionless shear $\Lambda=\tau_P \dot{\gamma}$ in a steady shear flow with intensity $\dot{\gamma}$. Symbols are the results of the numerical simulations~\cite{SbragagliaGuptaScagliarini,SbragagliaGupta} with different imposed shears, different $\tau_P$ and different $L^2$. All the numerical results collapse on different master curves, dependently on the value of $L^2$: $L^2=5 \times 10$ (squares), $L^2=5 \times 10^2$ (circles), $L^2=5 \times 10^3$ (triangles). The lines are the theoretical predictions based on Eqs. (\ref{lindner}) and (\ref{lindner2}). Panels (c)-(d):  we plot the dimensionless polymer shear viscosity and the first normal stress coefficient extracted from data in the top panels. \label{fig:0}}
\end{figure*}

Numerical modeling of viscoelastic fluids often relies on the coupling of constitutive relations for the stress tensor, typically obtained via approximate representations of some underlying micro-mechanical model for the polymer molecules, with a Navier-Stokes (NS) description for the solvent. The FENE-P constitutive model is obtained via a pre-averaging approximation applied to a suspension of non interacting finitely extensible non-linear elastic (FENE) dumbbells. FENE-P is well-adapted for dilute (and semi-dilute) polymer solutions, and has been used previously to analyze filament thinning of viscoelastic fluids in macroscopic experiments~\cite{Wagner05,Lindner03}, as well as the effects of viscoelasticity on the dynamics of filament thinning and break-up processes in microchannels~\cite{Arratia08,Arratia09}. In this paper we provide quantitative details on how the FENE-P model parameters affect the break-up properties of confined threads in microfluidic T-junctions, by analyzing separately the cases of {\it Droplet Viscoelasticity} (DV), where the viscoelastic properties are confined in the dispersed (d) phase undergoing the break-up process, as well as the cases with  {\it Matrix Viscoelasticity} (MV), where the viscoelastic properties are confined in the continuous (c) phase. A fluid described by the FENE-P model possesses the same dynamical properties as a fluid described by the much simpler Oldroyd-B model, which assumes that polymers can be modeled as Hookean springs which relax to the equilibrium configuration with a characteristic time $\tau_P$. The main difference is that the Oldroyd-B model allows for infinite extension of polymer molecules, while the FENE-P model uses a spring-force law in which the polymer molecules can be stretched only by a finite amount in the flow field~\cite{bird,Herrchen}. Thus we can explore systematically both the effects of the polymer relaxation times as well as their finite extensibility.\\
The solvent part of the model is obtained with lattice Boltzmann models (LBM)~\cite{Zhang11,Aidun10}, which proved to be extremely valuable tools for the simulation of droplet deformation problems~\cite{Xi99,vandersman08,Komrakova13,Liuetal12}, droplets dynamics in open~\cite{Moradi,Thampi} and confined~\cite{Gupta09,Gupta10,Liuetal12} microfluidic geometries. LBM is instrumental to solve the diffuse-interface hydrodynamic equations of a binary mixture of two components~\cite{Yue04,Yueetal05,Yueetal06a,Yueetal06b,Yueetal08,Yueetal12}: the resulting physical domain can be partitioned into different subdomains, each occupied by a ``pure'' fluid, with the interface between the two fluids described as a thin layer where the fluid properties change smoothly. The FENE-P constitutive equations are solved with a  finite difference scheme which is coupled with the solvent LBM as described in~\cite{SbragagliaGuptaScagliarini,SbragagliaGupta}. The numerical approach has been extensively validated in our previous works~\cite{SbragagliaGuptaScagliarini,SbragagliaGupta}, where we have provided evidence that the model is able to capture quantitatively rheological properties of dilute suspensions as well as deformation and orientation of single viscoelastic droplets in confined shear flows. The main essential features of the model are recalled in Appendix~\ref{sec:appendix}.\\
In the MV case, the equations we solve in the continuous phase are the Navier-Stokes (NS) equations coupled to the FENE-P constitutive equations
\be\label{NSc}
\begin{split}
\rho_{c} & \left[ \partial_t \bm u_{c} + ({\bm u}_{c} \cdot {\bm \nabla}) \bm u_{c} \right]  =  - {\bm \nabla}P_{c}+\\ &  {\bm \nabla} \left(\eta_{c} ({\bm \nabla} {\bm u}_{c}+({\bm \nabla} {\bm u}_{c})^{T})\right)  +\frac{\eta_{P}}{\tau_{P}}{\bm \nabla} \cdot [f(r_{P}){\bm {\bm {C}}}].
\end{split}
\ee
\be\label{FENEP}
\begin{split}
\partial_t {\bm {C}} + (\bm u_{c} \cdot {\bm \nabla}) {\bm {C}}  =  {\bm {C}} \cdot ({\bm \nabla} {\bm u}_{c}) + & {({\bm \nabla} {\bm u}_{c})^{T}} \cdot {\bm {C}}  \\  -& \left(\frac{{f(r_{P}){\bm {C}} }- {{\bm I}}}{\tau_{P}}\right).
\end{split}
\ee
Here, ${\bm u}_c$ and $\eta_c$ are the velocity and the dynamic viscosity of the continuous phase, respectively. $\rho_c$ is the solvent density, $P_c$ the solvent bulk pressure, and $({\bm \nabla} {\bm u}_c)^T$ the transpose of $({\bm \nabla} {\bm u}_c)$. As for the polymer details, $\eta_{P}$ is the viscosity parameter for the FENE-P solute, $\tau_P$ the polymer relaxation time,  ${\bm {C}}$ the polymer-conformation tensor, ${\bm I}$ the identity tensor, $f(r_P)\equiv{(L^2 -3)/(L^2 - r_P^2)}$ the FENE-P potential that ensures finite extensibility, $ r_P \equiv \sqrt{Tr({\bm {C}})}$ and $L$ is the maximum possible extension of the polymers~\cite{bird,Herrchen}. In the dispersed phase we just consider the NS equations 
\begin{eqnarray}\label{NSd}
\rho_d \left[ \partial_t \bm u_{d} + ({\bm u}_{d} \cdot {\bm \nabla}) \bm u_{d} \right] 
&=&  - {\bm \nabla}P_{d} \nonumber \\ && + {\bm \nabla} \left(\eta_{d} ({\bm \nabla} {\bm u}_{d}+({\bm \nabla} {\bm u}_{d})^{T})\right)                                              
\end{eqnarray}
where the different fields have the same physical meaning but they refer to the dispersed phase. Immiscibility between the dispersed phase and the continuous phase is introduced using the so-called ``Shan-Chen'' model~\cite{SbragagliaGuptaScagliarini,SC93,SC94} which ensures phase separation with the formation of stable interfaces between the two phases characterized by a positive surface tension $\sigma$.\\
For the DV case, we consider the reversed case, where the FENE-P constitutive equations are integrated in the dispersed phase (i.e. \eqref{NSc}-\eqref{FENEP} with c $\rightarrow$ d), while only the NS equations are considered in the continuous phase (i.e. \eqref{NSd} with d $\rightarrow$ c).\\
As for the geometry used, the T-junction is embedded in a rectangular parallelepiped with size $L_{x}  \times L_y \times L_z$, and channels have a square cross-section with edge $H=L_{z}$. The square cross-section is resolved with $H \times H = 32 \times 32$ grid points. The main channel and the side channel lengths are resolved with a variable number of grid points (see also table \ref{table:para}), depending on the characteristic regime analyzed and the characteristic size of the droplet after break-up.\\
Besides the geometrical parameters, the Newtonian problem is described by six parameters characterizing the flow and material properties of the fluids. These parameters are the mean speeds of the continuous and dispersed phases, $v_c$ and $v_d$, respectively; the viscosities of the two fluids $\eta_c$ and $\eta_d$ of Eqs. (\ref{NSc}) and (\ref{NSd}), the interfacial tension $\sigma$, and the total density $\rho_c=\rho_d=\rho$ (the same for the dispersed and continuous phases). We will assume perfect wetting for the continuous phase, while the dispersed fluid does not wet the walls. Wetting properties are introduced at the boundaries declaring the stress of the density fields~\cite{Benzi06,sbragaglia08}. We then choose the following groups~\cite{Demenech07,LiuZhang09,LiuZhang11}: the Capillary number calculated for the continuous phase,
\be\label{eq:Ca}
\Ca = \frac{(\eta_{\mbox{\tiny{TOT}},c}) v_c}{\sigma}
\ee the Reynolds number $\Ren =\rho v_c H/ (\eta_{\mbox{\tiny{TOT}},c})$, the viscosity ratio $\lambda$, and the flow rate ratio
\be\label{eq:Q}
Q=\frac{v_d}{v_c} = \frac{Q_d}{Q_c}
\ee
where $Q_d=v_d H^2$ and $Q_c =v_c H^2$ are the flow rates at the two inlets.  For the flow regimes under consideration, the Reynolds number is small ($\Ren \approx 0.01-0.1$), and does not influence the droplet size, which leaves us with the three governing parameters: $\Ca$, $\lambda$ and $Q$. Notice that the total viscosity in the continuous phase $\eta_{\mbox{\tiny{TOT}},c}$ is either $\eta_{\mbox{\tiny{TOT}},c}=\eta_c+\eta_P$ (for MV) or $\eta_{\mbox{\tiny{TOT}},c}=\eta_c$ (for DV). In the outlet, we impose pressure boundary conditions and use Neumann boundary conditions for the velocity field. A Dirichlet boundary condition is imposed at the inlets by specifying the pressure gradient that is compatible with the analytical solution of a Stokes flow in a square duct~\cite{vandersman08}. As for the polymer boundary conditions, we impose a Dirichlet type boundary conditions by linearly extrapolating the conformation tensor at the boundaries.\\
Our numerical approach offers the possibility to tune the viscosity ratio of the two Newtonian phases~\cite{SbragagliaGuptaScagliarini,SbragagliaGupta}. This will allow us to work with unitary viscosity ratio, defined in terms of the total (fluid + polymer) shear viscosity $\lambda=\eta_d/(\eta_c+\eta_{P})=1.0$ for MV and  $\lambda=(\eta_d+\eta_P)/\eta_c=1.0$ for DV. Consistently, we will compare the non-Newtonian simulations with the corresponding Newtonian case at $\lambda=\eta_d/\eta_c=1.0$. The ratio between the polymer viscosity and the total viscosity is set to $\eta_P/(\eta_{c,d}+\eta_P) \approx 0.265$. Similarly to problems involving single droplet deformation and dynamics~\cite{Greco02,Greco02b,Minale10,Minale04,Minale10b}, we choose to quantify the degree of viscoelasticity with the Deborah number that we define as $\De=\frac{N_1 H}{2 \sigma}\left(\frac{\sigma}{(\eta_{d,c}+\eta_P) H \dot{\gamma}}\right)^2$, where $N_1$ is the first normal stress difference which develops in the viscoelastic phase in presence of a homogeneous steady shear~\cite{bird,Lindner03}. In the definition of the Deborah number, the viscosity is obviously indicated in the viscoelastic phase, either $\eta_{c}+\eta_P$ for MV or $\eta_{d}+\eta_P$ for DV. The shear rheology of the model can be quantitatively verified in the numerical simulations. There are indeed exact analytical results one can get by solving the constitutive equations for the hydrodynamical problem of steady shear flow, $u_x=\dot{\gamma} y$, $u_y=u_z=0$: both the polymer shear stress and the first normal stress difference $N_1$ for the FENE-P model~\cite{bird,Lindner03} follow
\begin{eqnarray}\label{lindner}
\frac{\eta_P}{\tau_P} f(r_P) {C}_{xy} &=& \frac{2 \eta_P}{\tau_P} \left(\frac{L^2}{6} \right)^{1/2} \times \nonumber \\ 
&& \sinh \left(\frac{1}{3} \mbox{arcsinh} \left(\frac{\dot{\gamma} \tau_P L^2}{4} \left(\frac{L^2}{6}\right)^{-3/2}\right) \right) \label{S}
\end{eqnarray}
\begin{eqnarray}\label{lindner2}
N_1 &=& \frac{\eta_P}{\tau_P} f(r_P) ({C}_{xx}-{C}_{yy})=8 \frac{\eta_P}{\tau_P} \left(\frac{L^2}{6} \right) \times \nonumber \\ 
&& \sinh^2 \left(\frac{1}{3} \mbox{arcsinh} \left(\frac{\tau_P \dot{\gamma} L^2}{4} \left(\frac{L^2}{6}\right)^{-3/2}\right) \right) \label{N1}.
\end{eqnarray}
The validity of both Eqs. (\ref{lindner}) and (\ref{lindner2}) is benchmarked in Fig. \ref{fig:0}: numerical simulations have been carried out in three dimensional domains with $H \times H \times H=  20 \times 20 \times 20$ cells. Periodic boundary conditions are applied in the stream-flow (x) and in the transverse-flow (z) directions while two walls are located at $y=0$ and $y=H$. The linear shear flow $u_x=\dot{\gamma} y$, $u_y=u_z=0$ is imposed in the numerics by applying two opposite velocities in the stream-flow direction ($u_x(x,y=0,z)=-u_x(x,y=H,z)=U_w$) at the upper ($y=H$) and lower wall ($y=0$) with the bounce-back rule~\cite{Gladrow00}. We next change the shear in the range $10^{-6} \le 2U_w/H \le 10^{-2}$ lbu (lattice Boltzmann units) and the polymer relaxation time in the range $10^1 \le \tau_P \le 10^5$ lbu for different values of the finite extensibility parameter ranging in the interval $L^2=5 \times 10 - 5 \times 10^3$, and fixed $\eta_P$. The various quantities are made dimensionless with the viscosity $\eta_P$ and the relaxation time $\tau_P$, and they are plotted as a function of the dimensionless shear $\Lambda=\tau_P \dot{\gamma}$. The values of the conformation tensor are taken when the simulation has reached the steady state. As we can see from the figures, all the numerical simulations collapse on different master curves, dependently on the value of $L^2$. In particular, both the  stress \eqref{lindner} and first normal stress difference \eqref{lindner2} increase at large $\Lambda$ to exhibit variable levels depending on $L^2$, and consistently with the theoretical predictions~\cite{bird,Herrchen,Lindner03}.  The dependence from $L^2$ reflects in thinning effects visible in the dimensionless polymer shear viscosity, $f(r_P) {C}_{xy}/\Lambda$, and first normal stress coefficient, $\Psi_1=f(r_P) ({C}_{xx}-{C}_{yy})/\Lambda^2$, which are analyzed in the bottom panel of figure \ref{fig:0}. Overall, the numerical simulations performed to quantify the shear rheology reveal a very good agreement with the theoretical predictions both in the polymer shear viscosity and in the first normal stress difference. Similar analysis can be performed for extensional flows, showing that the increase of the extensional viscosity predicted by the theory~\cite{bird,Herrchen,Lindner03} is indeed found in the numerical simulations~\cite{SbragagliaGuptaScagliarini}. The coupling between normal stresses and single droplet dynamics under simple shear has also been extensively verified in the numerical simulations. In particular, in~\cite{SbragagliaGuptaScagliarini} we provided evidence that the model proposed captures quantitatively single droplet orientation and deformation in presence of viscoelastic stresses.\\
In the limit of Hookean dumbbells (Oldroyd-B limit, $L^2 \gg 1$) we can use the asymptotic expansion of the hyperbolic functions and we get $N_1 \approx 2 \tau_P \eta_P \dot{\gamma}^2$, so that 
\be\label{Desimple}
\De=\frac{\tau_P}{\tau_{\mbox{\tiny{H}}}} \frac{\eta_P}{\eta_{d,c}+\eta_P}.
\ee
Equation (\ref{Desimple}) shows that $\De$ is clearly dependent on the ratio between the polymer relaxation time $\tau_P$ and the time $\tau_H$ defined as
\be\label{emulsiontime}
\tau_{H}=\frac{H (\eta_{d,c}+\eta_P)}{\sigma}
\ee
which represents the relaxation time of a droplet with characteristic size $H$, determined by viscous and capillary forces. Clearly, definition \eqref{Desimple} is dependent on rheology and geometry. The values of $L^2$ we use in the numerical simulations of the confined T-Junctions are such that $L^2 \ge 10^2$, ruling out important thinning effects for the shears achieved in our simulations. We therefore choose to report results based on the definition of the Deborah number \eqref{Desimple} together with the finite extensibility parameter $L^2$. All the various parameters are summarized in Table \ref{table:para}. An interesting point of discussion emerges from the attempt of connecting results from numerical simulations with experimental data, and in particular how appropriate is the choice of the parameters $\eta_P$, $\tau_P$ and $L^2$. Some of these information are available from the literature (see~\cite{Arratia08,Arratia09} and references therein). Arratia {\it et al.}~\cite{Arratia08,Arratia09} performed experiments on filament thinning and break-up of viscoelastic fluids in microchannels: for a viscoelastic fluid made by adding 100 ppm of polyacrylamide (PAA) with MW (molecular weight) of $10^5$, a concentration of $\eta_P/(\eta_{d,c}+\eta_P) \approx 10^{-1}$, a finite extensibility parameter $L^2 \approx 10^3$ and fluid relaxation time $\tau_P =0.05 s$ are found to best fit the experimental rheological data. The polymer relaxation time decreases at decreasing the molecular weight, down to $\tau_P \approx 10^{-3}s$, for MW of $1 \times 10^3$. In the present study, we choose to use different $L^2$, so as to study the enhancement of viscoelastic effects up to the value above cited. As for the polymer relaxation time $\tau_P$, we notice that a Newtonian droplet with characteristic size of the order of $10^{-4} m $ would result in a $\tau_{H}$ ($\eta_d \approx 0.2 \hspace{.03in} Pa \hspace{.03in} s$ and $\sigma=10^{-2} N /m$~\cite{Arratia08,Arratia09}) of the order of $\tau_{H}=\eta_d H /\sigma \approx 10^{-3} s$, hence $\tau_P/\tau_{H}$ ranges from $1$ to a few tens. Such a range can actually be explored in the numerics by tuning $\tau_P$ in the range $250-4000$ lbu ($\tau_P/\tau_{H}$ in the range $1-25$).

%%%%%%%%%%%%%%%%%%%%%%%%%%%%%%%%%%%%%%%%%%%%%%%%%%%%%%%%%%%%%%%%%%%%%%%%%%%%%%%%
%%%%%%%%%%%%%%%%%%%%%%%%%%%%%%%%%%%%%%%%%FIG 1%%%%%%%%%%%%%%%%%%%%%%%%%%%%%%%%%%%%%%%%%%%%%%%%%%%%%%%%%%%%%%%%%%%%%%%%%%%%%%%%%%%%%%%%%%%%%%%%%%%%%%%%%%%%%%%%%%

%%%%%%%%%%%%%%%%%%%%%%%%%%%%%%%%%%%%%%%%%%%%%%%%%%%%%%%%%%%%%%%%%%%%%%%%%%%%%%%%%%%%%%%%%%%%%%%%%%%%%%%%%%%%%%%%%%%%%%%%%%%%%%%%%%%%%%%%%%%%%%%%%%%%%%%%%%%%%%%%
%%%%%%%%%%%%%%%%%%%%%%%%%%%%%%%%%%%%%%%%%%%%%%%%%%%%%%%%%%%%%%%%%%%%%%%%%%%%%%%%

%%%%%%%%%%%%%%%%%%%%%%%%%%%%%%%%%%%%%%%%%%%%%%%%%%%%%%%%%%%%%%%%%%%%%%%%%%%%%%%%
%%%%%%%%%%%%%%%% TABLE OF PARAMETERS  %%%%%%%%%%%%%%%%%%%%%%%%%%%%%%%%%%%%%%%%%%
%%%%%%%%%%%%%%%%%%%%%%%%%%%%%%%%%%%%%%%%%%%%%%%%%%%%%%%%%%%%%%%%%%%%%%%%%%%%%%%%
\begin{table*}[t!]\label{table}
\begin{center}
   \begin{tabular}{@{\extracolsep{\fill}} |c|c|c|c|c|c|c|c|c|}
    \hline
    $\Ca$ & $Q$ &$L_{x} \times L_y \times L_z$ &  $\eta_d$ & $\eta_c$  & $\eta_P$  & $\tau_P$ & $\De$ & $L^2$ \\
   & & cells & lbu & lbu & lbu & lbu &  &\\
   \hline \hline
    $0.002-0.02$ & $ 1.0$ & $640 \times 128 \times 32$ & $0.49$ & $0.49$ & $0.00$ &                 $   $ & $ $ & $ $ \\
    $0.002-0.02$ & $ 1.0$ & $896 \times 128 \times 32$ & $0.36$ & $0.49$ & $0.13$  &  $2-45 \times 10^2$ & $0.3-7.0$ & $10^2,10^3,10^4$ \\
    $0.002-0.02$ & $ 1.0$ & $896 \times 128 \times 32$ & $0.49$ & $0.36$ & $0.13$  &  $2-45 \times 10^2$ & $0.3-7.0$ & $10^2,10^3,10^4$ \\
\hline
    $0.002$ & $ 0.25-1.0$ & $640 \times 128 \times 32$ & $0.49$ & $0.49$ & $0.00$ &                 $   $ & $ $ & $ $ \\
    $0.002$ & $ 0.25-1.0$ & $896 \times 128 \times 32$ & $0.36$ & $0.49$ & $0.13$ &  $2-45 \times 10^2$ & $0.3-7.0$ & $10^2,10^3,10^4$ \\
    $0.002$ & $ 0.25-1.0$ & $896 \times 128 \times 32$ & $0.49$ & $0.36$ & $0.13$ &  $2-45 \times 10^2$ & $0.3-7.0$ & $10^2,10^3,10^4$ \\
\hline
   \end{tabular}
\end{center}
\caption{\small Parameters for the numerical simulations: $\Ca$ is the Capillary number (see Eq. \eqref{eq:Ca}), $Q=Q_d/Q_c$ is the flow-rate ratio between the dispersed (d) and continuous (c) phase. The T-junction is embedded in a rectangular parallelepiped with size $L_{x}  \times L_y \times L_z$, and channels have a square cross-section with edge $H=L_{z}$. $\eta_d$ is the dynamic viscosity of the Newtonian solvent inside the dispersed phase, $\eta_c$ is the dynamic viscosity of the Newtonian solvent inside the continuous phase, $\eta_P$ is the polymer viscosity, $\tau_P$ is the polymer relaxation time, $\De$ is Deborah number based on definition \eqref{Desimple}.} \label{table:para}
\end{table*}

%%%%%%%%%%%%%%%%%%%%%%%%%%%%%%%%%%%%%%%%%%%%%%%%%%%%%%%%%%%%%%%%%%%%%%%%%%%%%%%%

\section{Results and Discussions}\label{sec:results}

\begin{figure*}[t!]
%\makeatletter
%\def\@captype{figure}
%\makeatother
\begin{minipage}{0.325\textwidth}
\subfigure[{\scriptsize $t=t_0+5.6 \tshear$, $Q=0.5$, $\Ca = 0.0026$}]
{
\includegraphics[width = 0.8\linewidth]{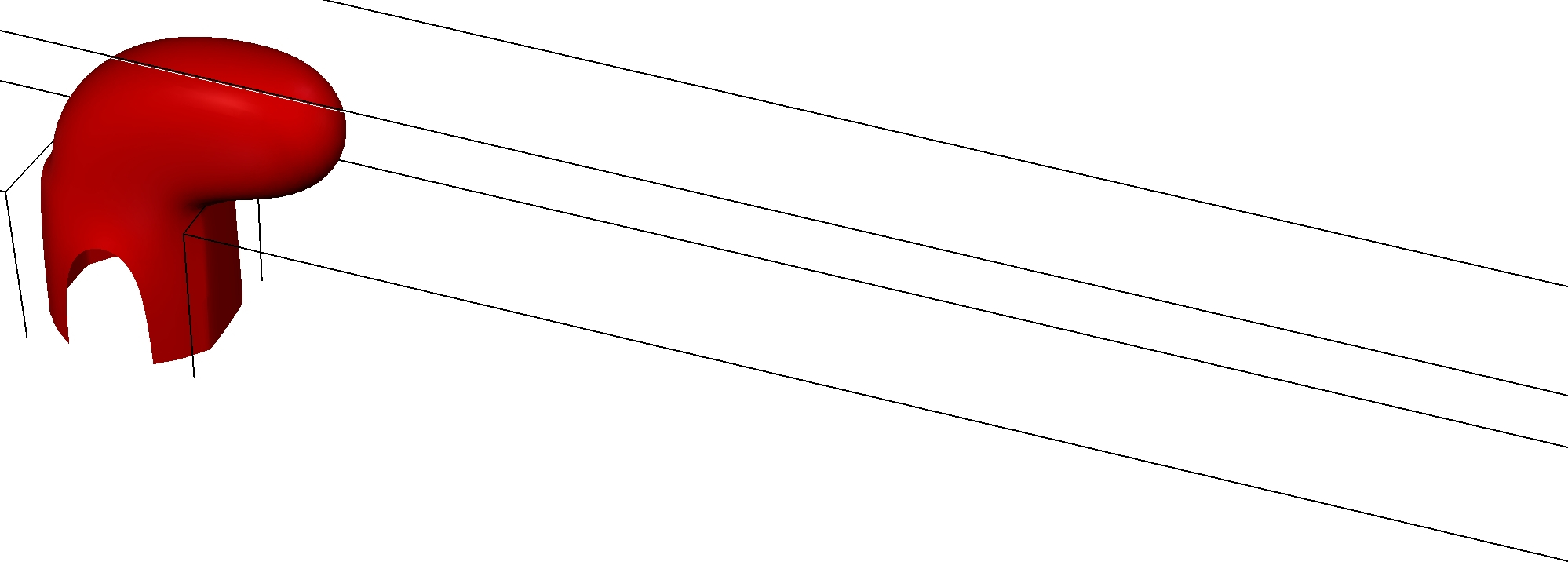}
}\\
\subfigure[{\scriptsize $t=t_0+6.7 \tshear$, $Q=0.5$, $\Ca = 0.0026$}]
{
\includegraphics[width = 0.8\linewidth]{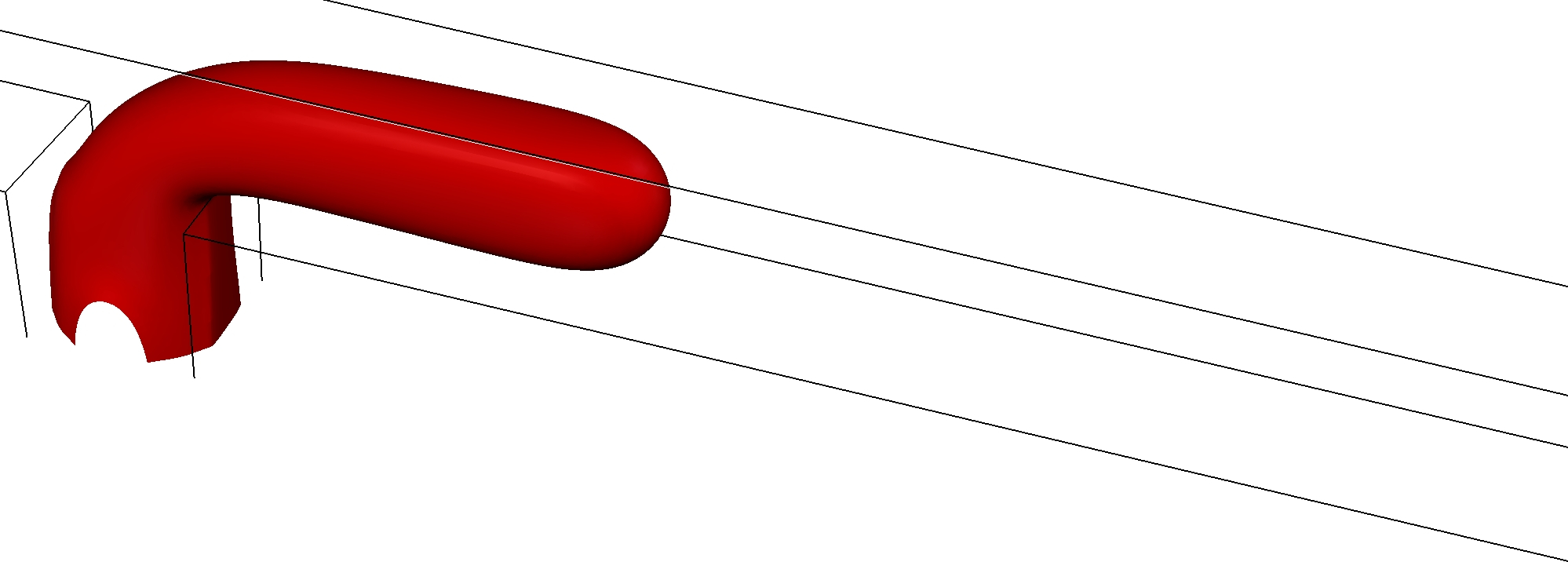}
}\\
\subfigure[{\scriptsize $t=t_0+7.4 \tshear$, $Q=0.5$, $\Ca = 0.0026$}]
{
\includegraphics[width = 0.8\linewidth]{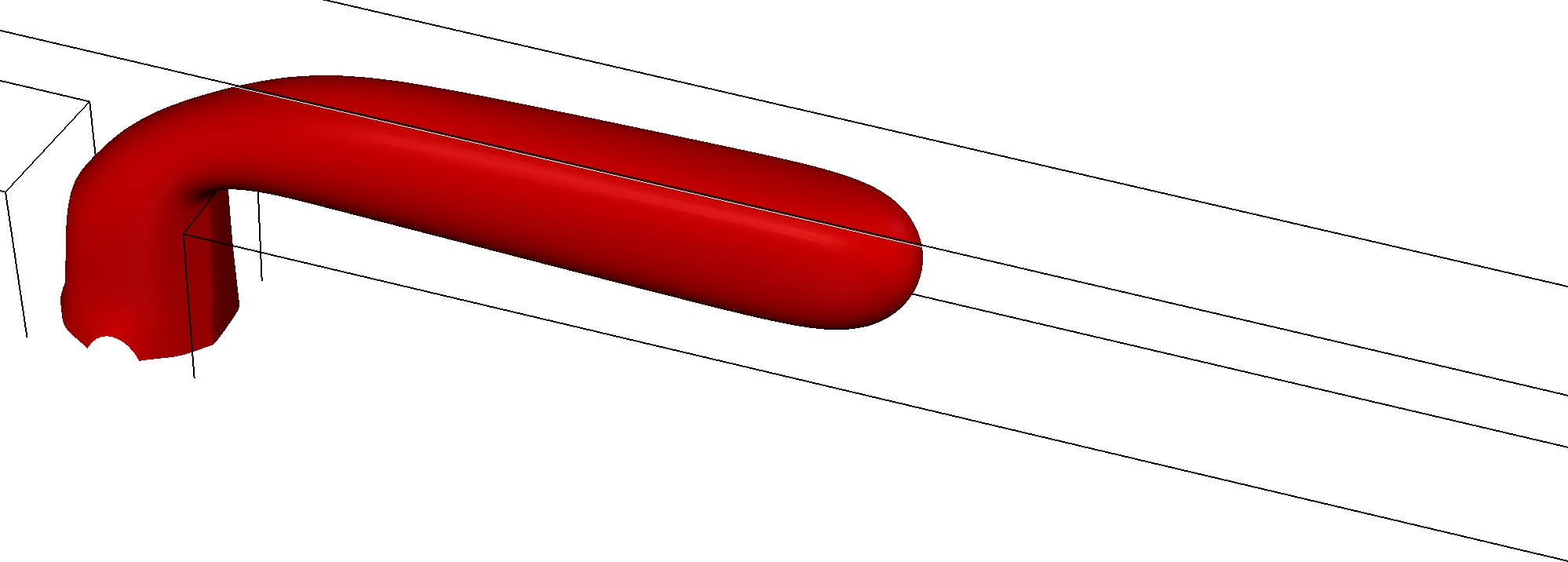}
}\\
\subfigure[{\scriptsize $t=t_0+7.8 \tshear$, $Q=0.5$, $\Ca = 0.0026$}]
{
\includegraphics[width = 0.8\linewidth]{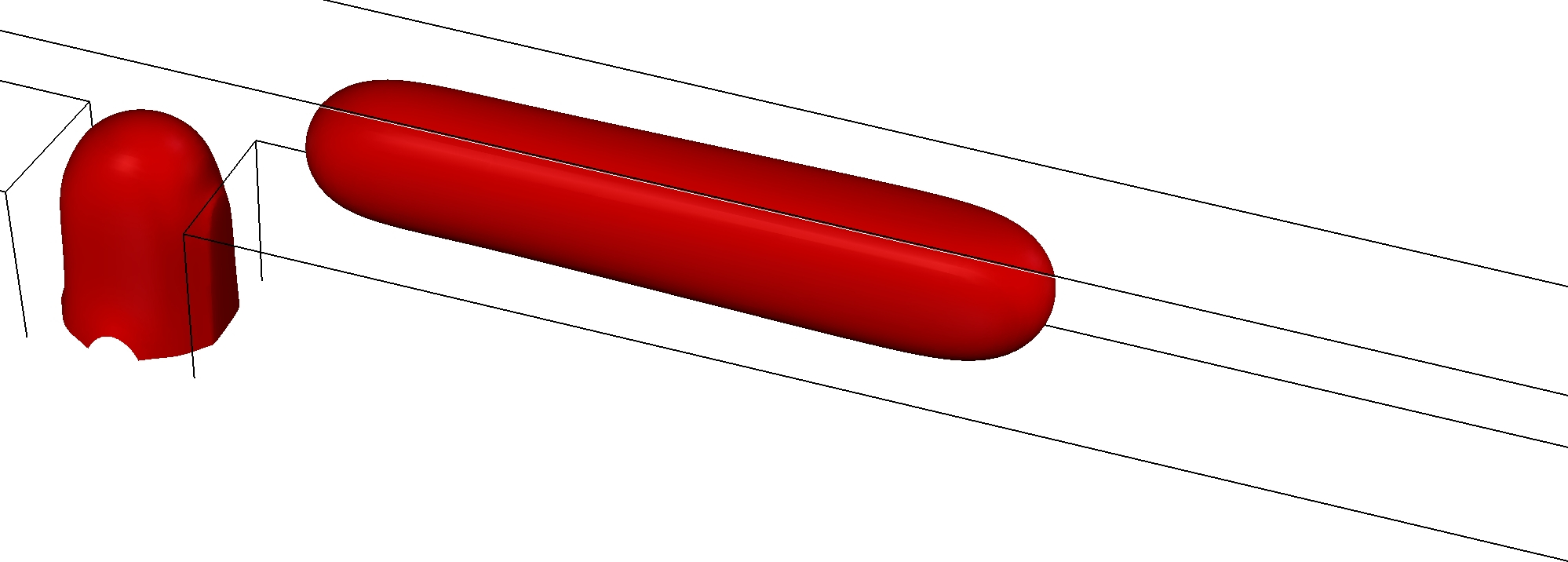}
}
\end{minipage}
\begin{minipage}{0.325\textwidth}
\subfigure[{\scriptsize $t=t_0+2.6 \tshear$, $Q=1.0$, $\Ca = 0.013$}]
{
\includegraphics[width = 0.8\linewidth]{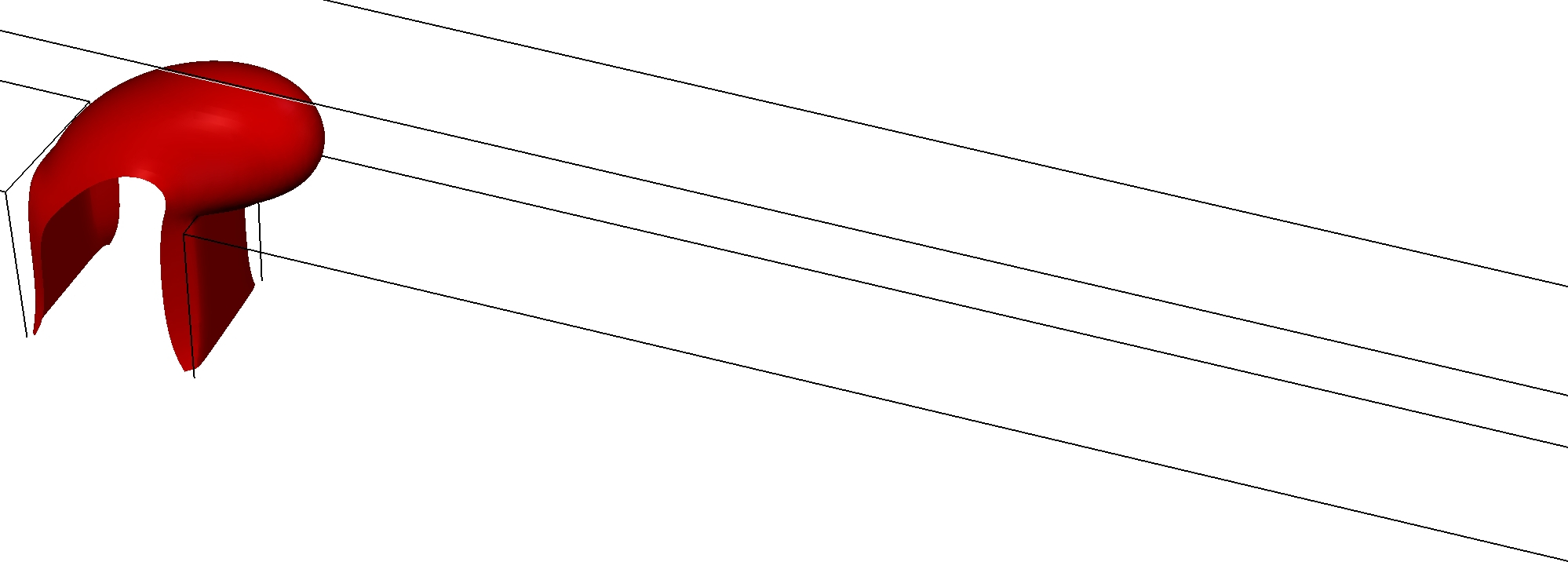}
}\\
\subfigure[{\scriptsize $t=t_0+3.2 \tshear$, $Q=1.0$, $\Ca = 0.013$}]
{
\includegraphics[width = 0.8\linewidth]{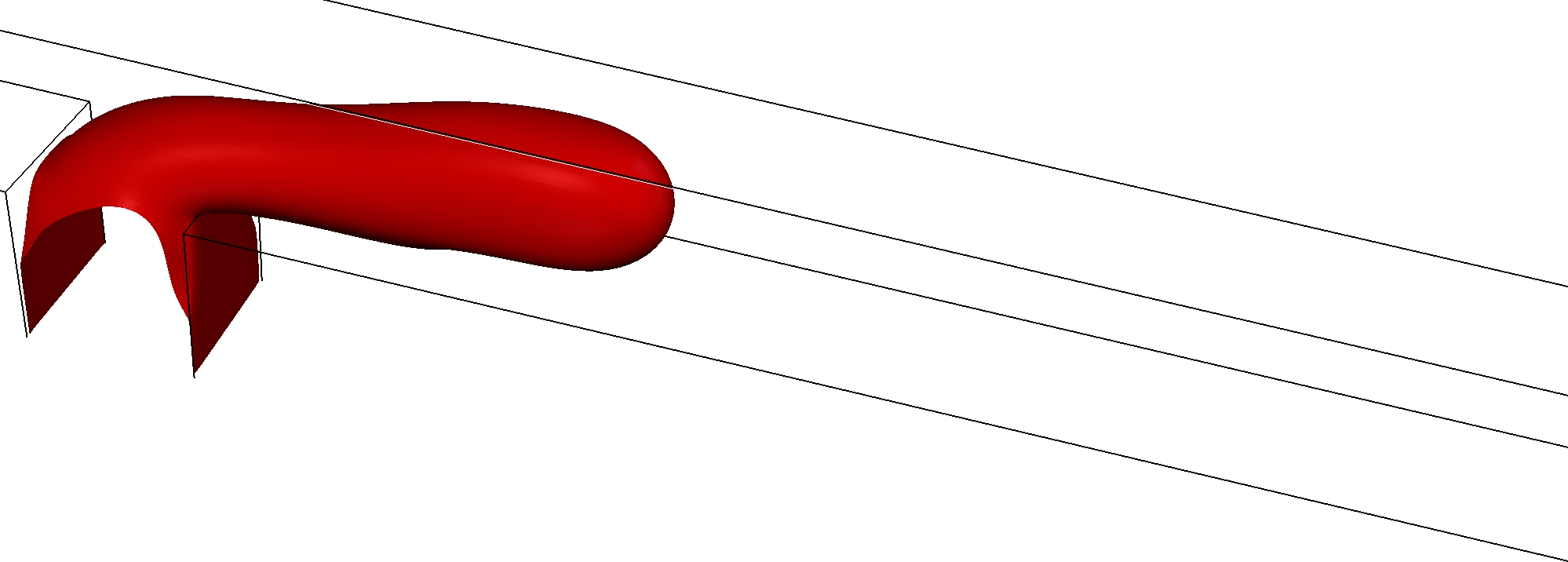}
}\\
\subfigure[{\scriptsize $t=t_0+4.0 \tshear$, $Q=1.0$, $\Ca = 0.013$}]
{
\includegraphics[width = 0.8\linewidth]{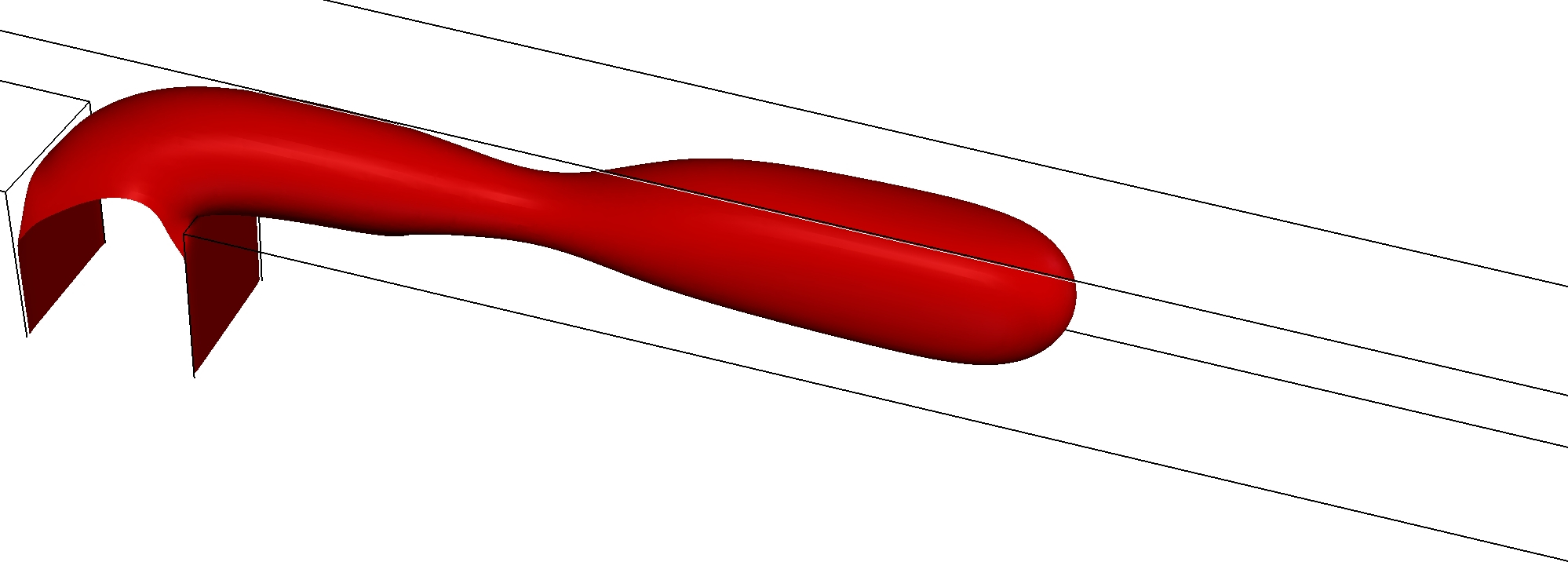}
}\\
\subfigure[{\scriptsize $t=t_0+5.6 \tshear$, $Q=1.0$, $\Ca = 0.013$}]
{
\includegraphics[width = 0.8\linewidth]{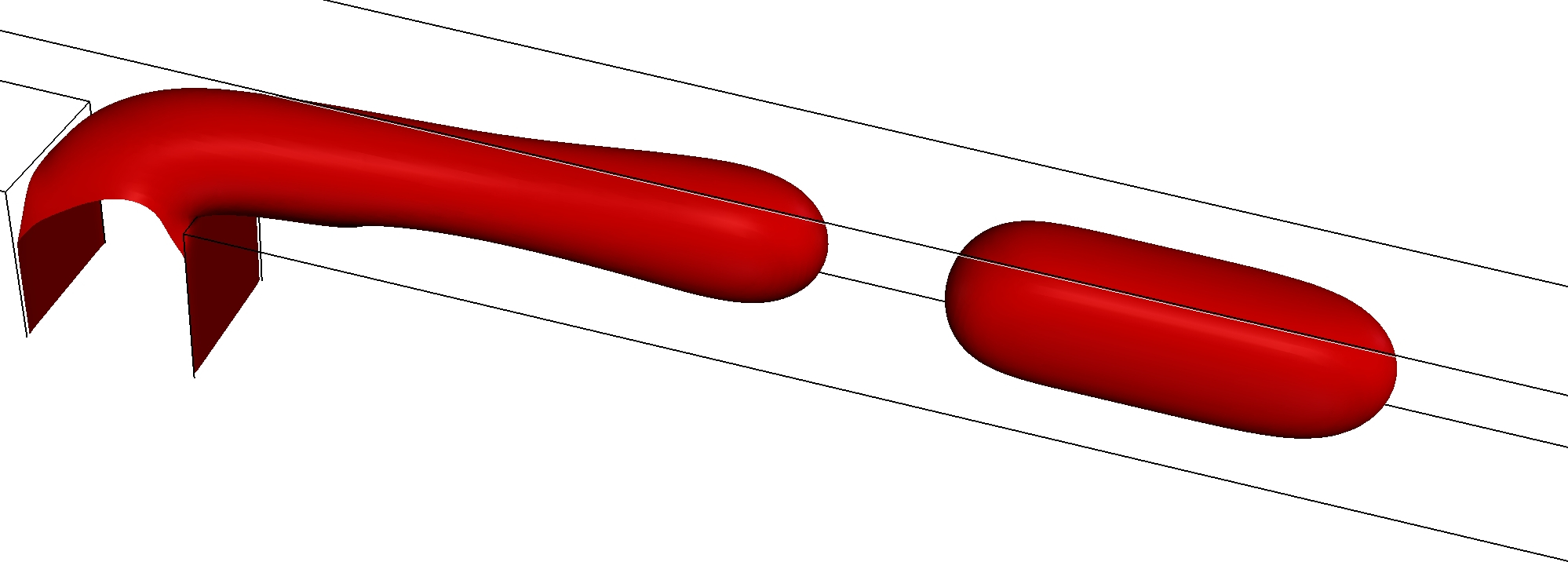}
}\\
\end{minipage}
\begin{minipage}{0.410\textwidth}
\subfigure[{\scriptsize $t=t_0+2.8 \tshear$, $Q=1.0$, $\Ca = 0.026$}]
{
\includegraphics[width = 0.8\linewidth]{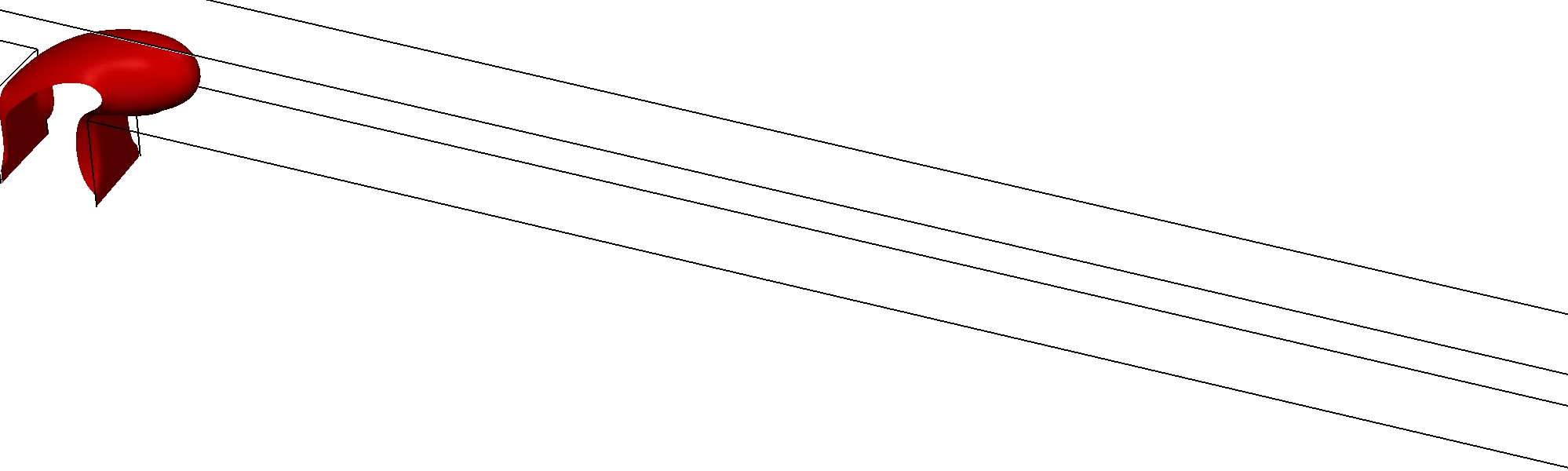}
}\\
\subfigure[{\scriptsize $t=t_0+3.8 \tshear$, $Q=1.0$, $\Ca = 0.026$}]
{
\includegraphics[width = 0.8\linewidth]{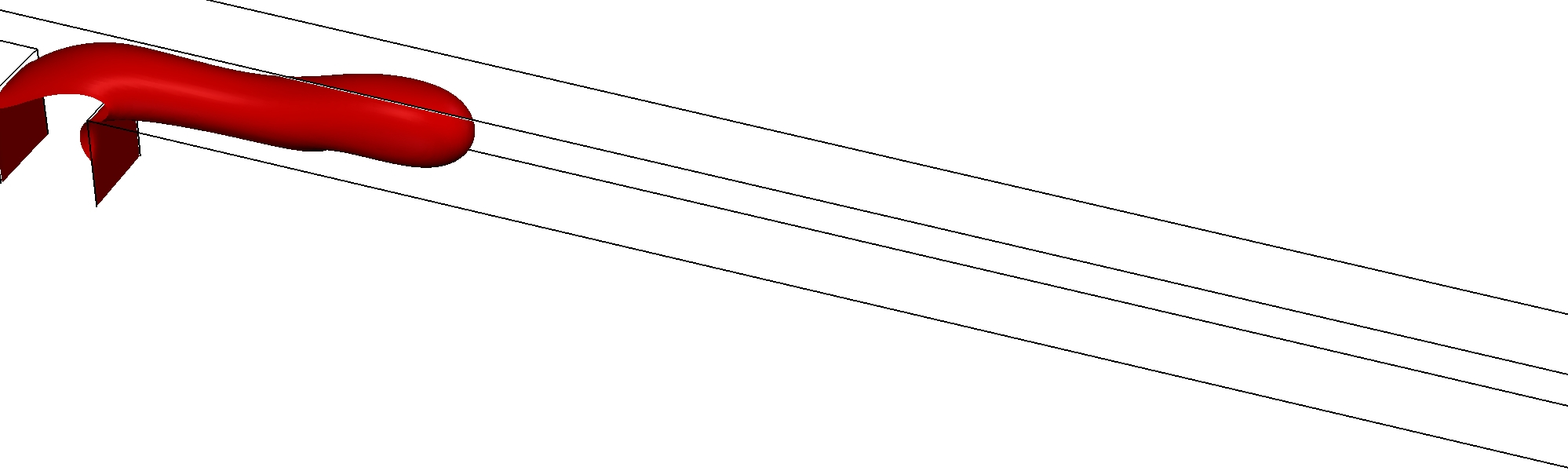}
}\\
\subfigure[{\scriptsize $t=t_0+5.2 \tshear$, $Q=1.0$, $\Ca = 0.026$}]
{
\includegraphics[width = 0.8\linewidth]{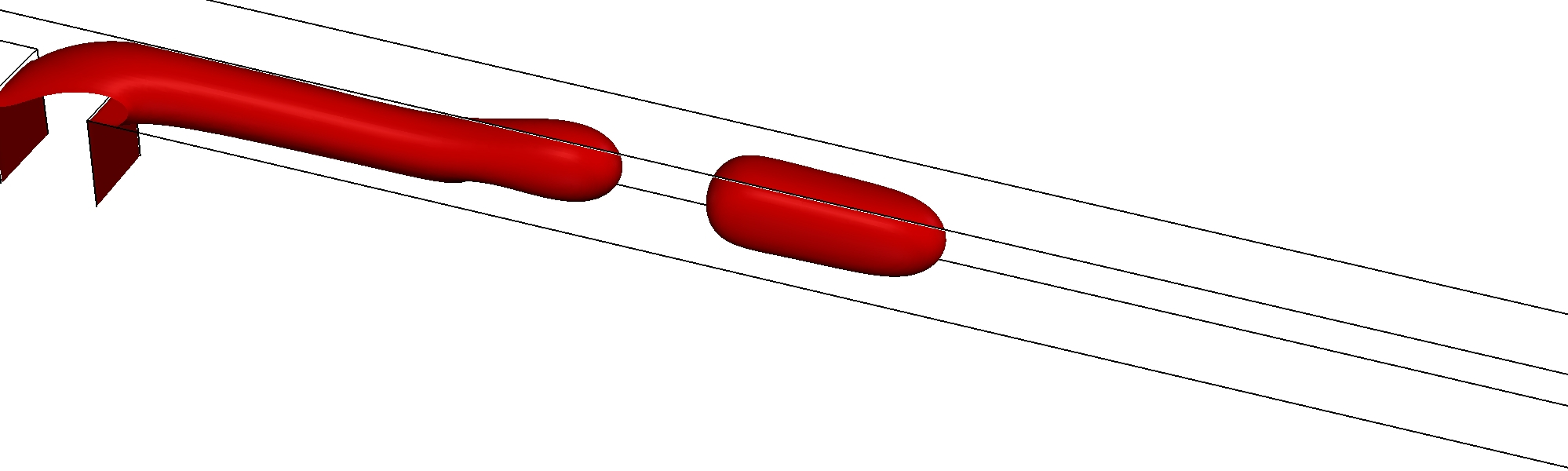}
}\\
\subfigure[{\scriptsize $t=t_0+6.6 \tshear$, $Q=1.0$, $\Ca = 0.026$}]
{
\includegraphics[width = 0.8\linewidth]{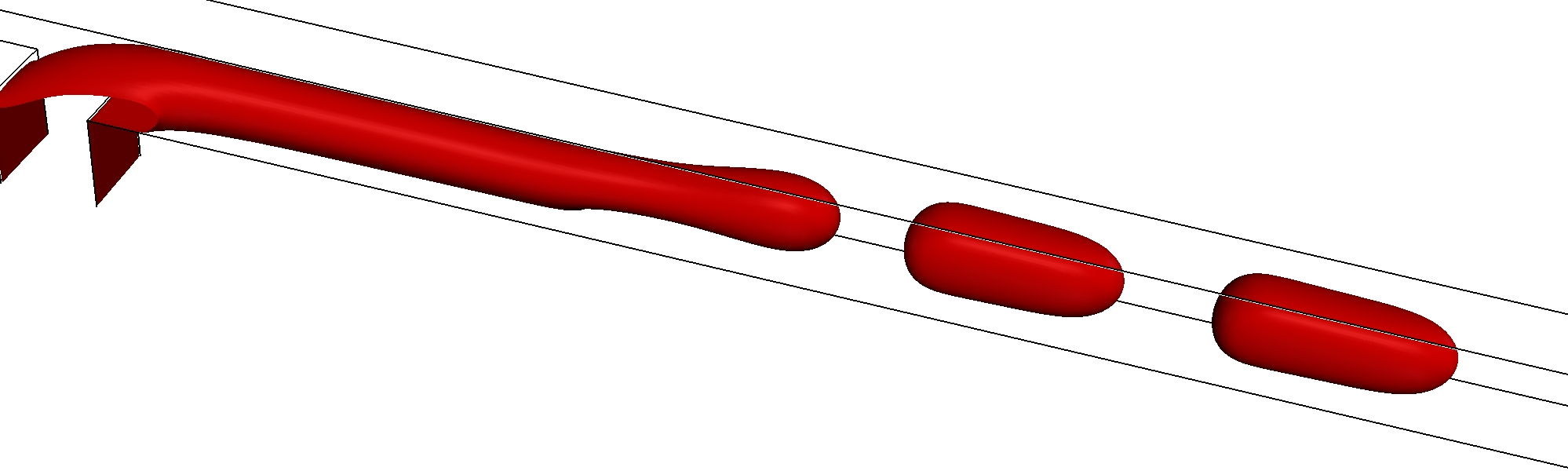}
}\\
\end{minipage}
\caption{Droplet formation in T-junction geometries for a Newtonian case with viscosity ratio $\lambda=1.0$. Panels (a)-(d): we illustrate the squeezing regime at $\Ca = 0.0026$ and flow-rate ratio $Q = 0.5$: the fluid thread enters and obstructs the main channel and break-up is mainly driven by the pressure build-up upstream of the emerging thread~\cite{Demenech07}. Both the dynamics of break-up and the scaling of the sizes of droplets are influenced weakly by viscous forces~\cite{Demenech07,Demenech06}. Panels (e)-(h) show typical features of the dripping regime at $\Ca = 0.013$ and $Q = 1.0$: the break-up process starts to be influenced by the shear forces, although the thread still occupy a significant portion of the main channel. This results in smaller droplets formed downstream of the T-junction. Panels (i)-(l) report snapshots from the jetting regime at larger Capillary number, $\Ca = 0.026$, and $Q = 1.0$: the dispersed phase develops a thread entering the main channel and the droplet detachment point gradually moves downstream, until a jet is formed. To better highlight the jetting regime, the associated figures display a larger portion of the main channel of the T-junction. In all cases we have used the characteristic shear time $\tau_{\mbox{\tiny{shear}}}=H/v_c$ as a unit of time, while $t_0$ is a reference time (the same for all simulations). \label{fig:00}}
\end{figure*}

In figure \ref{fig:00} we report 3D snapshots illustrating geometry mediated break-up in various scenarios depending on $\Ca$. These snapshots allow us to identify the various regimes which are known from the literature on droplet formation in confined T-junctions (see~\cite{Demenech07,LiuZhang09,LiuZhang11} and references therein): these will be used as ``reference'' Newtonian scenarios to quantify the importance of viscoelasticity. Notice that we have used the characteristic shear time $\tau_{\mbox{\tiny{shear}}}=H/v_c$ as a unit of time. At low $\Ca$ (Panels (a)-(d) of figure \ref{fig:00}), the incoming thread tends to occupy and obstruct the entire cross-section of the main channel, with the break-up occurring at the junction (Panel (d) in figure \ref{fig:00}). By increasing $\Ca$, a dripping scenario is entered (Panels (e)-(h) in figure \ref{fig:00}) in which the obstruction of the cross-section in the main channel is less visible and viscous shear forces start to influence the droplet break-up process immediately after the droplet enters into the main channel (Panel (f) in figure \ref{fig:00}). As a result of the combined effect of surface tension and viscous forces, smaller droplets are formed downstream of the T-junction (see Panels (g)-(h) in figure \ref{fig:00}). By further increasing $\Ca$, a critical value~\cite{Demenech07} exists above which the dispersed phase develops a thread entering the main channel and the droplet detachment point gradually moves downstream, until a jet is formed. The length of the jet is obviously limited by the size of the computational domain and simulations with large resolution are indeed necessary (see table \ref{table:para}) to make sure that the finite simulation domain does not play a role in the droplet formation inside the junction. A quantitative analysis on the influence of viscosity ratio and channel geometries on the above described physical scenarios has already been provided in the literature~\cite{Demenech07,LiuZhang09,LiuZhang11}. Here, instead, we aim to illustrate the effects of viscoelasticity. As already stressed in section \ref{sec:model}, our numerical approach offers the possibility to tune the viscosity ratio of the two Newtonian phases~\cite{SbragagliaGuptaScagliarini,SbragagliaGupta}. By fixing the polymer viscosity $\eta_P$, we can use such flexibility to achieve unitary viscosity ratio, defined in terms of the total (fluid + polymer) shear viscosity $\lambda=\eta_d/(\eta_c+\eta_{P})=1.0$ for MV and  $\lambda=(\eta_d+\eta_P)/\eta_c=1.0$ for DV. This will allow us to compare the non-Newtonian simulations with the corresponding Newtonian case at {\it the same} (unitary) viscosity ratio. We will explore systematically both the effects of the finite extensibility parameter $L^2$ and the polymer relaxation time $\tau_P$.

%%%%%%%%%%%%%%%%%%%%%%%%%%%%%%%%%%%%%%%%%%%%%%%%%%%%%%%%%%%%%%%%%%%%%%%%%%%%%%%%
%%%%%%%%%%%%%%%%%%%%%%%%%%%%%%%%%%%%FIG 2%%%%%%%%%%%%%%%%%%%%%%%%%%%%%%%%%%%%%%%%%%%%%%%%%%%%%%%%%%%%%%%%%%%%%%%%%%%%%%%%%%%%%%%%%%%%%%%%%%%%%%%%%%%%%%%%%%%%%%%

%%%%%%%%%%%%%%%%%%%%%%%%%%%%%%%%%%%%%%%%%%%%%%%%%%%%%%%%%%%%%%%%%%%%%%%%%%%%%%%%%%%%%%%%%%%%%%%%%%%%%%%%%%%%%%%%%%%%%%%%%%%%%%%%%%%%%%%%%%%%%%%%%%%%%%%%%%%%%%%%%%%%%%%%%%%%%%%%%%%%%%%%%%%%%%%%%%%%%%%%%%%%%%%%%%%%%%%%%%%%%%%%%%%%%%%%%%%%%%%%

\subsection{Squeezing Regime}\label{sec:squeezing}

%%%%%%%%%%%%%%%%%%%%%%%%%%%%%%%%%%%%%%%%%%%%%%%%%%%%%%%%%%%%%%%%%%%%%%%%%%%%%%%%%%%%%%%%%%%%%%%%%%%%%%%%%%%%%%%%%%%%%%FIG 3%%%%%%%%%%%%%%%%%%%%%%%%%%%%%%%%%%%%%%%%%%%%%%%%%%%%%%%%%%%%%%%%%%%%%%%%%%%%%%%%%%%%%%%%%%%%%%%%%%%%%%%%%%%%%%%%%%%%%

\begin{figure*}[th!]
\begin{center}
\begin{minipage}{0.2\textwidth}
\subfigure[{\scriptsize $Q=0.25$, $\Ca = 0.0026$}]
{
\includegraphics[width = 1.0\linewidth]{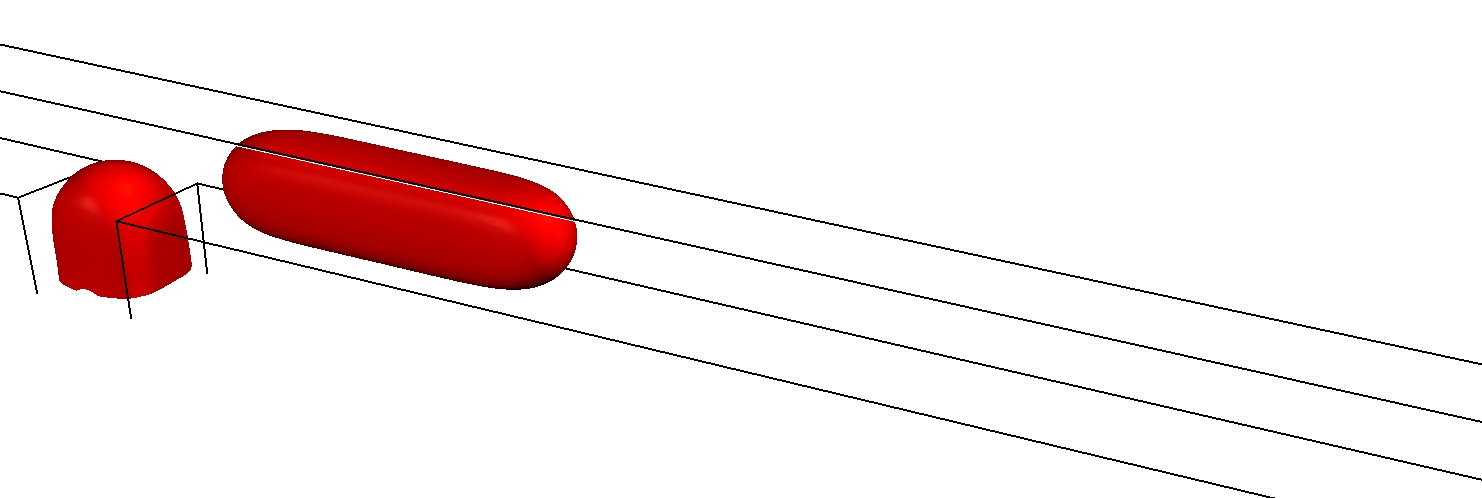}
}\\
\subfigure[{\scriptsize $Q=0.5$, $\Ca = 0.0026$}]
{
\includegraphics[width = 1.0\linewidth]{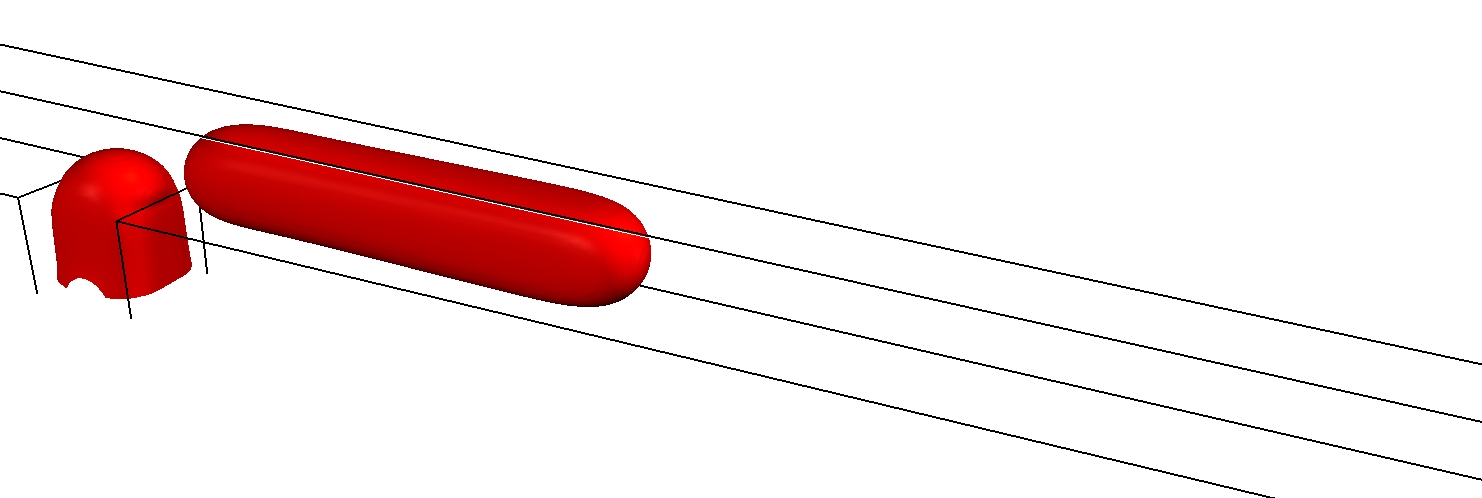}
}\\
\subfigure[{\scriptsize $Q=1.0$, $\Ca = 0.0026$}]
{
\includegraphics[width = 1.0\linewidth]{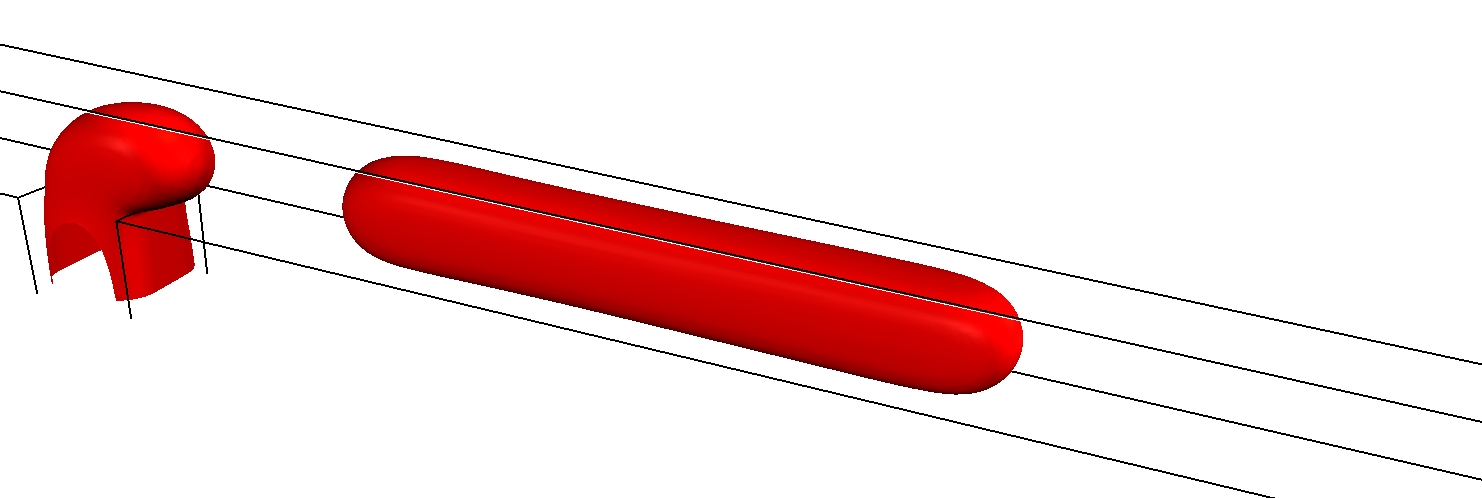}
}\\
\end{minipage}
\begin{minipage}{0.45\textwidth}
\subfigure[{\scriptsize}]
{
\includegraphics[width = 1.0\linewidth]{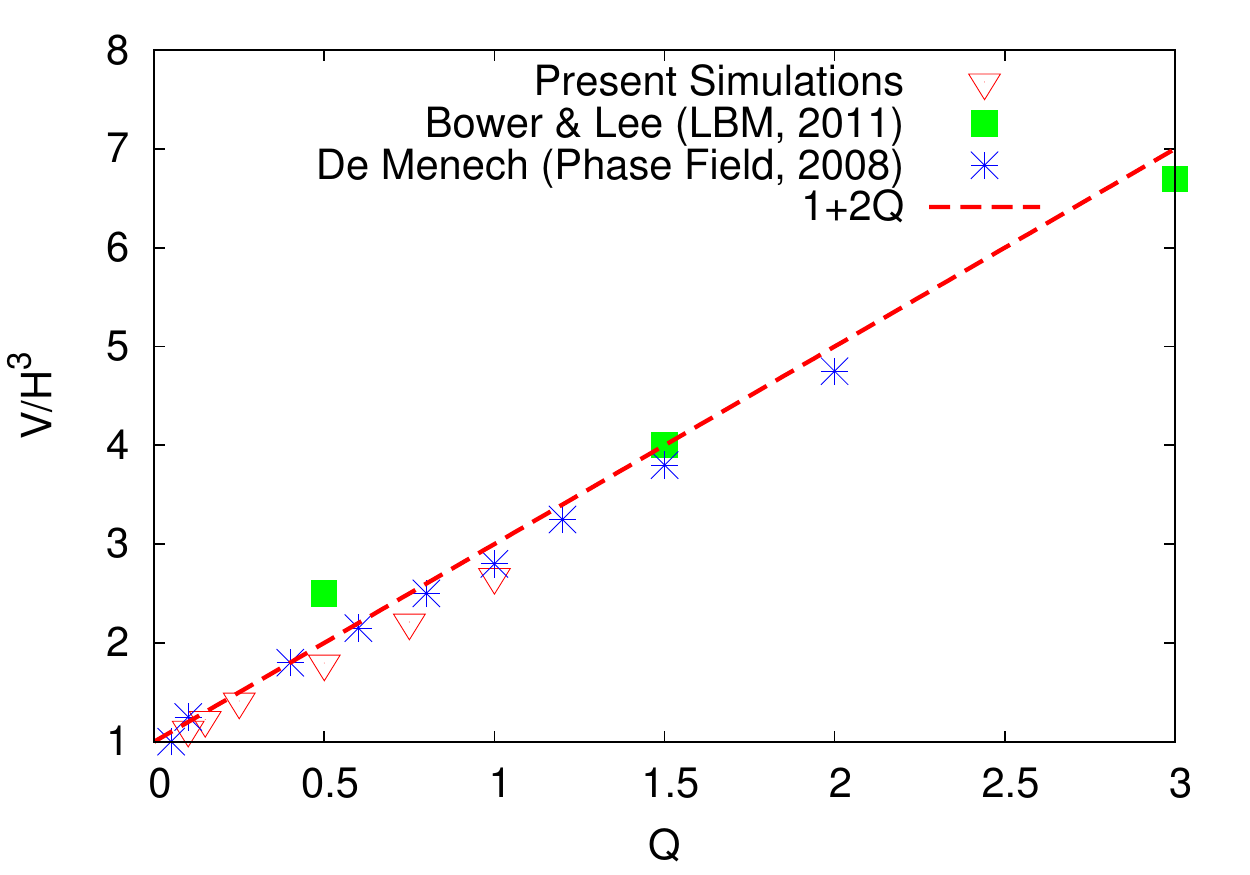}
}\\
\end{minipage}
\end{center}
\caption{Panels (a)-(c): Effect of the flow-rate ratio $Q$ in the squeezing regime with $\Ca = 0.0026$ and $\lambda=1.0$. In Panel (d) we report the dimensionless droplet volume as a function of the flow-rate ratio $Q$. Our data are compared with the phase field numerical simulations of De Menech {\it et al.}~\cite{Demenech07} and the LBM simulations of Bower \& Lee~\cite{BowerLee11}. Superimposed we report the linear fit predicted by Garstecki {\it et al.}~\cite{Garstecki06} (see Eq. (\ref{eq:Garstecki})), based on the assumption that the droplet size is greatly determined by the ratio of the volumetric flow-rates of the two immiscible fluids. Notice that the numerical simulations of Bower \& Lee~\cite{BowerLee11} are performed with a viscosity ratio $\lambda=0.02$ which differs from ours. However, in the squeezing regime good agreement is still found, since the droplet size is greatly affected by $Q$ and little effect is expected from a change in the fluid properties (i.e. change in $\lambda$). To test the robustness of our findings at changing the channel dimensionality, we repeated the numerical simulations in a 2d channel with viscosity ratio $\lambda=0.05$ (see also section \ref{sec:drippingjetting} for discussions). \label{fig:bench}}
\end{figure*}

%%%%%%%%%%%%%%%%%%%%%%%%%%%%%%%%%%%%%%%%%%%%%%%%%%%%%%%%%%%%%%%%%%%%%%%%%%%%%%%%
%%%%%%%%%%%%%%%%%%%%%%%%%%%%%%%%%%%%%%%%%%%%%%%%%%%%%%%%%%%%%%%%%%%%%%%%%%%%%%%%
%%%%%%%%%%%%%%%%%%%%%%%%%%%%%%%%%%%%%%%%%%%%%%%%%%%%%%%%%%%%%%%%%%%%%%%%%%%%%%%%

%%%%%%%%%%%%%%%%%%%%%%%%%%%%%%%%%%%%%%%%%%%%%%%%%%%%%%%%%%%%%%%%%%%%%%%%%%%%%%%%%%%%%%%%%%%%%%%%%%%%%%%%%%%%%%%%%%%%%%FIG 4%%%%%%%%%%%%%%%%%%%%%%%%%%%%%%%%%%%%%%%%%%%%%%%%%%%%%%%%%%%%%%%%%%%%%%%%%%%%%%%%%%%%%%%%%%%%%%%%%%%%%%%%%%%%%%%%%%%%%

\begin{figure*}[t!]
\subfigure[{\scriptsize Matrix Viscoelasticity (MV), $\Ca = 0.0026$}]
{
\includegraphics[width = 0.475\linewidth]{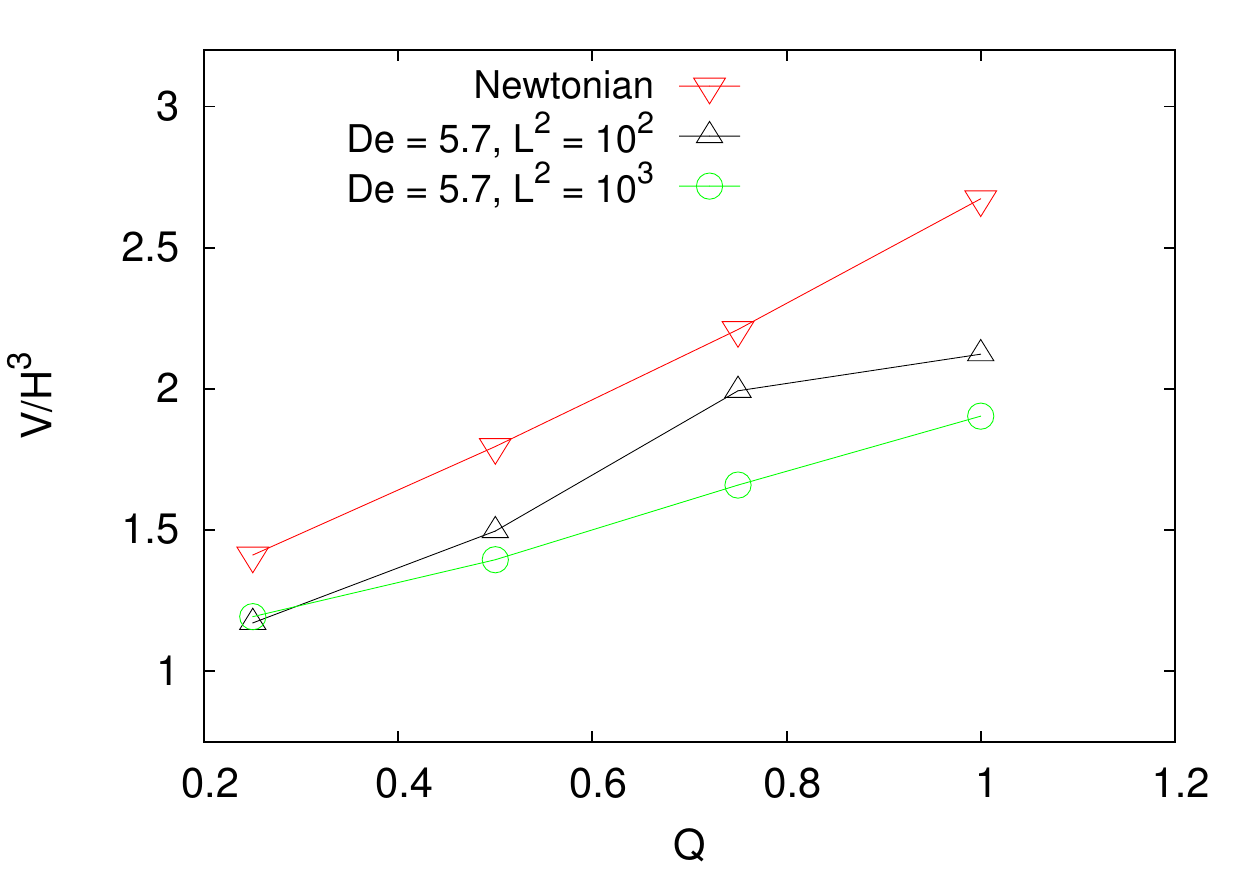}
}
\subfigure[{\scriptsize Droplet Viscoelasticity (DV), $\Ca = 0.0026$}]
{
\includegraphics[width = 0.475\linewidth]{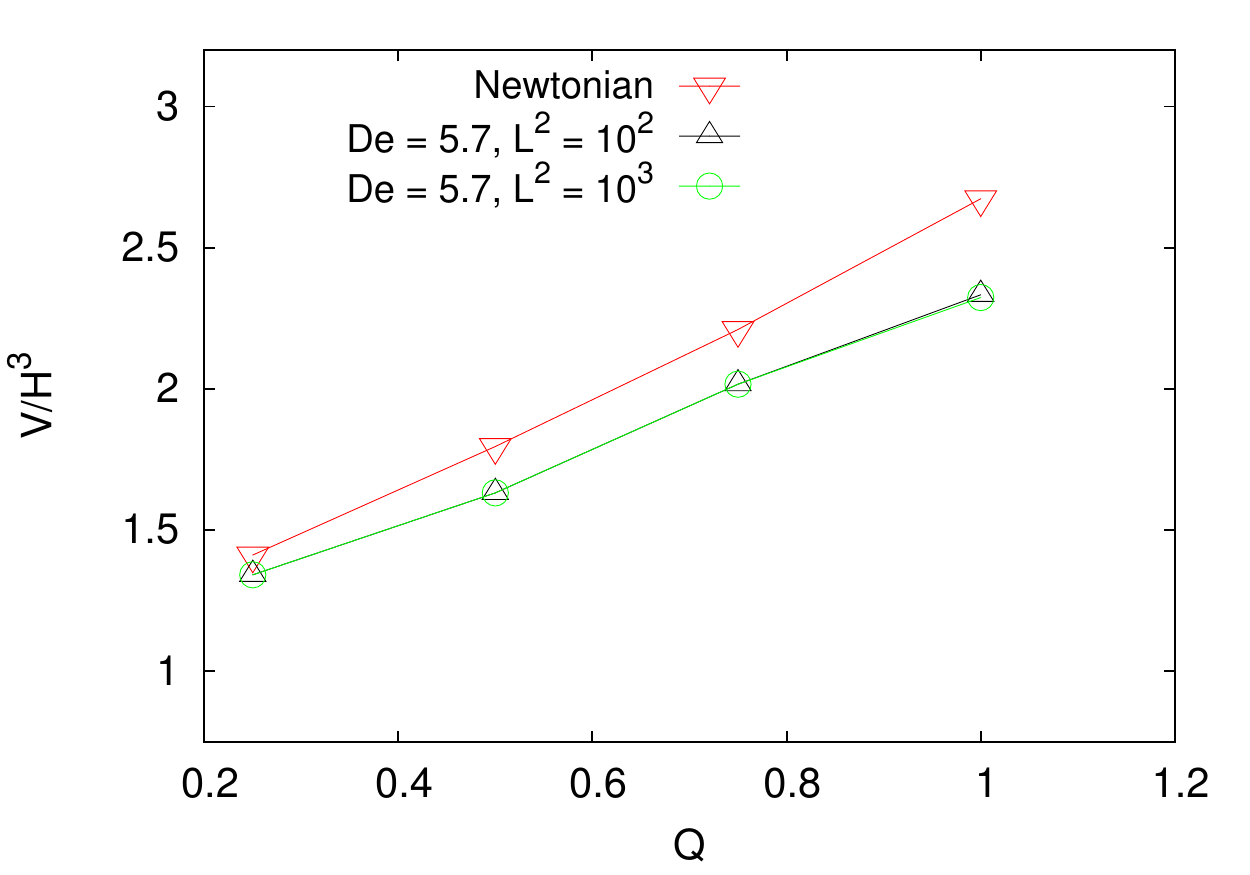}
}
\caption{Quantitative analysis of the break-up process in the squeezing regime at $\Ca=0.0026$. We report the dimensionless droplet volume $V/H^3$ soon after break-up for a case with matrix viscoelasticity (MV) and droplet viscoelasticity (DV). We choose the flow-rate ratio $Q$ and finite extensibility parameter $L^2$ ranging in the interval $Q=0.2-1.0$ and $L^2=10^2-10^3$, respectively. For the non-Newtonian cases, the polymer relaxation time has been kept fixed to $\tau_P=4000$ lbu, corresponding to a Deborah number $\De=5.7$, based on definition \eqref{Desimple}. Data for different $\tau_P$ at fixed flow-rate ratio $Q = 1.0$ are reported in figure \ref{fig:5-6}. \label{fig:1-4}}
\end{figure*}

%%%%%%%%%%%%%%%%%%%%%%%%%%%%%%%%%%%%%%%%%%%%%%%%%%%%%%%%%%%%%%%%%%%%%%%%%%%%%%%%
%%%%%%%%%%%%%%%%%%%%%%%%%%%%%%%%%%%%%%%%%%%%%%%%%%%%%%%%%%%%%%%%%%%%%%%%%%%%%%%%
%%%%%%%%%%%%%%%%%%%%%%%%%%%%%%%%%%%%%%%%%%%%%%%%%%%%%%%%%%%%%%%%%%%%%%%%%%%%%%%%

%%%%%%%%%%%%%%%%%%%%%%%%%%%%%%%%%%%%%%%%%%%%%%%%%%%%%%%%%%%%%%%%%%%%%%%%%%%%%%%%%%%%%%%%%%%%%%%%%%%%%%%%%%%%%%%%%%%%%%FIG 5%%%%%%%%%%%%%%%%%%%%%%%%%%%%%%%%%%%%%%%%%%%%%%%%%%%%%%%%%%%%%%%%%%%%%%%%%%%%%%%%%%%%%%%%%%%%%%%%%%%%%%%%%%%%%%%%%%%%%

\begin{figure*}[t!]
\begin{center}
\includegraphics[width = 0.5\linewidth]{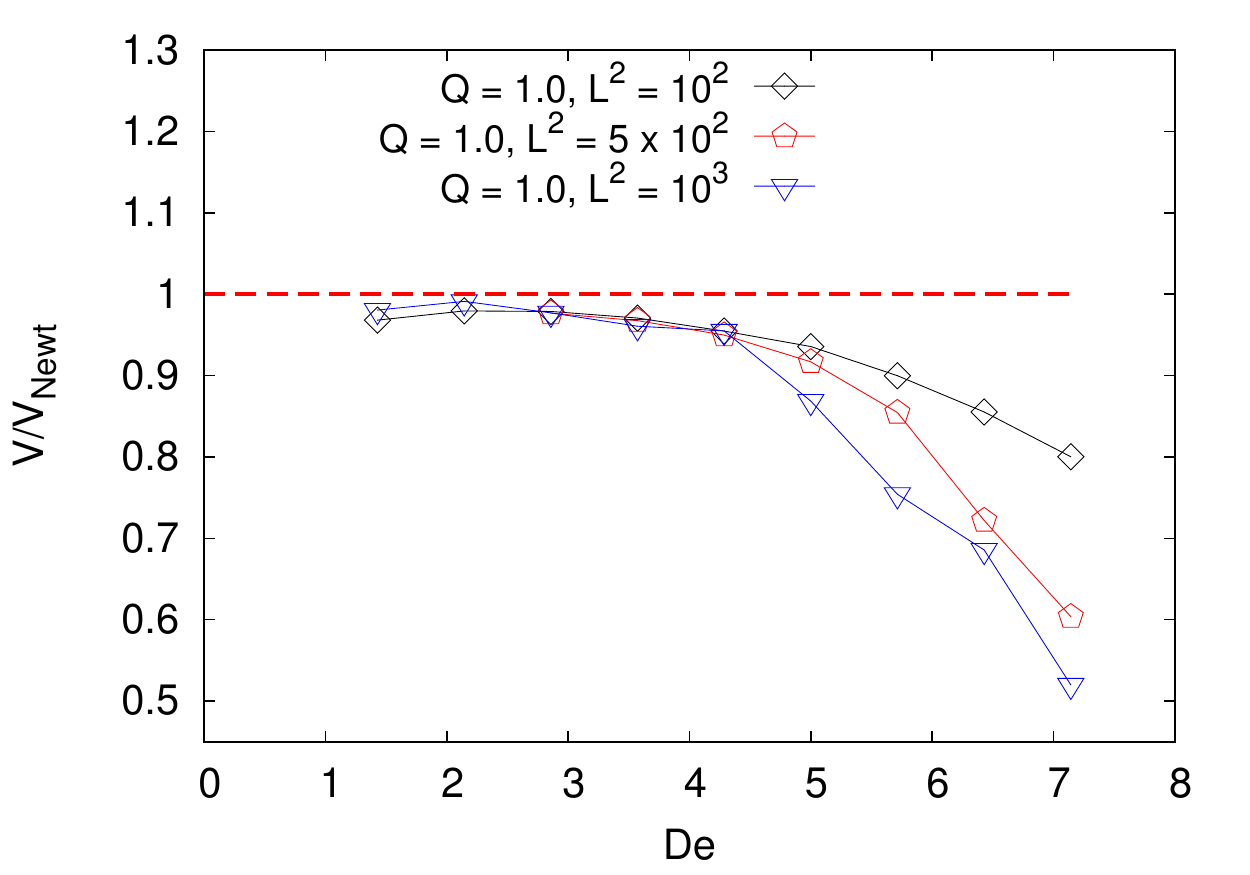}
\caption{Quantitative analysis of the break-up process in the squeezing regime at $\Ca=0.0026$. We report the dimensionless droplet volume $V/V_{\mbox{\tiny{Newt}}}$ soon after break-up for a case with matrix viscoelasticity (MV). The droplet volume has been made dimensionless with respect to the Newtonian volume ($V_{\mbox{\tiny{Newt}}}$) for the same $\Ca$. The finite extensibility parameter $L^2$ and the polymer relaxation time $\tau_P$ are ranging in the interval $L^2=10^2-10^3$ and $\tau_P =250-4000$ lbu, respectively. Correspondingly, the Deborah number \eqref{Desimple} is reported. In all cases, the flow-rate ratio $Q$ and the viscosity ratio between the two fluids have been kept fixed to $Q=\lambda=1.0$. \label{fig:5-6}}
\end{center}
\end{figure*}

%%%%%%%%%%%%%%%%%%%%%%%%%%%%%%%%%%%%%%%%%%%%%%%%%%%%%%%%%%%%%%%%%%%%%%%%%%%%%%%%
%%%%%%%%%%%%%%%%%%%%%%%%%%%%%%%%%%%%%%%%%%%%%%%%%%%%%%%%%%%%%%%%%%%%%%%%%%%%%%%%
%%%%%%%%%%%%%%%%%%%%%%%%%%%%%%%%%%%%%%%%%%%%%%%%%%%%%%%%%%%%%%%%%%%%%%%%%%%%%%%%

%%%%%%%%%%%%%%%%%%%%%%%%%%%%%%%%%%%%%%%%%%%%%%%%%%%%%%%%%%%%%%%%%%%%%%%%%%%%%%%%
%%%%%%%%%%%%%%%%%%%%%%%%%%%%%%%%FIG 6%%%%%%%%%%%%%%%%%%%%%%%%%%%%%%%%%%%%%%%%%%%
\begin{figure*}[th!]
\begin{center}
\subfigure[{\scriptsize $t=t_0+3.9 \tshear$, $L^2 = 10^2$}]
{
\includegraphics[width = 0.54\linewidth]{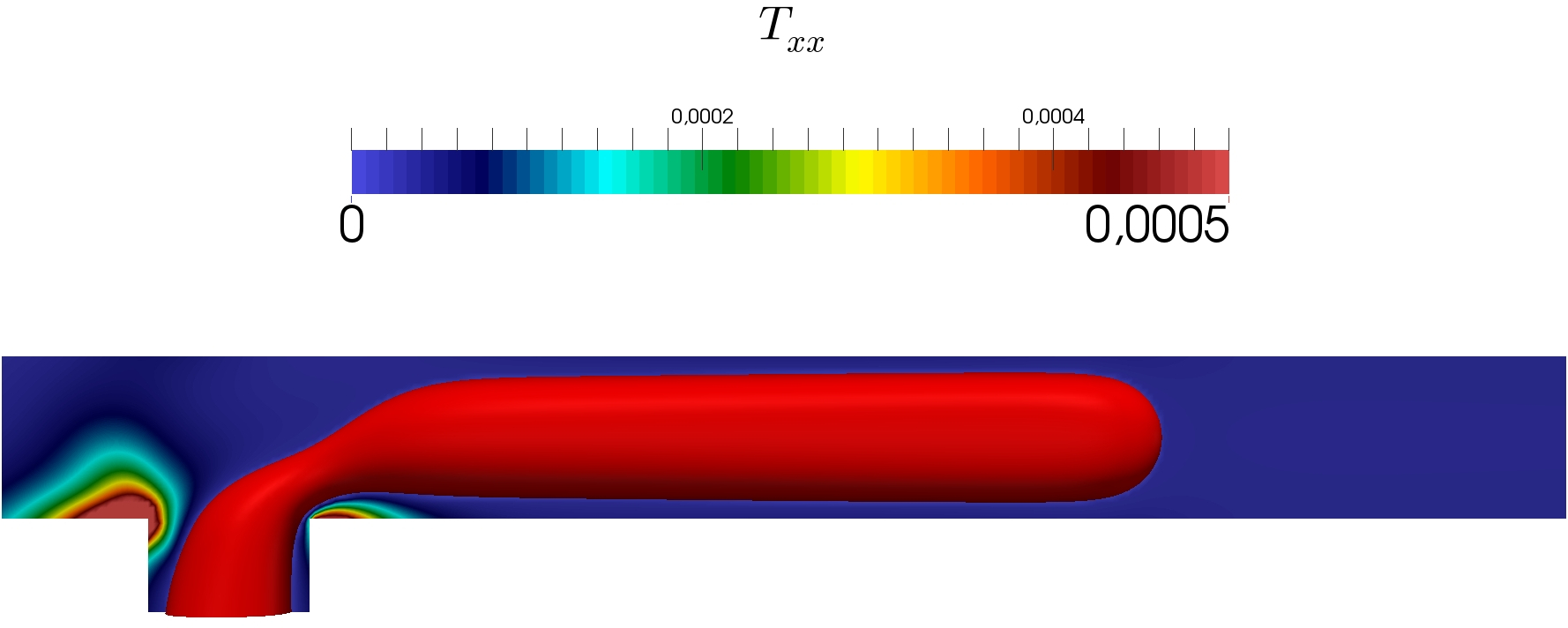}
}\\
\subfigure[{\scriptsize $t=t_0+3.6 \tshear$, $L^2 = 5 \times 10^2$}]
{
\includegraphics[width = 0.54\linewidth]{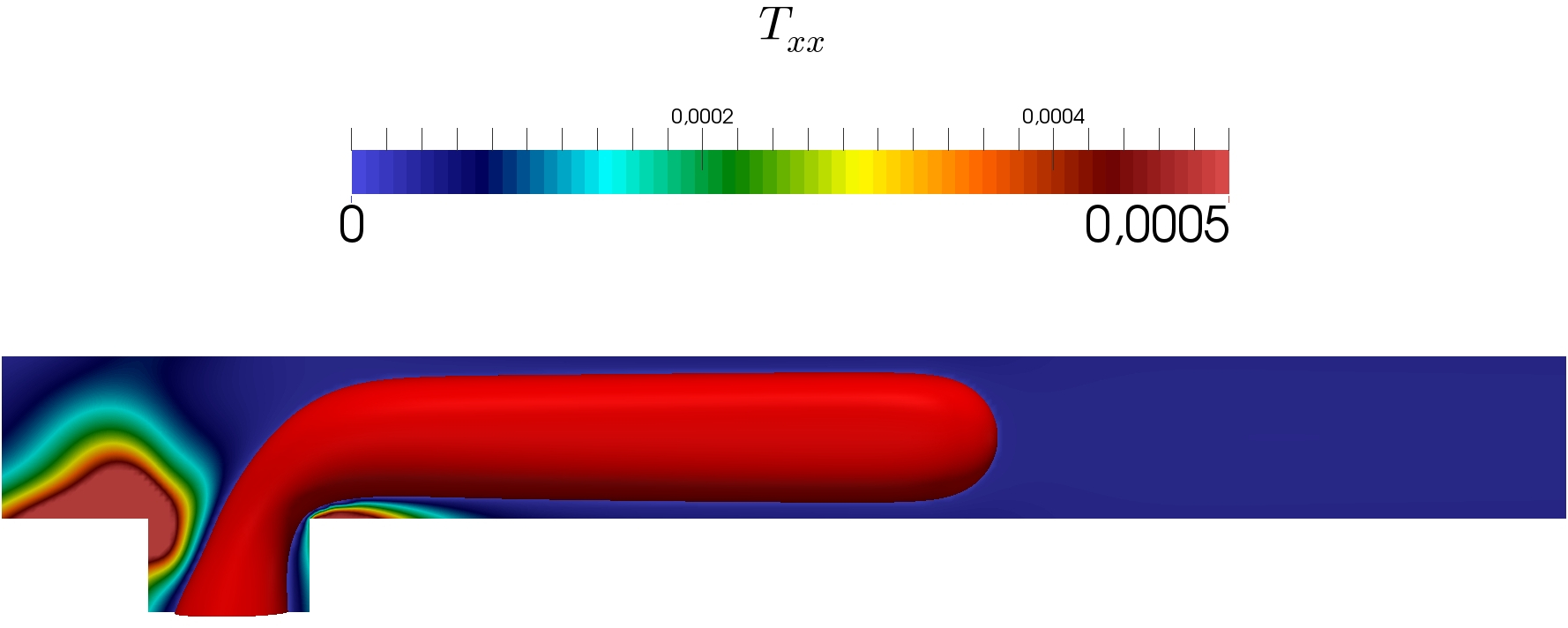}
}\\
\subfigure[{\scriptsize $t=t_0+3.4 \tshear$, $L^2 = 10^3$}]
{
\includegraphics[width = 0.54\linewidth]{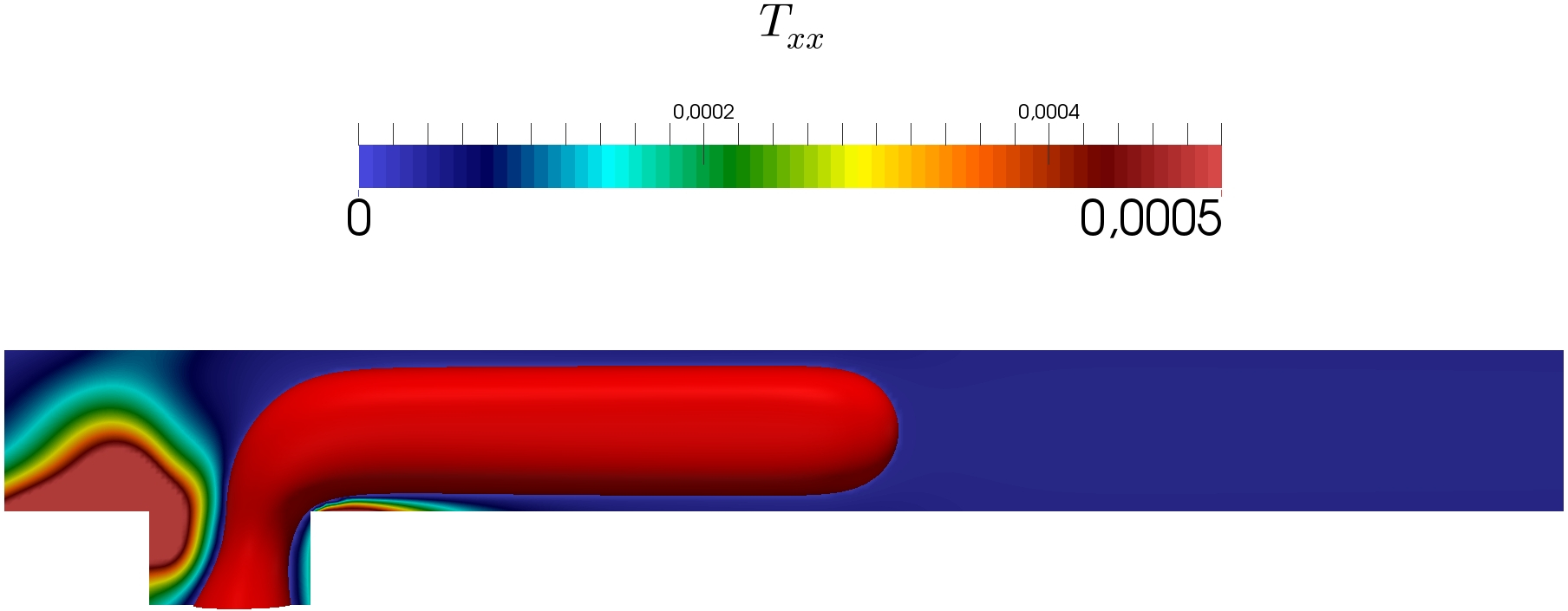}
}
\caption{Panels (a)-(c): density contours of the dispersed phase overlaid on the polymer feedback stress in the stream-flow direction (see Eqs. \eqref{NSc} and \eqref{eq:Txx}) for a case with matrix viscoelasticity (MV) with three different $L^2$: $L^2 = 10^2$ (Panel (a)), $L^2 = 5 \times 10^2$ (Panel (b)) and $L^2 = 10^3$ (Panel (c)). All the other parameters are kept fixed: $\De=5.7$, $\Ca=0.0026$, $\lambda=1.0$ and $Q=1.0$. As $L^2$ is increased, we see that the flow in the continuous phase develops enhanced polymer feedback stresses upstream of the emerging thread. In all cases we have used the characteristic shear time $\tau_{\mbox{\tiny{shear}}}=H/v_c$ as a unit of time, while $t_0$ is a reference time (the same for all simulations). Notice that the colorbar of the feedback stress \eqref{eq:Txx} is the same. \label{fig_choreography}}
\end{center}
\end{figure*}

%%%%%%%%%%%%%%%%%%%%%%%%%%%%%%%%%%%%%%%%%%%%%%%%%%%%%%%%%%%%%%%%%%%%%%%%%%%%%%%%
%%%%%%%%%%%%%%%%%%%%%%%%%%%%%%%%%%%%%%%%%%%%%%%%%%%%%%%%%%%%%%%%%%%%%%%%%%%%%%%%

%%%%%%%%%%%%%%%%%%%%%%%%%%%%%FIG 12%%%%%%%%%%%%%%%%%%%%%%%%%%%%%%%%%%%%%%%
%%%%%%%%%%%%%%%%%%%%%%%%%%%%%%%%%%%%%%%%%%%%%%%%%%%%%%%%%%%%%%%%%%%%%%%%%%

\begin{figure*}[t!]
\subfigure[{\scriptsize $L^2=100$, $\Ca = 0.0026$ }]
{
\includegraphics[width = 0.475\linewidth]{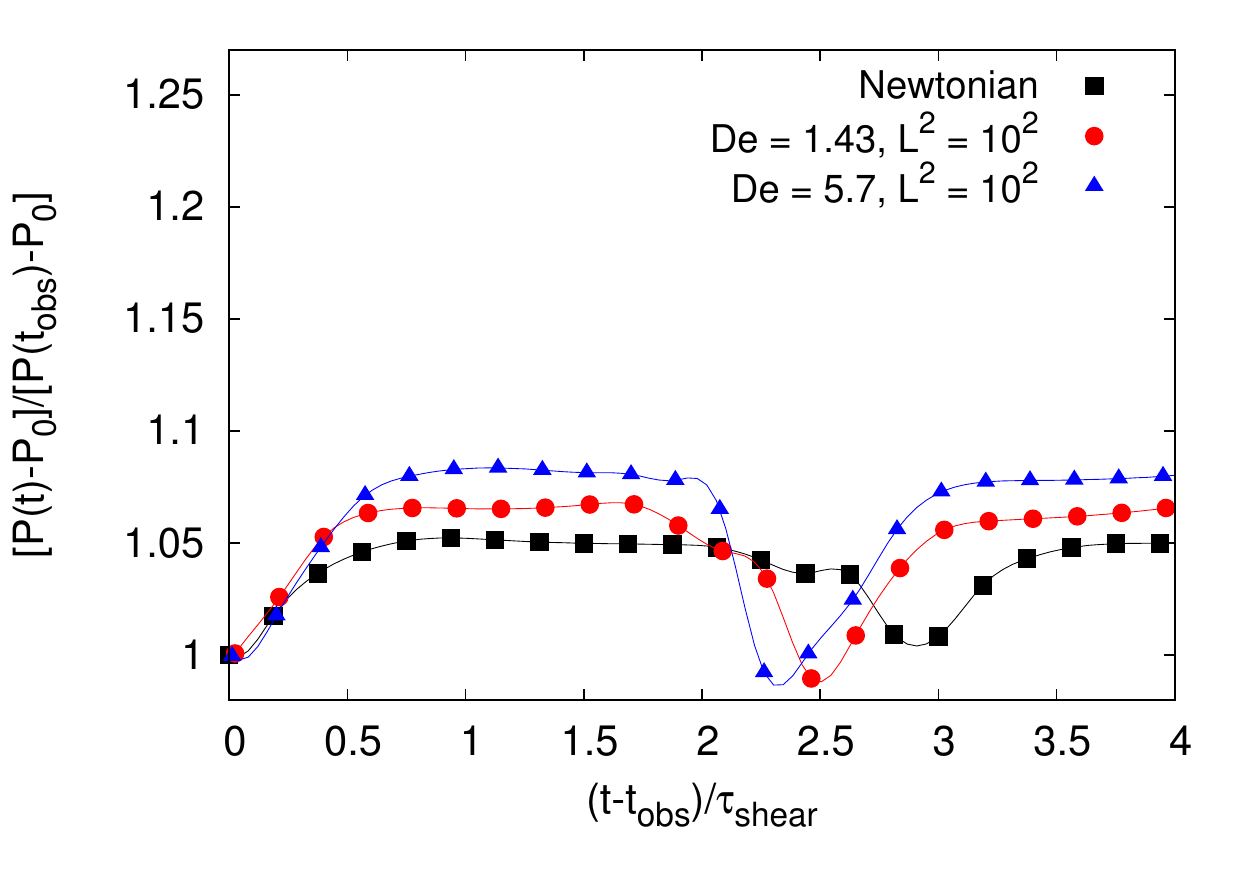}
}
\subfigure[{\scriptsize $L^2=1000$, $\Ca = 0.0026$ }]
{
\includegraphics[width = 0.475\linewidth]{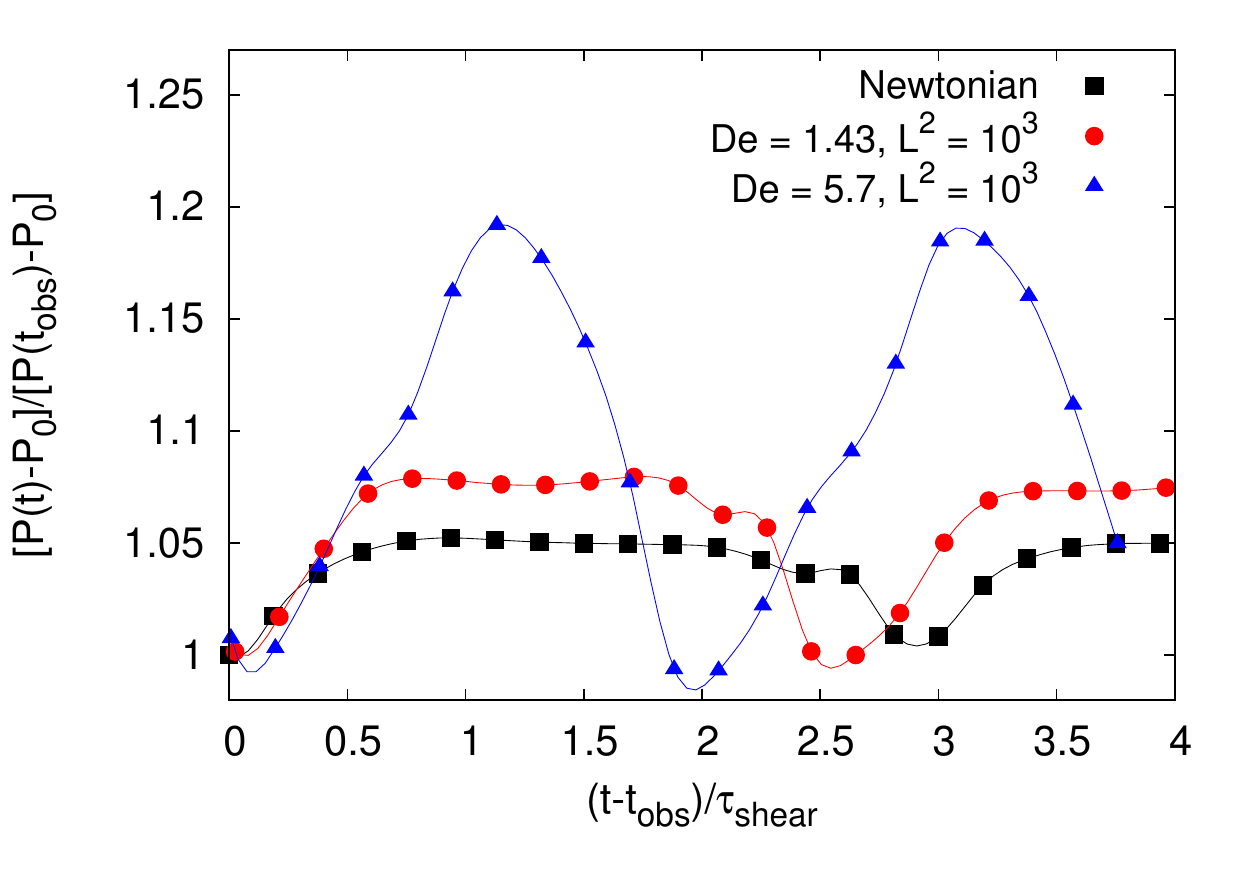}
}
\caption{Analysis of the pressure ($P$) versus time in the squeezing regime. The pressure is computed upstream of the T-junction and $P_0$ is a constant reference pressure computed in the static case. Panel (a): we report the normalized pressure versus time with fixed  $\Ca = 0.0026$ for the Newtonian case (black squares) and two cases with matrix viscoelasticity (MV) at fixed $L^2 = 100$: $\De=1.43$ (red circles) and $\De=5.7$ (blue triangles). Panel (b): same as panel (a) with $L^2 = 10^3$. In all cases we have used the characteristic shear time $\tau_{\mbox{\tiny{shear}}}=H/v_c$ as a unit of time, while $\tobs$ is the time when the thread starts to obstruct the channel. \label{fig_pressure}}
\end{figure*}

%%%%%%%%%%%%%%%%%%%%%%%%%%%%%%%%%%%%%%%%%%%%%%%%%%%%%%%%%%%%%%%%%%%%%%%%%%%%%%%%%%%%%%%%%%%%%%%%%%%%%%%%%%%%%%%%%%%%%%%%%%%%%%%%%%%%%%%%%%%%%%%%%%%%%%%%%%%%%%%%

%the effective force has a positive stream-flow component (panel (a) figure \ref{fig_effectiveforce}) which then changes sign closer to the droplet interface. 

%%%%%%%%%%%%%%%%%%%%%%%%%%%%%%%%%%%%%%%%%%%%%%%%%%%%%%%%%%%%%%%%%%%%%%%%%%%%%%%%%%%%%%%%%%%%%%FIGURA STREAM%%%%%%%%%%%%%%%%%%%%%%%%%%%%%%%%%%%%%%%%%%%%%%%%%%%%%

\begin{figure*}[t!]
\begin{center}
\subfigure[{\scriptsize $t=t_0+0.38 \tshear$, $\Ca = 0.0052$, $\De = 5.7$, $L^2 = 10^3$}]
{
\includegraphics[width = 0.48\linewidth]{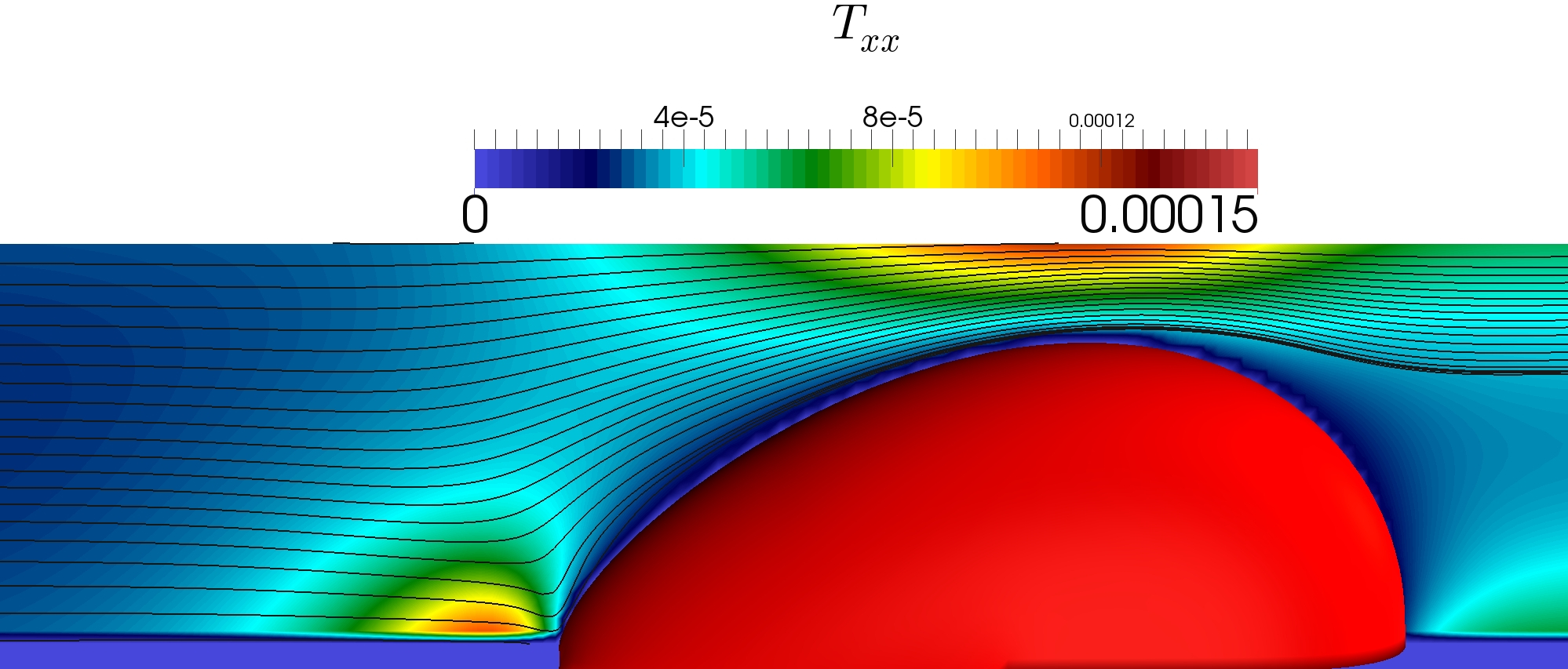}
\includegraphics[width = 0.5\linewidth]{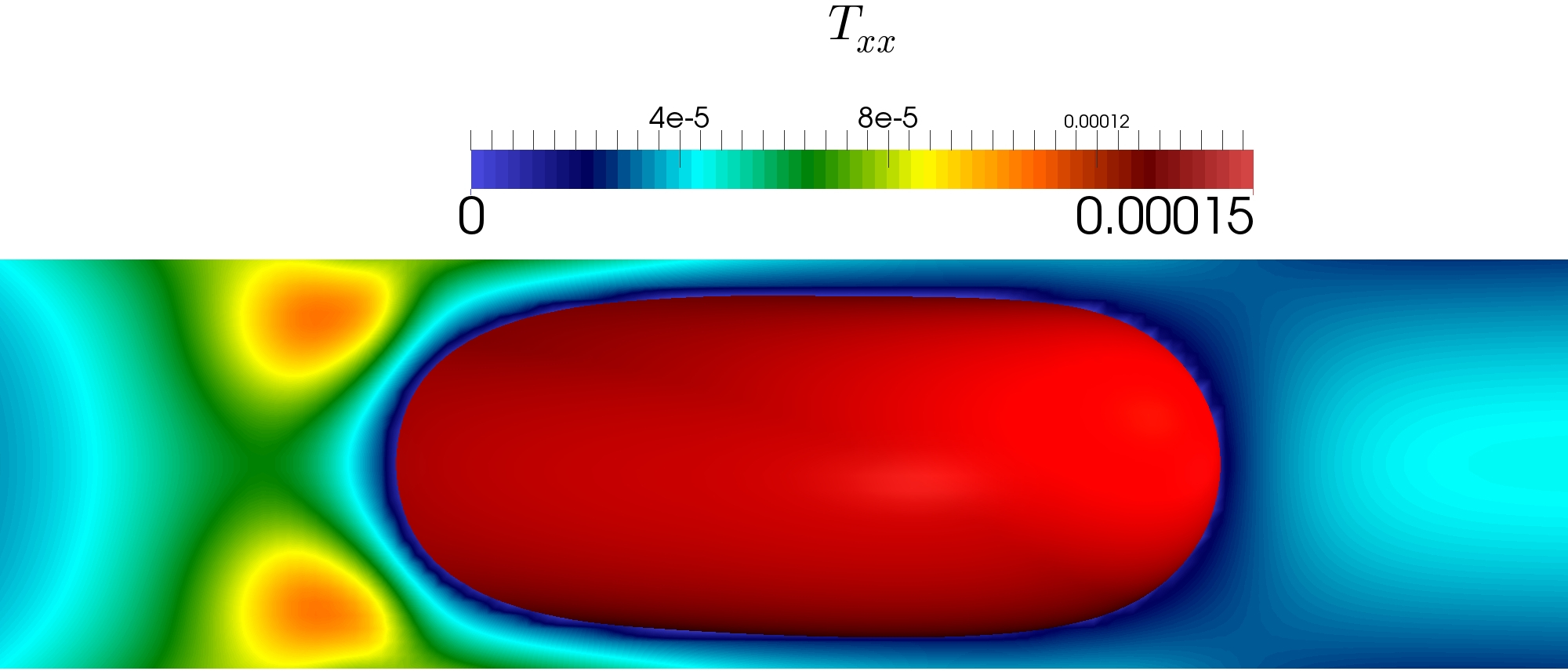}
}
\caption{Left Panel: Velocity streamlines (black lines) overlaid on the polymer feedback stress in the streamflow direction (see Eqs. \eqref{NSc} and \eqref{eq:Txx}) for a case with matrix viscoelasticity (MV). The obstruction provided by the thread forces the flow to converge into the gap, hence triggering an extensional response in the fluid region upstream of the emerging thread. Right Panel: a top view of the polymer feedback stress at a distance $\approx H/6$ from the bottom wall of the main channel.  \label{fig_stream}}
\end{center}
\end{figure*}

%%%%%%%%%%%%%%%%%%%%%%%%%%%%%%%%%%%%%%%%%%%%%%%%%%%%%%%%%%%%%%%%%%%%
%%%%%%%%%%%%%%%%%%%%%%%%%%%%%%%%%%%%%%%%%%%%%%%%%%%%%%%%%%%%%%%%%%%%

%%%%%%%%%%%%%%%%%%%%%%%%%%%%%%%%%%%%%%%%%%%%%%%%%%%%%%%%%%%%%%%%%%%%%%%%%%%%%%%%
%%%%%%%%%%%%%%%%%%%%%%%%%%%%%%%%%%%%%FIG 7%%%%%%%%%%%%%%%%%%%%%%%%%%%%%%%%%%%%%%%%%%%

\begin{figure*}[t!]
\begin{center}
\subfigure{\includegraphics[width = 0.35\linewidth]{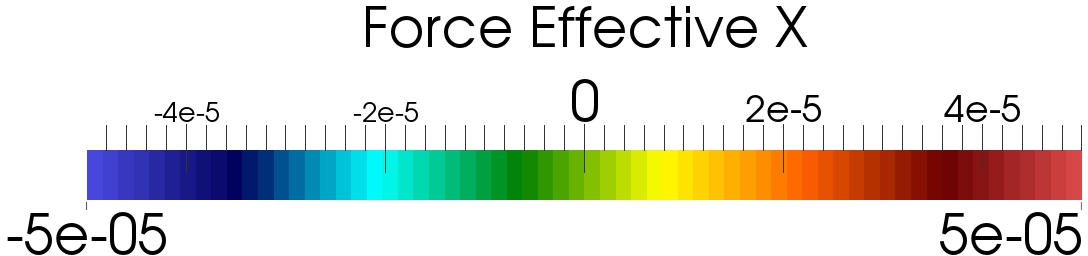}}
\hspace{3.5cm}
\subfigure{\includegraphics[width = 0.35\linewidth]{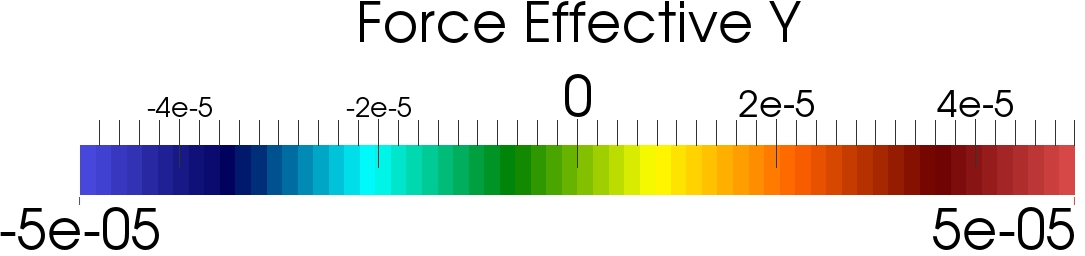}}\\\setcounter{subfigure}{0}% Reset subfigure counter
\subfigure[{\scriptsize $t=t_0+3.4 \tshear$, $\De = 3.1$, $L^2 = 5\times 10^3$}]
{
\includegraphics[width = 0.43\linewidth]{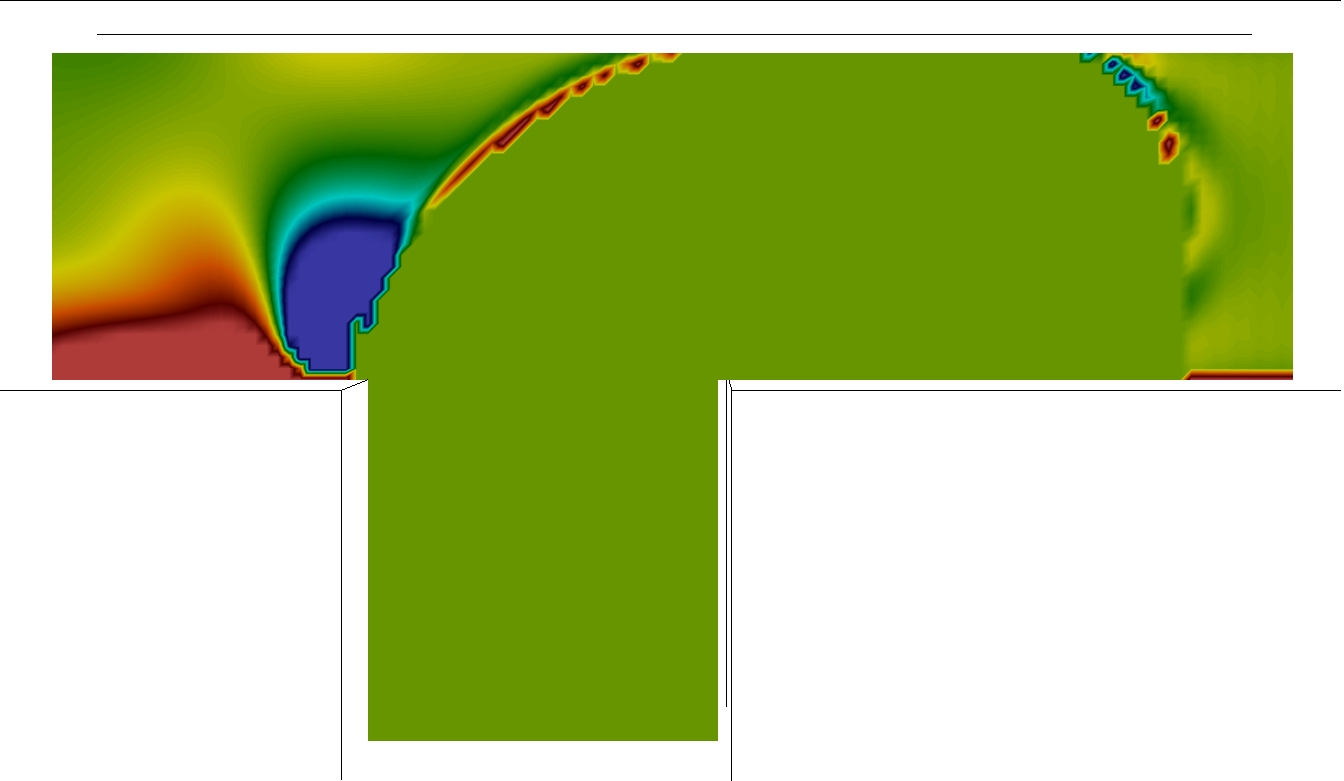}
}
\hspace{2.0cm}
\subfigure[{\scriptsize $t=t_0+3.4 \tshear$, $\De = 3.1$, $L^2 = 5\times 10^3$}]
{
\includegraphics[width = 0.43\linewidth]{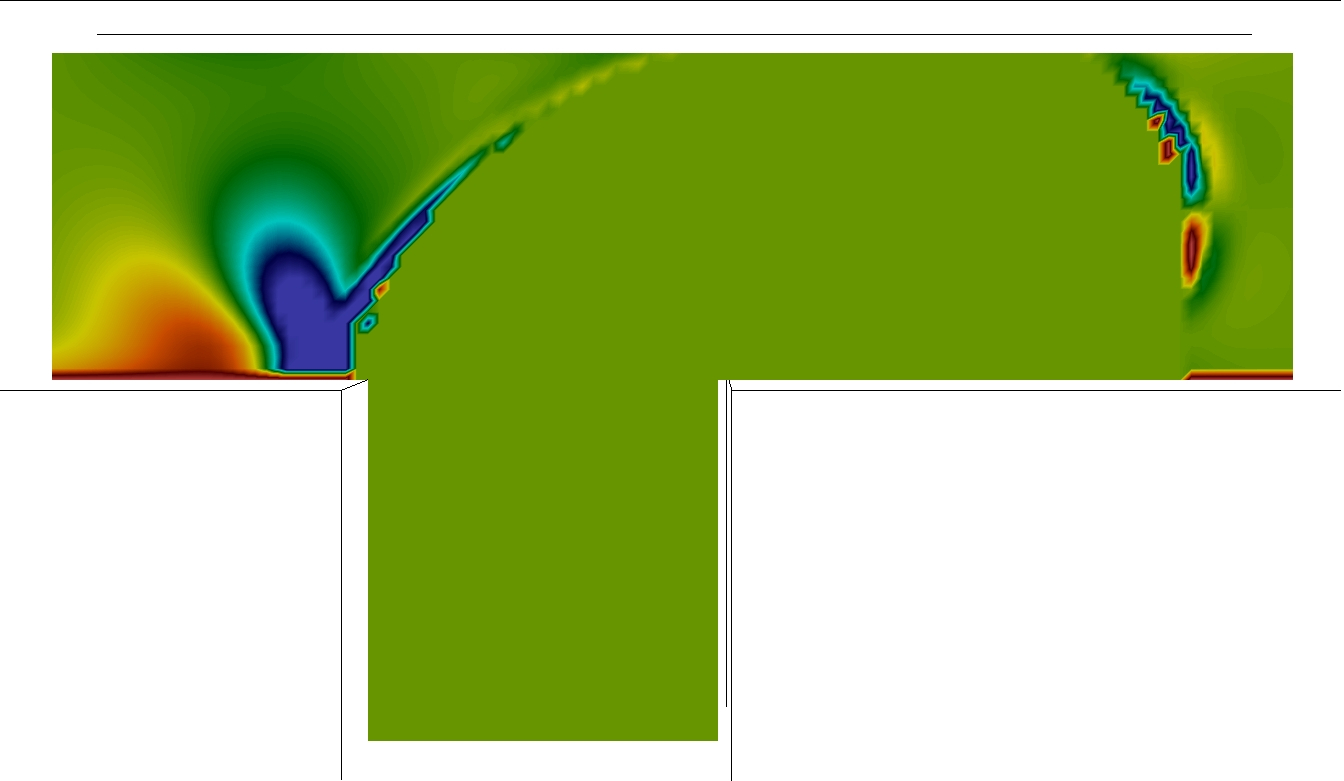}
}\\
\caption{Panels (a)-(b): x and y component of the effective force ${\bm F}_{\mbox{\tiny{eff}}}$ (see Eq. \eqref{eq:effectiveforce}) for a matrix viscoelasticity (MV) case at $t=t_0+3.4 \tshear$, $\De = 3.1$, $L^2 = 5\times 10^3$, $\Ca=0.0026$, $\lambda=1.0$ and $Q=1.0$. We have used the characteristic shear time $\tau_{\mbox{\tiny{shear}}}=H/v_c$ as a unit of time, while $t_0$ is a reference time (the same for all simulations). \label{fig_effectiveforce}}
\end{center}
\end{figure*}

%%%%%%%%%%%%%%%%%%%%%%%%%%%%%%%%%%%%%%%%%%%%%%%%%%%%%%%%%%%%%%%%%%%%%%%%%%%%%%%%
%%%%%%%%%%%%%%%%%%%%%%%%%%%%%%%%%%%%%%%%%%%%%%%%%%%%%%%%%%%%%%%%%%%%%%%%%%%%%%%%

%%%%%%%%%%%%%%%%%%%%%%%%%%%%%%%%%%%%%%%%%%%%%%%%%%%%%%%%%%%%%%%%%%%%%%%%%%%%%%%%
%%%%%%%%%%%%%%%%%%%%%%%%%%%%%%%%%%%%%%%%%%%%%%%%%%%%%%%%%%%%%%%%%%%%%%%%%%%%%%%%

To go deeper and be more quantitative on the characterization of the various regimes, we start by investigating the droplet size as a function of the flow-rate ratio $Q$ in the squeezing regime. The characteristic droplet size $L_d$ in the squeezing regime is only weekly affected by the viscosity ratio and mainly determined by the ratio of the volumetric flow-rates of the two immiscible fluids as
\be\label{eq:Garstecki}
L_d=\alpha_1+\alpha_2 \frac{Q_{d}}{Q_{c}}=\alpha_1+\alpha_2 Q.
\ee
The constants $\alpha_1$ and $\alpha_2$, which are of the order one, are determined by the channel geometry~\cite{LiuZhang11}. The linear scaling law \eqref{eq:Garstecki} has already been verified in experiments~\cite{Christopher08,Glawdeletal,Garstecki06} and also in numerical simulations~\cite{Demenech07,LiuZhang09,LiuZhang11,BowerLee11}. Our Newtonian data in the squeezing regime are quantitatively analyzed in figure \ref{fig:bench}, where we report the dimensionless droplet volume $V/H^3= L_d/H$. The linear behaviour of Eq. (\ref{eq:Garstecki}) is indeed reproduced by our simulations ($\alpha_1=1$ and $\alpha_2=2$) which are well in agreement with other existing numerical data in the literature, obtained with phase field numerical simulations~\cite{Demenech07} and LBM simulations~\cite{BowerLee11}.  Notice that the numerical simulations of Bower \& Lee~\cite{BowerLee11} are performed with a viscosity ratio $\lambda=0.02$. Nevertheless, their results agree with the others (including ours), which is a distinctive feature of the squeezing regime, where the droplet size is greatly affected by $Q$ and little effect is expected from a change in the fluid properties (i.e. change in the viscosity ratio $\lambda$).\\
To proceed further, we compute the droplet size for the two distinct cases of MV and DV. Panel (a) in figure \ref{fig:1-4} refers to a case with MV, with flow-rate ratio and finite extensibility parameter ranging in the interval $Q=0.2-1.0$ and $L^2=10^2-10^3$, respectively. For the non-Newtonian cases, the polymer relaxation time has been kept fixed to $\tau_P=4000$ lbu: this is a value at which the characteristic Deborah number \eqref{Desimple} is of order $1$ and viscoelastic effects are clearly visible. Panel (b) of figure \ref{fig:1-4} reports the same quantities as Panel (a) for a case with DV. The scaling relation \eqref{eq:Garstecki}, which is peculiar of the Newtonian cases, is a result of continuity. The analysis of the droplet size as a function of the flow-rate ratio $Q$ reveals that such relation needs to be modified to account for the effects of viscoelasticity: for increasing finite extensibility parameters, the droplet size is manifestly decreased by matrix viscoelasticity. Overall, figure \ref{fig:1-4} conveys the message that viscoelastic effects are more pronounced in the case of MV, whereas cases with DV only show smaller deviation with respect to the Newtonian reference case. This is not surprising, in view of the fact that the break-up process in the squeezing regime is driven by the action of the flow upstream of the emerging thread. More quantitatively, the linear scaling law \eqref{eq:Garstecki} is the result of two distinct physical processes: first, the dispersed phase grows until it effectively blocks the cross–section of the main channel and obstructs the flow of the continuous fluid (see also Panel (a) in figure \ref{fig:00}). At this particular moment, the ``blocking length'' $L_{block}$ is of the order of the channel width, say  $\alpha_1 H$ (with $\alpha_1$ a constant of order unity). Afterwards, the increased pressure in the continuous phase begins to squeeze the neck of dispersed phase (see also Panels (b)-(d) in figure \ref{fig:00}). For a neck with a characteristic width $\alpha_2 H$ ($\alpha_2$ is a constant, again, of order unity) and squeezing at a rate approximately equal to the average velocity ($Q_c/H^2$), it takes a time $\tau_{squeeze} \approx \alpha_2 H H^2/Q_c$ to complete the squeezing process. During this time, the thread continues to elongate at rate $Q_d/H^2$. The resulting ``squeezing length'' is therefore $L_{squeeze} \approx \tau_{squeeze} Q_d/H^2 = \alpha_2 H Q_d/Q_c$. Consequently, the final dimensionless size $L_d/H$ of the droplet can be expressed as $L_d/H \approx \alpha_1+ \alpha_2 Q$. Panel (a) of figure \ref{fig:1-4} actually reveals a change in the slope at increasing $L^2$: while the slope at $L^2=100$ is still almost same to that of the Newtonian case, the slope at $L^2=1000$ is visibly different. This points to the fact that the largest elastic effects may effectively perturb the region of the fluid upstream of the junction. \\
To better complement the results of figure \ref{fig:1-4}, in figure \ref{fig:5-6} we study the droplet size for the same values of $L^2$ analyzed in figure \ref{fig:1-4} and different values of $\tau_P$ ranging in the interval $\tau_P =250-4000$ lbu, resulting in a Deborah number ranging in the interval $\De = 0.4 - 6.2$. For the all $L^2$ studied, the droplet size shows a decreasing behaviour at increasing the Deborah number, which is more pronounced at larger $L^2$. Consistently with the expectations, when $\De \rightarrow 0$ we observe minor deviations with respect to the Newtonian case. We notice that the same analysis (data not shown) for DV reveals only a minor effect of non-Newtonian rheology in the dispersed phase, stressing once more the fact that viscoelastic effects in the upstream of the emerging thread are more efficient in perturbing the break-up process.\\
That viscoelastic effects are more pronounced in presence of larger $L^2$ is qualitatively understood because, by increasing $L^2$, the polymer dumbbell becomes more extensible and the maximum level of stress attainable is increased~\cite{bird,Herrchen}. Consistently, we expect an increased effect of the polymer feedback stresses on the Newtonian solvent. However, results of figures \ref{fig:1-4}-\ref{fig:5-6} only support this statement indirectly, i.e. without any information on the distribution of polymer feedback stresses and their action on the droplet formation process. To go deeper into this point, in figure \ref{fig_choreography} we report a simultaneous view of the droplet shape just before break-up and the polymer feedback stress that develops in the non-Newtonian phase. In particular, we focus on the polymer feedback stress in the stream-flow direction
\begin{equation}\label{eq:Txx}
T_{xx}=\frac{\eta_P}{\tau_P}f(r_{P})C_{xx}.
\end{equation}
We observe that $T_{xx}$ is enhanced in the region upstream of the emerging thread, providing extra viscoelastic forces which combine to change the droplet break-up process.\\
To make progress, we have monitored the time evolution of the pressure in the continuous fluid immediately upstream of the T-junction. In Panels (a)-(b) of figure \ref{fig_pressure} we report the pressure as a function of time for $\Ca=0.0026$, with $L^2=100$ (Panel (a)), $L^2=1000$ (Panel (b)), and different $\De$. It is evident that the obstruction of the main channel leads to an increase of the pressure upstream of the T-junction. As the dispersed phase enters the junction, the pressure rises gradually until the channel is blocked. The presence of viscoelasticity actually proves instrumental to enhance the pressure build-up and the effect is more pronounced at increasing both $\De$ and $L^2$. One may attempt to explain the observed behaviour by arguing that the obstruction provided by the thread forces the viscoelastic matrix fluid to ``converge'' and flow into a constriction, hence to develop a high extensional viscosity~\cite{bird,Herrchen}. This viscous response increases the dissipation and hence the pressure drop. This interpretation is actually supported by a direct observation of the velocity streamlines in the moment of the obstruction as reported in figure \ref{fig_stream}. Moreover, we have analyzed the force balance in the whole region upstream of the emerging thread. In particular, we have defined an {\it effective} force (${\bm F}_{\mbox{\tiny{eff}}}$) as~\cite{SbragagliaGupta}
\be\label{eq:effectiveforce}
{\bm F}_{\mbox{\tiny{eff}}}=\frac{\eta_P}{\tau_P}{\bm \nabla} \cdot [f(r_P){\bm {\bm C}}]-{\bm \nabla} \left(\eta_{P} ({\bm \nabla} {\bm u}_c+({\bm \nabla} {\bm u}_c)^{T})\right).
\ee
Indeed, we remark that viscoelastic forces provide a contribution to the shear forces. This happens in simple shear flows and also for weak viscoelasticity~\cite{bird,Herrchen}, where we expect that the viscoelastic stresses closely follow the viscous stresses, i.e. $\frac{\eta_P}{\tau_P}{\bm \nabla} \cdot [f(r_P){\bm {\bm C}}] \approx {\bm \nabla} \cdot \left(\eta_{P} ({\bm \nabla} {\bm u}_c+({\bm \nabla} {\bm u}_c)^{T})\right)$. Obviously, this cannot be the case when viscoelasticity is enhanced and the Deborah number is above unity. Since all our simulations are performed with the same shear viscosity, the effective force gives an idea of how much the viscoelastic system differs from the corresponding Newtonian system with the same viscosity. If present (${\bm F}_{\mbox{\tiny{eff}}} \neq 0$), this change is attributed to viscoelasticity. For a case with $L^2 = 5\times 10^3$, we analyze the effective force in the xy-plane at $z=L_z/2$ in the moment when the thread obstructs the main channel. Results are reported in figure \ref{fig_effectiveforce}, where we show two distinct plots for the x and y component of ${\bm F}_{\mbox{\tiny{eff}}}$. Upstream of the emerging thread, and close to the bottom wall, we indeed observe a resistance force which opposes to the flowing through the constriction, and we believe is responsible for the build-up in the pressure.\\
The tendency of viscoelastic stresses to promote a smaller droplet volume soon after break-up may be provisionally thought of as an anticipation of the dripping regime, thus echoing the work by Derzsi {\it et al.}~\cite{Garstecki13} in the flow-focusing geometry, where the authors found that the viscoelasticity leads to transitions between various regimes at lower ratios of flow and at lower values of the Capillary numbers in comparison to the Newtonian focusing liquids. However, upon entering the dripping regime, viscous shear forces will become relevant and since the shear viscosity is kept the same in all the simulations, one should expect to find a less pronounced effect of viscoelasticity at larger Capillary numbers. These expectations are indeed borne out by numerical simulations in the next section.

\subsection{Dripping and Jetting Regimes}\label{sec:drippingjetting}

\begin{figure*}[t!]
\subfigure[{\scriptsize $\Ca = 0.0026$ }]
{
\includegraphics[width = 0.457\linewidth]{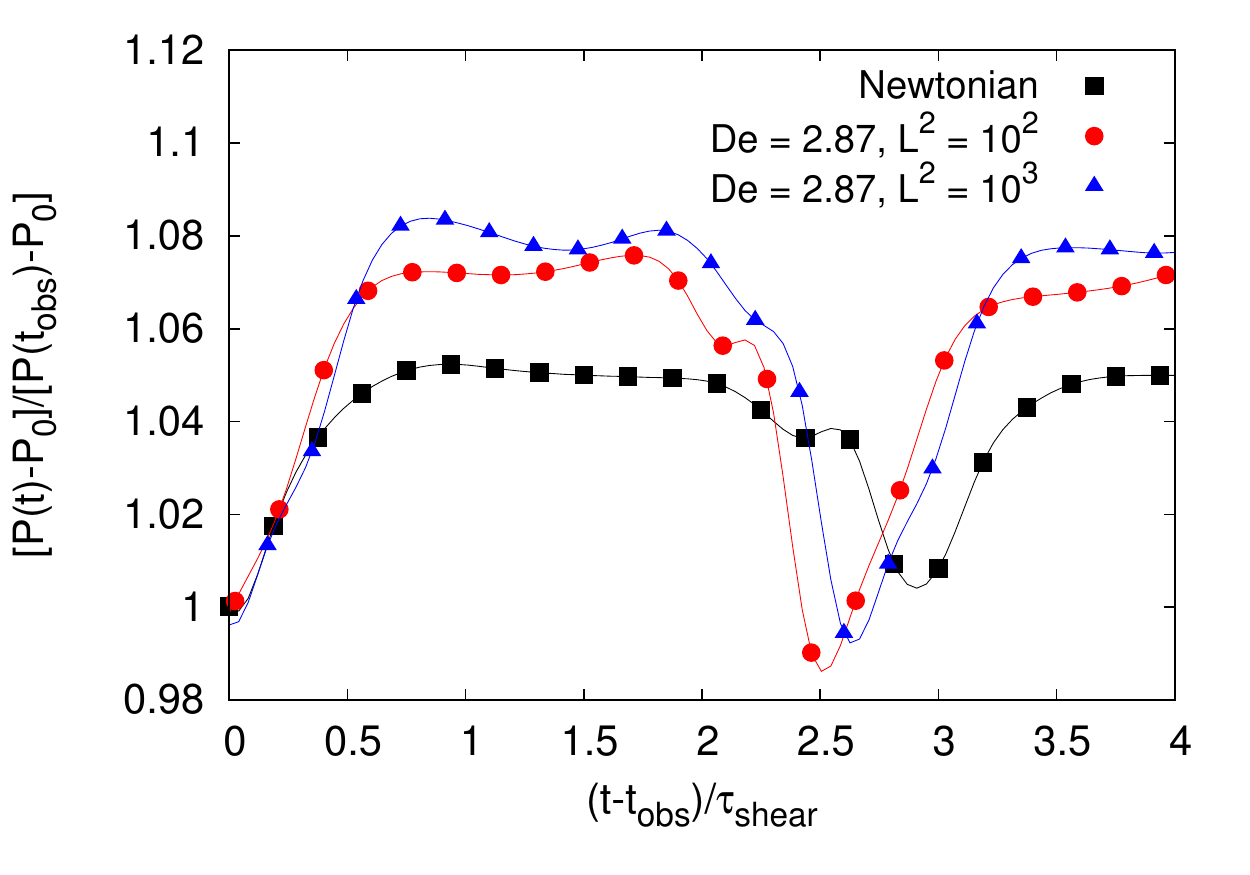}
}
\subfigure[{\scriptsize $\Ca = 0.0052$ }]
{
\includegraphics[width = 0.457\linewidth]{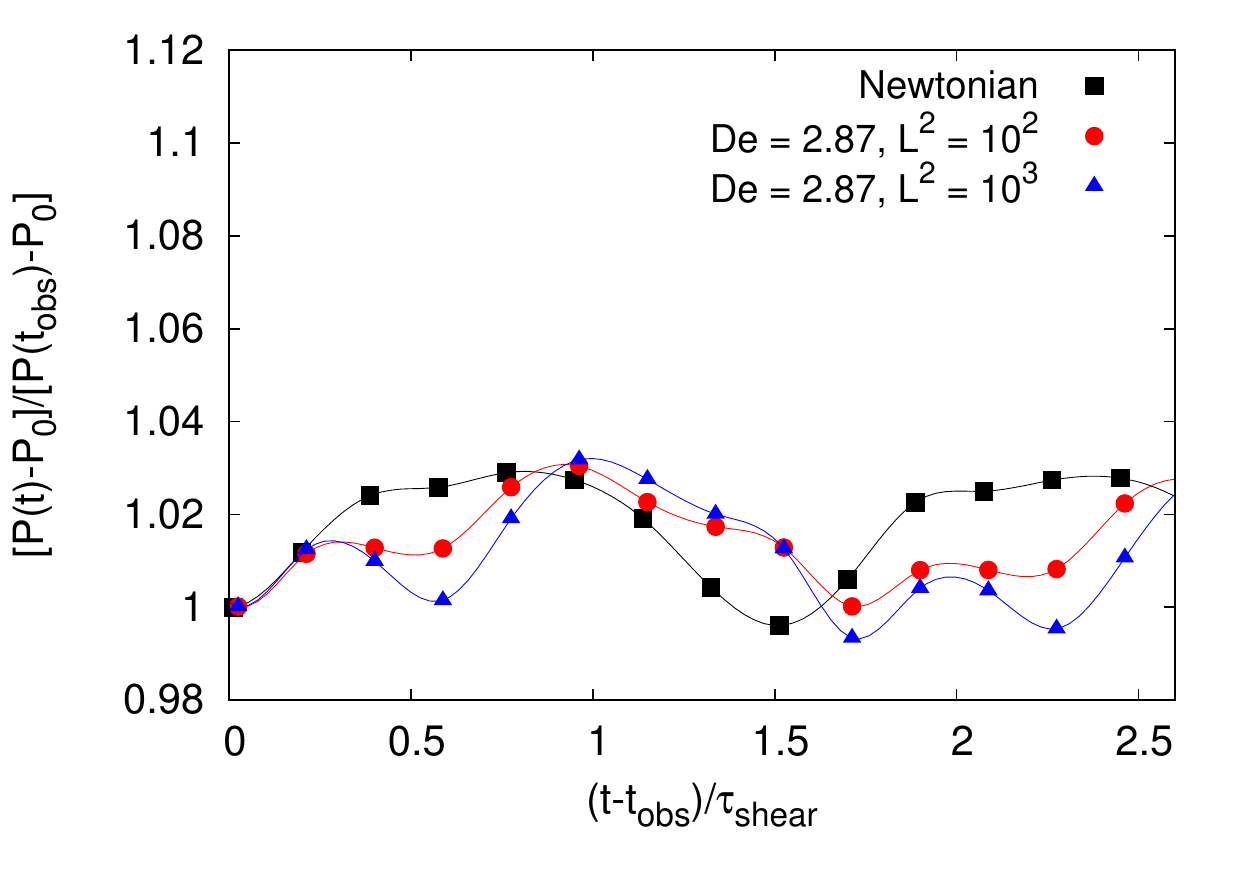}
}
%\subfigure[{\scriptsize $\Ca = 0.0078$ }]
%{
%\includegraphics[width = 0.33\linewidth]{figure12/figure12_Ca_c.pdf}
%}
\caption{Analysis of the pressure ($P$) versus time for different $\Ca$. The pressure is computed upstream of the T-junction and $P_0$ is a constant reference pressure computed in the static case. Panel (a): we report the normalized pressure versus time with fixed $\Ca = 0.0026$ for the Newtonian case (black squares) and two cases with matrix viscoelasticity (MV) at fixed $\De=5.7$ and different $L^2$: $L^2 = 100$ (red circles) and $L^2 = 1000$ (blue triangles). Panel (b): same as panel (a) for $\Ca = 0.0052$. In all cases we have used the characteristic shear time $\tau_{\mbox{\tiny{shear}}}=H/v_c$ as a unit of time, while $\tobs$ is the time when the thread starts to obstruct the channel. \label{fig_pressureb} }
\end{figure*}

\begin{figure*}[t!]
\subfigure[{\scriptsize Matrix Viscoelasticity (MV), Droplet Size}]
{
\includegraphics[width = 0.475\linewidth]{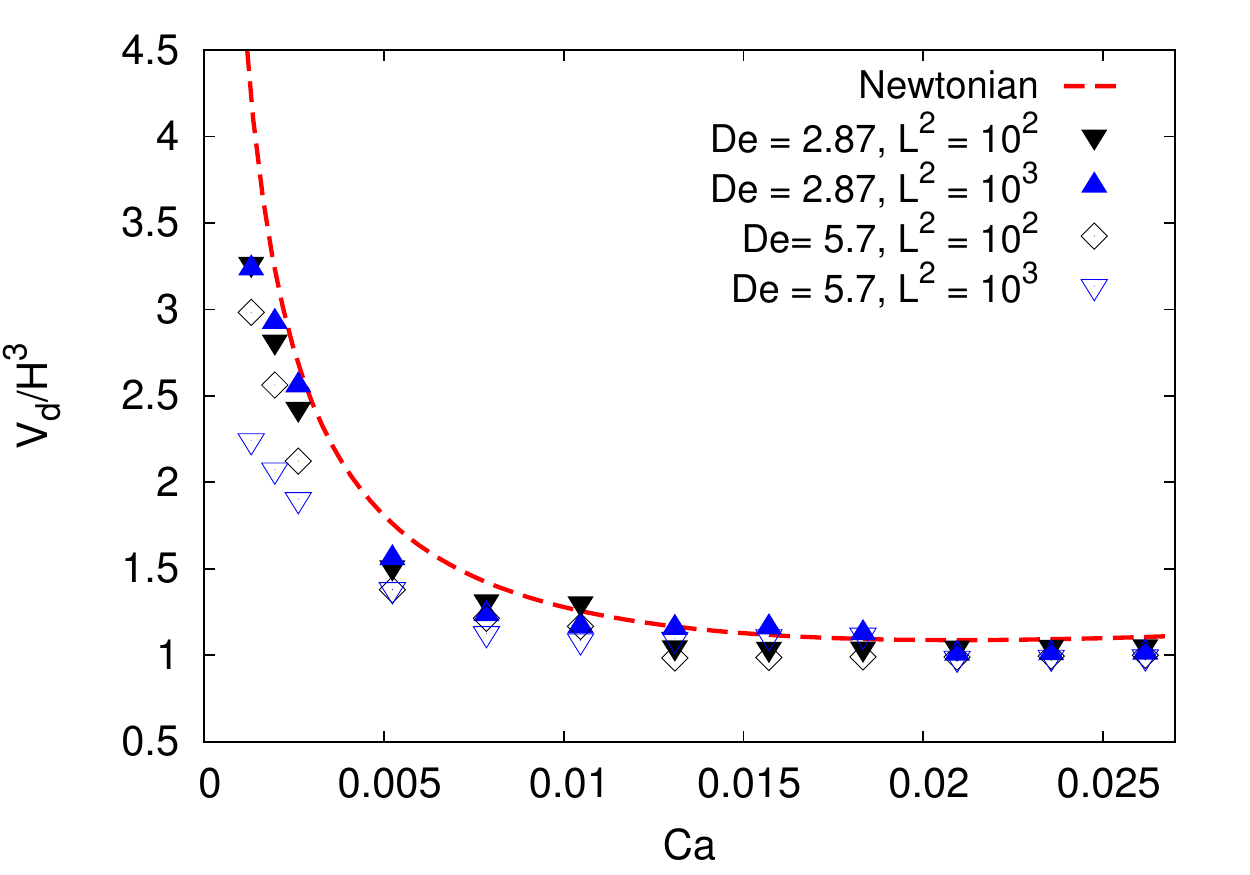}
}
\subfigure[{\scriptsize Matrix Viscoelasticity (MV), Break-up time}]
{
\includegraphics[width = 0.475\linewidth]{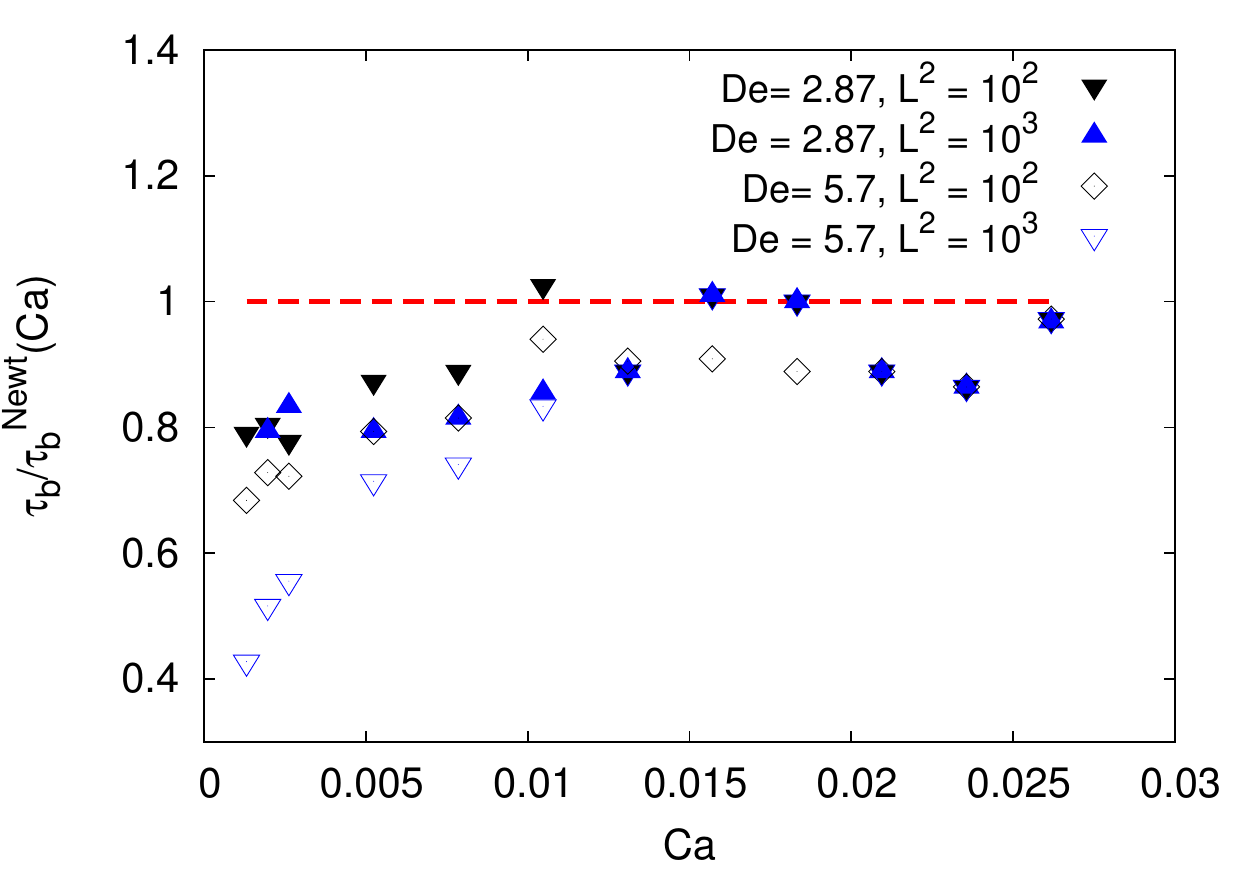}
}
\caption{Quantitative analysis of the break-up process for different $\Ca$. Panel (a): we report the dimensionless droplet volume $V/H^3$ soon after break-up for a case with matrix viscoelasticity (MV). We choose the polymer relaxation time $\tau_P$ and the finite extensibility parameter $L^2$ ranging in the interval $\tau_P=2000-4000$ lbu ($\De=2.85-5.71$ based on \eqref{Desimple}) and $L^2=10^2-10^3$, respectively. The flow rate ratio is kept fixed to $Q=1.0$ and the Capillary number is changed in the range $\Ca=0.001-0.03$. Panel (b): we report the break-up time $\tau_b$ normalized to the break-up time of the corresponding Newtonian case $\tau_b^{\mbox{\tiny{Newt}}}$. \label{fig:7-8}}
\end{figure*}

\begin{figure*}[t!]
\begin{center}
\subfigure[{\scriptsize $t=t_0+ 11.3 \tshear$, $\De = 1.42$, $L^2 = 10^3$}]
{
\includegraphics[width = 0.47\linewidth]{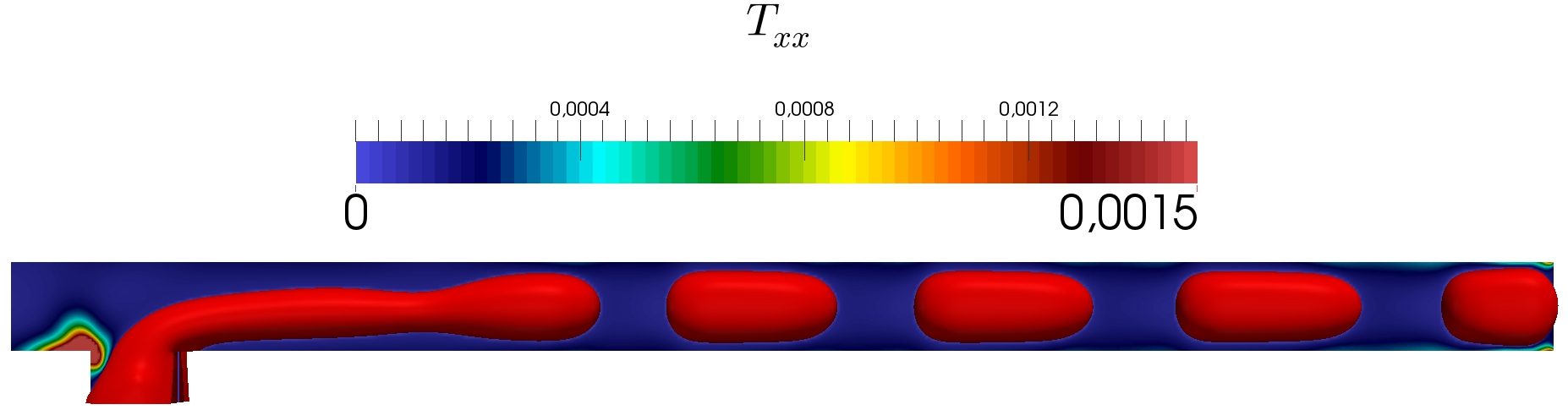}
}
\subfigure[{\scriptsize $t=t_0+ 10.9 \tshear$, $\De = 7.14$, $L^2 = 10^3$}]
{
\includegraphics[width = 0.47\linewidth]{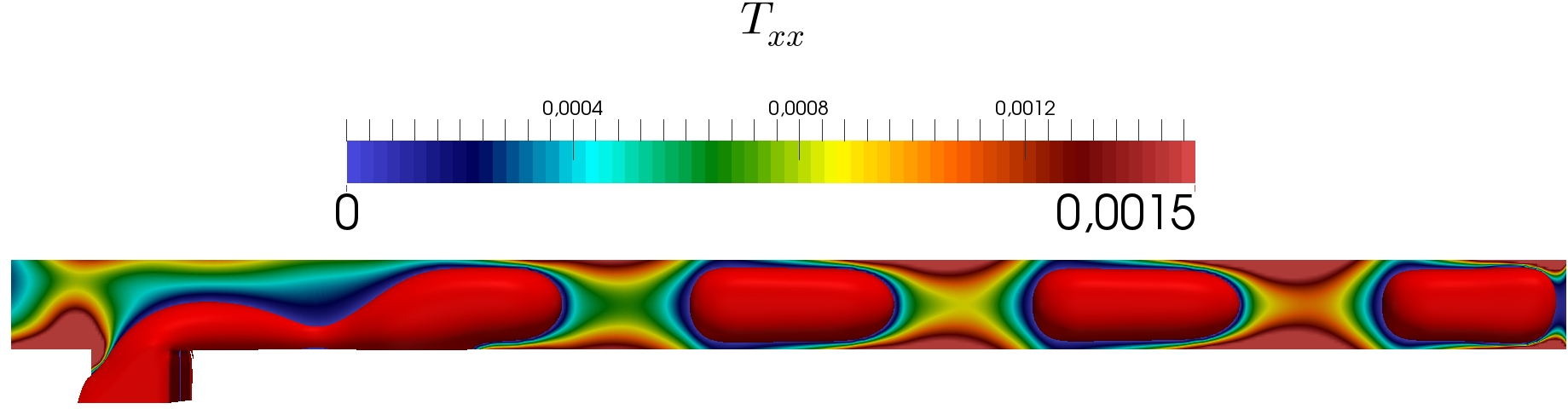}
}
\caption{Panels (a)-(b): density contours of dispersed phase overlaid on the polymer feedback stress in the stream-flow direction (see Eqs. \eqref{NSc} and \eqref{eq:Txx}) for a case with matrix viscoelasticity (MV) with $L^2=10^3$ and two different values of $\De$:  $\De = 1.42$ (Panel (a)) and $\De = 7.14$ (Panel (b)). The other parameters are kept fixed to $\Ca=0.013$, $\lambda=1.0$, $Q=1.0$.  In all cases we have used the characteristic shear time $\tau_{\mbox{\tiny{shear}}}=H/v_c$ as a unit of time, while $t_0$ is a reference time (the same for all simulations). Notice that the colorbar of the feedback stress \eqref{eq:Txx} is the same. \label{fig_choreographybb}}
\end{center}
\end{figure*}

\begin{figure*}[t!]
\begin{center}
\subfigure[{\scriptsize $t=t_0+8.8 \tshear$, $\De = 7.14$, $L^2 = 10^3$}]
{
\includegraphics[width = 0.47\linewidth]{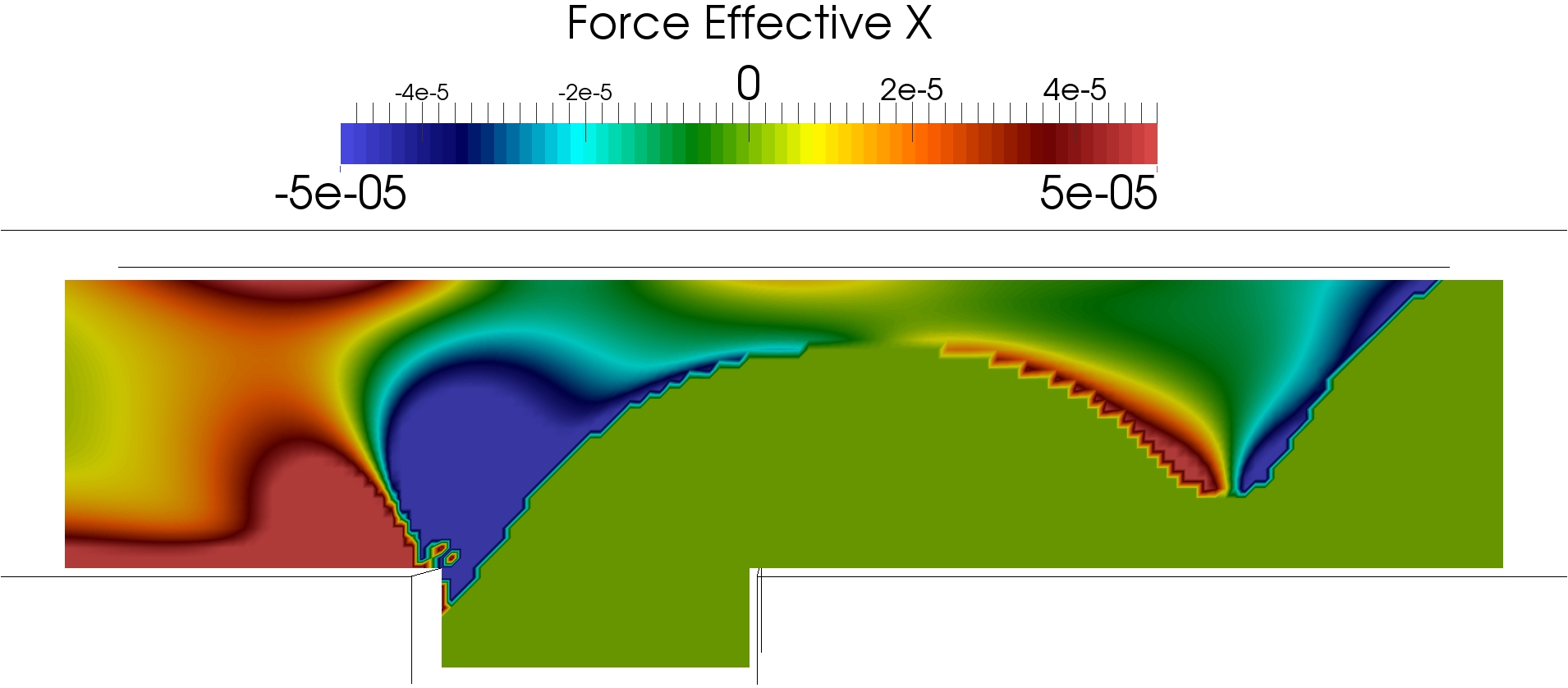}
}
\subfigure[{\scriptsize $t=t_0+8.8 \tshear$, $\De = 7.14$, $L^2 = 10^3$}]
{
\includegraphics[width = 0.47\linewidth]{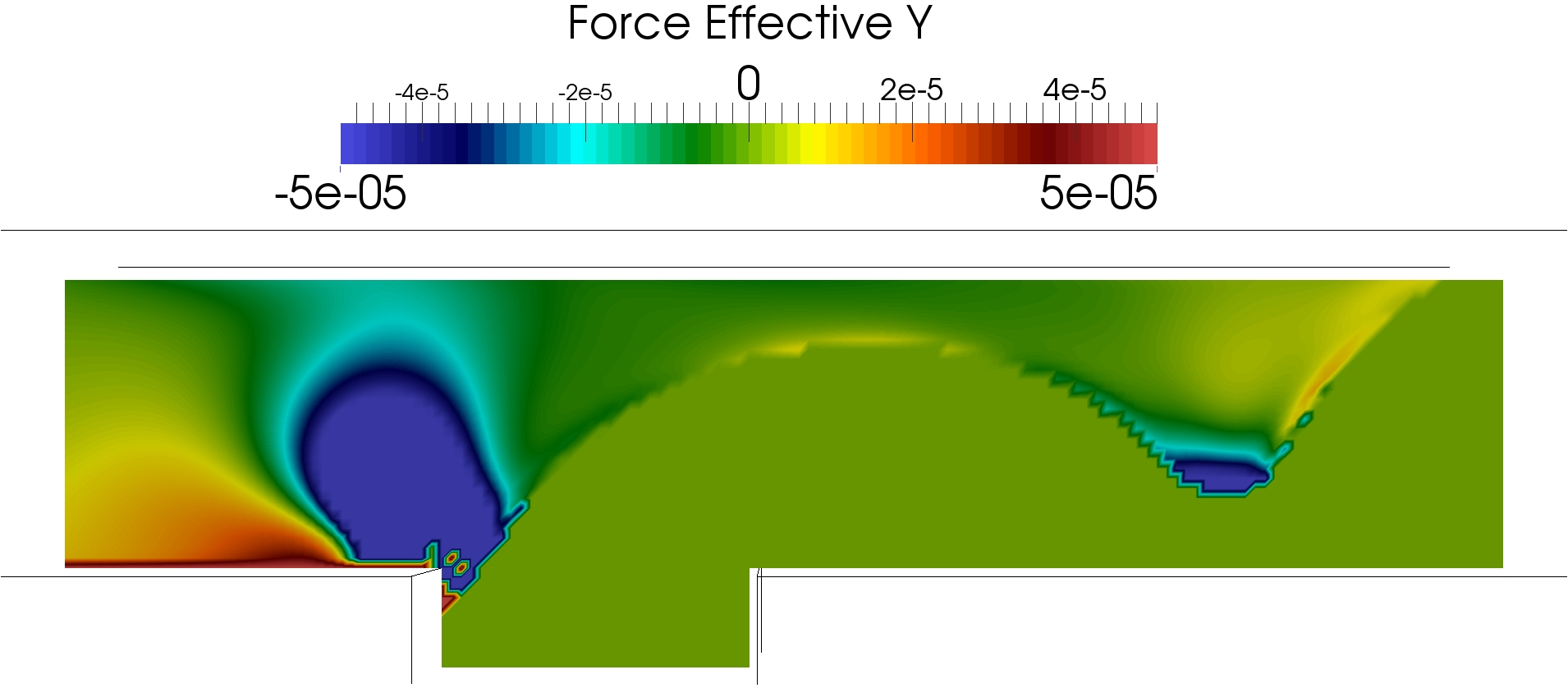}
}\\
\caption{Panels (a)-(b): x and y component of effective force ${\bm F}_{\mbox{\tiny{eff}}}$ (see Eq. \eqref{eq:effectiveforce}) for a matrix viscoelasticity (MV) case at $t=t_0+ 8.8 \tshear$, $\De = 7.14 $, $L^2 = 10^3$, $\Ca=0.013$, $\lambda=1.0$ and $Q=1.0$. At the break-up point we see that there is a net effective force which forces the necking process towards the boundary. We have used the characteristic shear time $\tau_{\mbox{\tiny{shear}}}=H/v_c$ as a unit of time, while $t_0$ is a reference time (the same for all simulations). \label{fig_effectiveforce_dripping}}
\end{center}
\end{figure*}

\begin{figure*}[t!]
\begin{center}
\subfigure[{\scriptsize $t=t_0+ 6.8 \tshear$, $\De = 1.42$, $L^2 = 10^3$}]
{
\includegraphics[width = 0.47\linewidth]{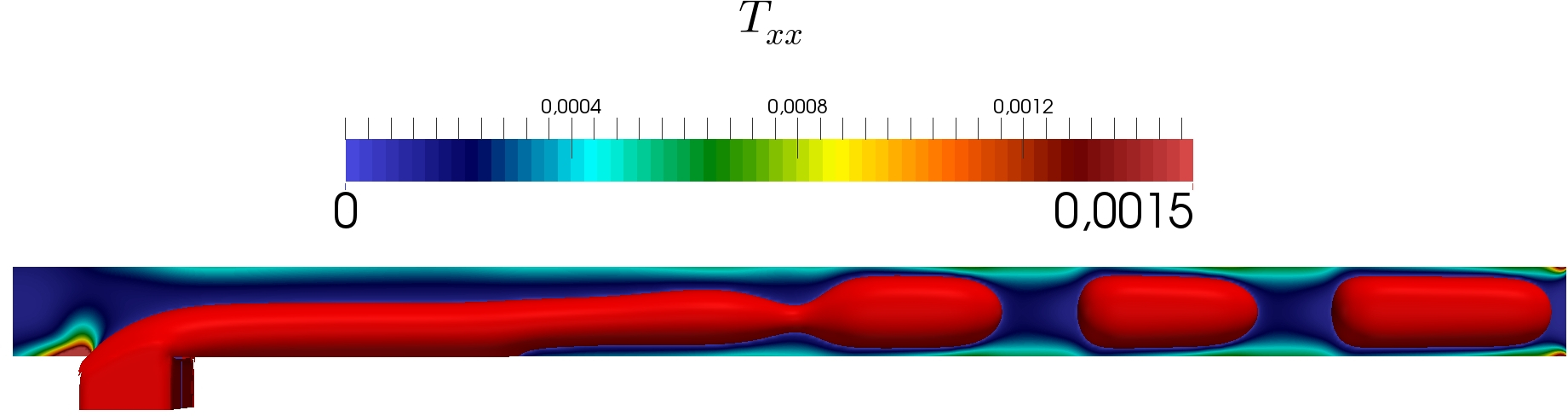}
}
\subfigure[{\scriptsize $t=t_0+ 6.8 \tshear$, $\De = 7.14$, $L^2 = 10^3$}]
{
\includegraphics[width = 0.47\linewidth]{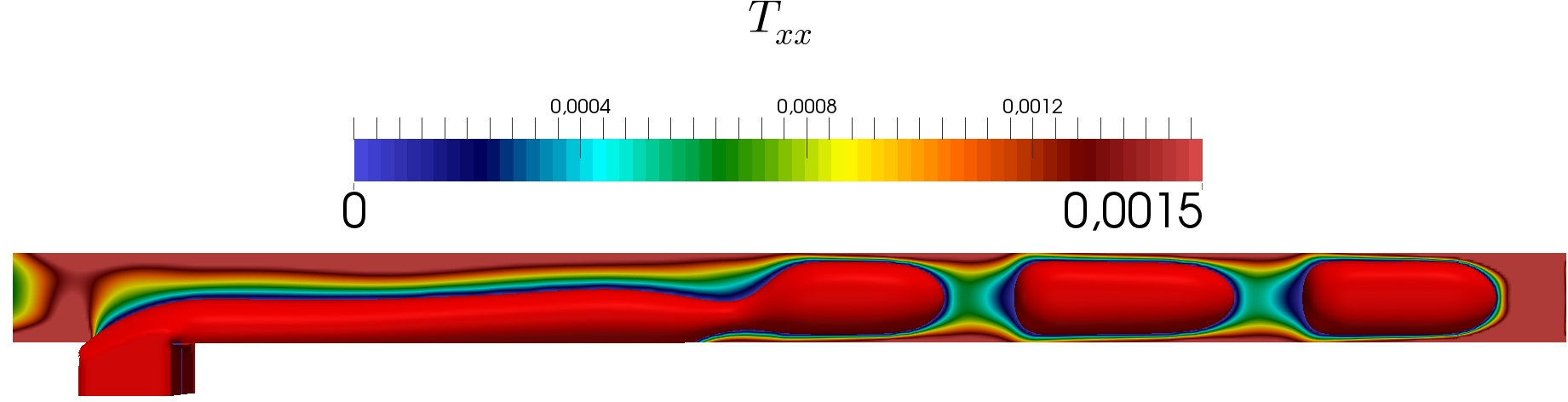}
}
\caption{Panels (a)-(b): density contours of dispersed phase overlaid on the polymer feedback stress in the stream-flow direction (see Eqs. \eqref{NSc} and \eqref{eq:Txx}) for a case with matrix viscoelasticity (MV) with $L^2=10^3$ and two different values of $\De$:  $\De = 1.42$ (Panel (a)) and $\De = 7.14$ (Panel (b)). The other parameters are kept fixed to $\Ca=0.026$, $\lambda=1.0$, $Q=1.0$.  In all cases we have used the characteristic shear time $\tau_{\mbox{\tiny{shear}}}=H/v_c$ as a unit of time, while $t_0$ is a reference time (the same for all simulations). Notice that the colorbar of the feedback stress \eqref{eq:Txx} is the same. \label{fig_choreographybbb}}
\end{center}
\end{figure*}

The analysis in the squeezing regime has evidenced the non trivial role of the polymer feedback stresses in changing the dynamics and break-up properties in a situation where $\Ca$ is moderately small. Consequently, an interesting point of discussion emerges on the role of viscoelasticity on scenarios which are different from the squeezing regime. As we have seen in figure \ref{fig:00}, by increasing $\Ca$ at fixed flow-rate ratio we move from the squeezing regime to the dripping and jetting regimes. Also, as already stressed before, the effect of MV is more pronounced with respect to the effect of DV, a conclusion that still holds for the $\Ca$ and flow parameters used in both the dripping and jetting regime. We therefore choose to report on the effects of viscoelasticity in the transition from squeezing to dripping/jetting by reporting data only for the case of MV. \\
In figure \ref{fig_pressureb} we report the analysis for the pressure upstream of the emerging thread for two different Capillary numbers: while for the smaller Capillary number the pressure build-up is clearly influenced by viscoelasticity (see also figure \ref{fig_pressure}), by increasing the Capillary number, this effect is less pronounced. We remark that the shear viscosity is kept the same in all the simulations, so one actually expects to find a less pronounced effect of viscoelasticity at larger Capillary numbers, where the viscous shear forces start to influence the droplet break-up process (see also figure \ref{fig:00}). This is also quantitatively supported by the results of Panel (a) of figure \ref{fig:7-8}, where we report the dimensionless droplet volume $V/H^3$ as a function of $\Ca$ for the same values of $L^2$ considered in the previous figures. The flow-rate ratio is kept fixed to $Q=1.0$ and $\Ca$ is changed in the range $\Ca=0.001-0.03$. The Deborah number is ranging in the interval $\De=2.85-5.71$ ($\tau_P=2000-4000$ lbu). At increasing the Capillary number, we observe that the tendency of viscoelastic stresses to promote a smaller volume soon after break-up is somehow less evident. This is also complemented by the results in Panel (b) of figure \ref{fig:7-8}, where we report the break-up time $\tau_b$ normalized to the break-up time of the corresponding Newtonian case $\tau_b^{\mbox{\tiny{Newt}}}$. Other non trivial effects, however, are present in the morphology of break-up: while for small $\De$ we observe that the detachment point shifts downstream of the junction (Panel (a) of figure \ref{fig_choreographybb}), the increase of the Deborah number favors a stabilization of the break-up point closer to the junction (Panel (b) of figure \ref{fig_choreographybb}). Another interesting feature found is that viscoelasticity favors the necking process to take place closer to the channels walls (see Panel (b) in figure \ref{fig_choreographybb}). To go deeper into this point, similarly to what we have done in figure \ref{fig_effectiveforce} for the squeezing regime, in figure \ref{fig_effectiveforce_dripping} we analyze the effective force \eqref{eq:effectiveforce} in the xy-plane at $z=L_z/2$ for a case with $L^2 = 5\times 10^3$ and $\Ca=0.013$. Again, an ``elastic'' region upstream of the emerging thread is observed. The straining of the fluid upstream of the emerging thread causes a storing of elastic energy which is released with an elastic expansion downstream of the emerging thread (negative y component of the effective force in Panel (b) of figure \ref{fig_effectiveforce_dripping}). This release of elastic energy forces the necking process towards the boundary.\\ 
We notice that in the plot of the normalized break-up time (Panel (b) of figure \ref{fig:7-8}), a Capillary number of the order of $\Ca  = \Cacr \approx 10^{-2}$ exists, above which the normalized break-up time is very close to unity and does not sensibly change with $\De$ and/or $L^2$. We attribute this behaviour to the emergence of the jetting regime. The corresponding Newtonian dynamics for such Capillary numbers (see figure \ref{fig:00}) indeed reveals that the dripping regime is not stable and the droplet detachment point gradually moves downstream, until a jet is formed~\cite{Demenech07,LiuZhang09}. Similarly to figure \ref{fig_choreographybb}, density contours of the dispersed phase overlaid on the polymer feedback stresses at these larger Capillary numbers are reported in figure \ref{fig_choreographybbb}. Panel (a) of figure \ref{fig_choreographybbb} reports the liquid thread just before break-up for a slightly viscoelastic case, corresponding to $\De=1.42$. The break-up point actually detaches from the wall as it moves progressively downstream (see also figure \ref{fig:00}). Panel (b) of figure \ref{fig_choreographybbb} reports a case with increased Deborah number $\De=7.14$: we observe that due to the presence of the feedback stresses, the break-up point shows a slight tendency to move towards the wall, which echoes the effects already found in the dripping regime. Notice that due to the increase of the Capillary number ($\Ca = 0.026$), the feedback stresses are more intense than situations at smaller $\Ca$.\\
The small effects on droplet size observed at the higher Capillary numbers in our simulations somehow echo the numerical work by Shonibare {\it et al.}~\cite{Shonibare} on T-junctions with viscoelastic phases. In particular, Shonibare {\it et al.} used 2d numerical simulations, using the Volume of Fluid (VOF) method, to predict the size and detachment point of a viscoelastic droplet in a Newtonian Matrix. The authors explored both pressure driven flows as well as plane Couette flows in the continuous phase: for the Newtonian problem they report smaller droplet sizes when the cross shear rate is increased, also in agreement with experimental work~\cite{Steinhaus}. However, the introduction of viscoelasticity was found to have minimal effects on the droplet size. In comparison to our work, some issues are worth being mentioned and discussed. Our numerical simulations on droplet viscoelasticity (data only partially shown, see also section \ref{sec:squeezing}) acknowledge a small effect on the droplet size as well. As already stressed earlier, we find that the elastic effects are more pronounced with matrix viscoelasticity and sensibly perturb the droplet size when the dispersed phase obstructs the main channel. Even if we were dealing with matrix viscoelasticity in the geometry of Shonibare {\it et al.}~\cite{Shonibare}, we believe that the elastic effects that we discussed in section 3.1  would be sensibly reduced as well, since the geometry of Shonibare {\it et al.}~\cite{Shonibare} does not allow a considerable obstruction of the main channel, but rather trigger droplet detachment and pinch-off based on forces generated by the cross-shear rate.\\ 
Another point to be discussed is the importance of the dimensionality in our numerical simulations. In need of an extensive study to quantify the importance of the various model parameters, we preliminarly explored the possibility to use two dimensional numerical simulations. In some situations, when the viscosity ratio $\lambda$ is smaller than one and moderate flow rate ratios are considered, 2d break-up is actually found to quantitatively well compare with 3d break-up (see also figure \ref{fig:bench}). However, other 2d numerical simulations with viscosity ratio of order 1 did not capture the underlying physics quantitatively: when the squeezing process is about to conclude, long filaments may be stabilized at the detachment point where the side channel meets the main channel, thus producing a break-up dynamics which is quantitatively different in 2d and 3d. This is a pathology of the 2d model, possibly related to the stability of filaments in 2d which would be absent in a 3d simulation. Other studies~\cite{LiuZhang09} based on LBM in 2d do not report on droplet break-up with those parameters where we observe such pathology. These facts said, and not to spoil the correctness of the 3d case, we decided to carry out simulations in 3d, while leaving a detailed comparison between the 2d and 3d simulations to a future study, possibly to identify the correct range of parameters where both can be matched

%%%%%%%%%%%%%%%%%%%%%%%%%%%%%FIG 12%%%%%%%%%%%%%%%%%%%%%%%%%%%%%%%%%%%%%%%
%%%%%%%%%%%%%%%%%%%%%%%%%%%%%%%%%%%%%%%%%%%%%%%%%%%%%%%%%%%%%%%%%%%%%%%%%%

%%%%%%%%%%%%%%%%%%%%%%%%%%%%%%%%%%%%%%%%%%%%%%%%%%%%%%%%%%%%%%%%%%%%%%%%%%%%%%%%%%%%%%%%%%%%%%%%%%%%%%%%%%%%%%%%%%%%%%%%%%%%%%%%%%%%%%%%%%%%%%%%%%%%%%%%%%%%%%%%

%%%%%%%%%%%%%%%%%%%%%%%%%%%%%%%%%%%%%%%%%%%%%%%%%%%%%%%%%%%%%%%%%%%%%%%%%%%%%%%%%%%%%%%%%%%%%%%%%%%%%%%%%%%%%%%%%%%%%%FIG 8%%%%%%%%%%%%%%%%%%%%%%%%%%%%%%%%%%%%%%%%%%%%%%%%%%%%%%%%%%%%%%%%%%%%%%%%%%%%%%%%%%%%%%%%%%%%%%%%%%%%%%%%%%%%%%%%%%%%%

%%%%%%%%%%%%%%%%%%%%%%%%%%%%%%%%%%%%%%%%%%%%%%%%%%%%%%%%%%%%%%%%%%%%%%%%%%%%%%%%
%%%%%%%%%%%%%%%%%%%%%%%%%%%%%%%%%%%%%%%%%%%%%%%%%%%%%%%%%%%%%%%%%%%%%%%%%%%%%%%%

%%%%%%%%%%%%%%%%%%%%%%%%%%%%%%%%%%%%%%%%%%%%%%%%%%%%%%%%%%%%%%%%%%%%%%%%%%%%%%%
%%%%%%%%%%%%%%%%%%%%%%%%%%%%%%%%%%%%%%FIG 9%%%%%%%%%%%%%%%%%%%%%%%%%%%%%%%%%%%%%%%%%%%%%%%%%%%%%%%%%%%%%%%%%%%%%%%%%%%%%%%%%%%%%%%%%%%%%%%%%%%%%%%%%%%%%%%%%%%%

%%%%%%%%%%%%%%%%%%%%%%%%%%%%%%%%%%%%%%%%%%%%%%%%%%%%%%%%%%%%%%%%%%%%%%%%%%%%%%%%
%%%%%%%%%%%%%%%%%%%%%%%%%%%%%%%%%%%%%%%%%%%%%%%%%%%%%%%%%%%%%%%%%%%%%%%%%%%%%%%%

%%%%%%%%%%%%%%%%%%%%%%%%%%%%%%%%%%%%%%%%%%%%%%%%%%%%%%%%%%%%%%%%%%%%%%%%%%%%%%%%
%%%%%%%%%%%%%%%%%%%%%%%%%%%%%%%%%%%%%%%%%FIG 10%%%%%%%%%%%%%%%%%%%%%%%%%%%%%%%%%%%%%%%%%%%%%%%%%%%%%%%%%%%%%%%%%%%%%%%%%%%%%%%%%%%%%%%%%%%%%%%%%%%%%%%%%%%%%%%%%%

%%%%%%%%%%%%%%%%%%%%%%%%%%%%%%%%%%%%%%%%%%%%%%%%%%%%%%%%%%%%%%%%%%%%%%%%%%%%%%%%
%%%%%%%%%%%%%%%%%%%%%%%%%%%%%%%%%%%%%%%%%%%%%%%%%%%%%%%%%%%%%%%%%%%%%%%%%%%%%%%%

%%%%%%%%%%%%%%%%%%%%%%%%%%%%%%%%%%%%%%%%%%%%%%%%%%%%%%%%%%%%%%%%%%%%%%%%%%%%%%%
%%%%%%%%%%%%%%%%%%%%%%%%%%%%%%%%%%%%%%FIG 11%%%%%%%%%%%%%%%%%%%%%%%%%%%%%%%%%%%%%%%%%%%%%%%%%%%%%%%%%%%%%%%%%%%%%%%%%%%%%%%%%%%%%%%%%%%%%%%%%%%%%%%%%%%%%%%%%%%%

%%%%%%%%%%%%%%%%%%%%%%%%%%%%%%%%%%%%%%%%%%%%%%%%%%%%%%%%%%%%%%%%%%%%%%%%%%%%%%%%
%%%%%%%%%%%%%%%%%%%%%%%%%%%%%%%%%%%%%%%%%%%%%%%%%%%%%%%%%%%%%%%%%%%%%%%%%%%%%%%%

\section{Conclusions}\label{sec:conclusions}

Microfluidic technologies offer the possibility to generate small fluid volumes of dispersed phases (droplets) in continuous phases. One of the most common droplet generator is represented by T-junction geometries~\cite{Demenech07,Demenech06}, where the dispersed phase is injected perpendicularly into the main channel and the break-up process is induced by forces created by the cross-flowing continuous phase. The confinement that naturally accompanies these devices has an impact on droplet deformation and break-up, which are significantly different from those of unbounded droplets. The situation is further complicated by the complex properties of the bulk phases, whenever constituents have a viscoelastic - rather than Newtonian - nature. In this paper we have investigated  the effects of viscoelasticity on the dynamics and break-up of droplets in microfluidic T-junctions using numerical simulations of dilute polymer solutions at small Capillary numbers up to $\Ca \approx 3 \times 10^{-2}$ and moderate flow-rate ratios $Q \approx {\cal O}(1)$. Our numerical model builds upon our previous studies~\cite{SbragagliaGuptaScagliarini,SbragagliaGupta} and is based on a Navier-Stokes (NS) description of the solvent based on the lattice Boltzmann models (LBM) coupled to constitutive equations for finite extensible non-linear elastic dumbbells with the closure proposed by Peterlin (FENE-P model). We have used three-dimensional simulations to characterize the various characteristic mechanisms of breakup in the confined T-junction.  Moreover, the various model parameters of the FENE-P constitutive equations, including the polymer relaxation time $\tau_P$ and the finite extensibility parameter $L^2$, have been changed to provide quantitative details on how the dynamics and break-up properties are affected by viscoelasticity, in cases where the viscoelastic properties are confined in the dispersed (d) phase (Droplet Viscoelasticity, DV), as well as cases where the viscoelastic properties are confined in the continuous (c) phase (Matrix Viscoelasticity, MV). At fixed flow conditions (i.e. the same $\Ca$ and $Q$) we find that the effects of viscoelasticity are more pronounced in the case with MV, which is quantitatively attributed to the fact that the flow driving the break-up process upstream of the emerging thread can be sensibly perturbed by the polymer feedback stresses. This has been evidenced by the analysis of simultaneous view of the droplet shape just before break-up and the polymer feedback stresses that develop in the non-Newtonian phase. In particular, the numerical simulations are crucial to elucidate the relative importance of the free parameters in the FENE-P model, and to visualize the distribution of the polymer feedback stresses. Thanks to these insights, it was possible to correlate the distribution of the stresses to the corresponding break-up morphology. We also tried some preliminary numerical simulations with viscoelastic behaviour simultaneously present in both the continuous and dispersed phase, and these seem to produce similar effects as the matrix viscoelasticity case, at least for the geometry and parameters that we explored in our simulations.\\
For future investigations, it is surely warranted a complementary study to highlight the role of viscoelasticity on the break-up properties of confined droplets in flow-focusing geometries~\cite{Garstecki13,Arratia08,Arratia09}. Complementing the available experimental results with the help of numerical simulations would be of extreme interest. Simulations can indeed be used to perform in-silico  comparative studies, at changing the model parameters, to shed lights on the complex properties of viscoelastic flows in confined geometries.\\ 

We kindly acknowledge funding from the European Research Council under the Europeans Community's Seventh Framework Programme (FP7/2007-2013) / ERC Grant Agreement  N. 279004. We acknowledge the computing hours from ISCRA B project (POLYDROP), CINECA Italy. We acknowledge Prof. K. Sugiyama for useful discussions and exchange of ideas on viscoelasticity during his visit in January 2014. We also acknowledge F. Bonaccorso for helpful visualizations of the T-junction geometries from the numerical simulations.

%%%%%%%%%%%%%%%%%%%%%%%%%%%%%%%%%%%%%%%%%%%%%%%%
%\appendix
\appendix

\section{Hybrid Lattice Boltzmann Models (LBM) - Finite Difference Scheme for dilute Polymer solutions}\label{sec:appendix}

In this appendix we report the essential details of the numerical scheme used. We refer the interested reader to a dedicated paper~\cite{SbragagliaGupta} where more extensive technical details are reported, together with benchmarks on the rheology of dilute homogeneous solutions (including steady shear flow, elongational flows, transient shear and oscillatory flows) and viscoelastic droplet deformation in confined geometries. We use a hybrid algorithm combining a multicomponent Lattice-Boltzmann model (LBM) with Finite Differences (FD) schemes, the former used to model the macroscopic hydrodynamic equations, and the latter used to model the polymer dynamics. The LBM equations evolve in time the discretized probability density function $f_{\alpha i}({\bm{r}},t)$ to find at position ${\bm{r}}$ and time $t$ a fluid particle of component $\alpha=A,B$ with velocity  ${\bm{c}}_{i}$. The dispersed (d) and the continuous (c) Newtonian phases in Eqs.~(\ref{NSc}) and (\ref{NSd}) are characterized by a majority of one of the two components, i.e. majority of $A$ ($B$) in the dispersed (continuous) phase. The LBM evolution scheme with a unitary time-step reads as follows~\cite{Succi01}:
\begin{equation}\label{EQ:LBapp}
f_{\alpha i} ({\bm{r}} + {\bm{c}}_{i} , t + 1)-f_{\alpha i} ({\bm{r}}, t) = \sum_{j} {\cal L}_{i j}(f_{\alpha j}-f^{(eq)}_{\alpha j}) + \Delta^{g}_{\alpha i}.
\end{equation}
The collisional operator in the rhs of Eq.~\eqref{EQ:LBapp} is linear and expresses the relaxation of $f_{\alpha i}$ towards the local equilibrium $f^{(eq)}_{\alpha i}$. We use the D3Q19 model with 19 velocities
\begin{equation}\label{velo}
{\bm{c}}_{i}=
\begin{cases}
(0,0,0) & i=0\\
(\pm 1,0,0), (0,\pm 1,0), (0,0,\pm 1) & i=1\ldots6\\
(\pm 1,\pm 1,0), (\pm 1,0,\pm 1), (0,\pm 1,\pm 1)  & i=7\ldots18
\end{cases}.
\end{equation}
The expression for the equilibrium distribution is a result of the projection onto orthogonal polynomials~\cite{Dunweg07,DHumieres02} 
\be\label{feq}
f_{\alpha i}^{(eq)}=w_{i} \rho_{\alpha} \left[1+\frac{{\bm{u}} \cdot {\bm{c}}_{i}}{c_s^2}+\frac{{\bm{u}}{\bm{u}}:({\bm{c}}_{i}{\bm{c}}_{i}-{\Id})}{2 c_s^4} \right]
\ee
and the weights $w_{i}$ are
\begin{equation}\label{weights}
w_{i}=
\begin{cases}
1/3 & i=0\\
1/18 & i=1\ldots6\\
1/36 & i=7\ldots18
\end{cases},
\end{equation}
where $c_s$ is the isothermal speed of sound (a constant in the model) and ${\bm{u}}$ is the fluid velocity.  The operator ${\cal L}_{i j}$ in Eq.~\eqref{EQ:LBapp} is the same for both components and is characterized by a diagonal representation in the {\it mode space}: the basis vectors ${\bm{H}}_{k}$ ($k=0,...,18$) of such mode space are constructed by orthogonalizing polynomials of the dimensionless velocity vectors~\cite{Dunweg07,DHumieres02,Premnath,SegaSbragaglia13}. The basis vectors are used to calculate a complete set of moments, the so-called ``modes'' $m_{\alpha k}=\sum_{i} {\bm{H}}_{k i} f_{\alpha i}$ ($k=0,...,18$). The lowest order modes are related to the hydrodynamic variables, in particular the density (of both components and the total one), $\rho_{\alpha}=m_{\alpha 0}=\sum_{i} f_{\alpha i}$,  $\rho=\sum_{\alpha}m_{\alpha 0} =\sum_{\alpha}\rho_{\alpha}$, while the next three moments $\tilde{\bm{m}}_{\alpha}=(m_{\alpha 1}, m_{\alpha 2}, m_{\alpha 3})$, are related to the velocity of the mixture 
\be\label{totmom}
{\bm{u}} \equiv \frac{1}{\rho}\sum_{\alpha} \tilde{\bm{m}}_{\alpha}   +\frac{\bm{g}}{2 \rho} = \frac{1}{\rho}\sum_{\alpha} \sum_{i} f_{\alpha i} {\bm{c}}_{i}+\frac{\bm{g}}{2 \rho}.
\ee
The higher order modes refer to the shear and bulk modes in the viscous stress tensor, and also other modes (the so called ``ghost modes'') which do not appear at the level of hydrodynamic equations. The operator ${\cal L}_{i j}$ possesses a diagonal representation in mode space, hence the collisional term describes a linear relaxation of the modes, $m^{post}_{\alpha k}=(1+\lambda_k)m_{\alpha k}+m_{\alpha k}^{g}$, where the ``post" indicates the post-collisional mode and where the relaxation frequencies $-\lambda_k$ are related to the transport coefficients of the modes. The term $m_{\alpha k}^{g}$ is the $k$-th moment of the forcing source $\Delta_{\alpha i}^{g}$ which embeds the effects of a forcing term with density ${\bm{g}}_{\alpha}$~\cite{Dunweg07,Premnath}. The term ${\bm{g}}=\sum_{\alpha} {\bm g}_{\alpha}$ in Eq.~\eqref{totmom} is the total (internal+external) force. Forces transfer an amount ${\bm{g}}_{\alpha}$ of total momentum to the fluid in one time step. The forcing term is determined in such a way that the hydrodynamic Eqs.~(\ref{eq:2})-(\ref{eq:3}) are obtained~\cite{Guo}
\begin{equation}
\Delta_{\alpha i}^{g}=\frac{w_{i}}{c_s^2} \left(\frac{2+\lambda_M}{2}\right) {\bm{g}}_{\alpha} \cdot {\bm{c}}_{i} +\frac{w_{i}}{c_s^2} \left[\frac{1}{2c_s^2} {\bm{G}} : ({\bm{c}}_{i} {\bm{c}}_{i}-c_s^2 {\Id} ) \right],
\end{equation}
\begin{equation}
{\bm G}=\frac{2+\lambda_s}{2}\left({\bm u} {\bm g} + ({\bm u}  {\bm g})^T-\frac{2}{3} {\Id} ({\bm u} \cdot {\bm g}) \right)+\frac{2+\lambda_b}{3} {\Id} ({\bm u} \cdot {\bm g})
\end{equation}
where the relaxation frequencies of the momentum ($-\lambda_M$), bulk ($-\lambda_b$) and shear ($-\lambda_s$) modes appear. LBM is able to reproduce the continuity equations and the NS equations for the total momentum~\cite{Dunweg07,Premnath,SegaSbragaglia13}
\be\label{eq:2}
\partial_t \rho_{\alpha}+ {\bm \nabla} \cdot (\rho_{\alpha} {\bm u}) = {\bm \nabla} \cdot {\bm D}_{\alpha},
\ee
\be\label{eq:3}
\begin{split}
\rho & \left[ \partial_t \bm u + ({\bm u} \cdot {\bm \nabla}) \bm u \right]= -{\bm \nabla}p \\ 
&+ {\bm \nabla} \left[ \eta_{s} \left( {\bm \nabla} {\bm u}+({\bm \nabla} {\bm u})^{T}-\frac{2}{3} {\Id} ({\bm \nabla} \cdot {\bm u}) \right) +\eta_{b}{\Id} ({\bm \nabla} \cdot {\bm u}) \right] + {\bm g}
\end{split}
\ee
where we have indicated with $\eta_s$, $\eta_b$ the shear and bulk viscosities, respectively. In Eq.~\eqref{eq:3}, $p=\sum_{\alpha} p_{\alpha}=\sum_{\alpha} c_s^2 \rho_{\alpha}$ represents the internal (ideal) pressure of the mixture. The quantity ${\bm D}_{\alpha}$ represents the inter-diffusion flux 
\begin{equation}\label{eq:comp_Pi}
{\bm D}_{\alpha}=\mu \left[\left({\bm \nabla} p_{\alpha}-\frac{\rho_{\alpha}}{\rho}{\bm \nabla} p\right)-\left({\bm g}_{\alpha}-\frac{\rho_{\alpha}}{\rho} {\bm g} \right) \right]
\end{equation}
with $\mu$ a mobility parameter. As for the internal forces, we will use the ``Shan-Chen'' model~\cite{SC93,CHEM09,sbragaglia12} for multicomponent fluids
\begin{equation}\label{eq:SCforce}
{\bm g}_{\alpha}({\bm{r}}) =  - {g}_{AB} \rho_{\alpha}({\bm{r}}) \sum_{i} \sum_{\alpha'\neq \alpha} w_{i} \rho_{\alpha^{\prime}} ({\bm{r}}+\bm{c}_{i}) {\bm c}_{i} \hspace{.2in} \alpha,\alpha^{\prime}=A,B
\end{equation}
where ${g}_{AB}$ is a parameter that regulates the interactions between the two components. When ${g}_{AB}$ is sufficiently large, the model can describe stable interfaces with a positive surface tension. The effect of interaction forces is to introduce an ``interaction'' pressure tensor ${\bm P}^{(\mbox{\tiny{int}})}$~\cite{SbragagliaBelardinelli}, which modifies the internal pressure, i.e. ${\bm P} = p \, {\Id}+{\bm P}^{(\mbox{\tiny{int}})}$ 
\be\label{PT}
\begin{split}
{\bm P}^{(\mbox{\tiny{int}})}({\bm r}) & =  \frac{1}{2} {g}_{AB} \rho_{A}({\bm r})\sum_{i} w_{i} \rho_{B}({\bm r}+{\bm c}_{i}) {\bm c}_{i} {\bm c}_{i} \\
& +\frac{1}{2} {g}_{AB} \rho_{B}({\bm r})\sum_{i} w_{i} \rho_{A}({\bm r}+{\bm c}_{i}){\bm c}_{i} {\bm c}_{i}.
\end{split}
\ee
A tuning of the density in contact with the wall allows for the modelling of the wetting properties~\cite{Benzi06,sbragaglia08}.\\ 
With regard to the transport coefficients of hydrodynamics, the relaxation frequencies of the momentum is related to the mobility coefficient
\begin{equation}
\mu=-\left(\frac{1}{\lambda_M}+\frac{1}{2} \right)
\end{equation}
while the relaxation frequencies of the bulk and shear modes in (\ref{EQ:LBapp}) are related to the viscosity coefficients as
\begin{equation}\label{TRANSPORTCOEFF}
\eta_s=-\rho c_s^2 \left(\frac{1}{\lambda_s}+\frac{1}{2} \right); \hspace{.1in} \eta_b=-\frac{2}{3}\rho c_s^2  \left(\frac{1}{\lambda_b}+\frac{1}{2} \right).
\end{equation}
The numerical simulations presented feature ${g}_{AB}=1.5$ lbu in (\ref{eq:SCforce}), corresponding to a surface tension $\sigma=0.1$ lbu and associated bulk densities $\rho_A=2.0$ lbu and $\rho_B=0.1$ lbu in the $A$-rich phase. The relaxation frequencies in (\ref{TRANSPORTCOEFF}) are set to $\lambda_M=-1.0$ lbu and $\lambda_s=\lambda_b$, thus reproducing the viscous stress tensor given in Eqs.~(\ref{NSc}) and (\ref{NSd})).  The viscosity ratio of the LBM fluid is changed by allowing $\lambda_s$ to depend on space 
\be
-\rho c_s^2 \left(\frac{1}{\lambda_s}+\frac{1}{2} \right)=\eta_s=\eta_d (f_+(\phi))+\eta_c (f_{-}(\phi))
\ee
where $\phi=\phi({\bm r})=\frac{(\rho_A({\bm r})-\rho_B({\bm r}))}{(\rho_A({\bm r})+\rho_B({\bm r}))}$. The functions $f_{\pm}(\phi)$ are chosen as
\be
f_{\pm}(\phi)=\left(\frac{1 \pm \tanh(\phi/\xi)}{2}\right)
\ee
which allows to recover the Newtonian part of the NS equations reported in Eqs.~(\ref{NSc}) and (\ref{NSd}) with shear viscosities $\eta_d$ and $\eta_c$. The smoothing parameter $\xi$ is chosen sufficiently small so as to match analytical predictions on droplet deformation in presence of viscoelastic stresses (see~\cite{SbragagliaGupta} for all details).\\
As for the polymer evolution given in Eq.~\eqref{FENEP}, we follow the two References~\cite{perlekar06,vaithianathan2003numerical} to solve the FENE-P equation. The polymer stress $f(r_P){\bm {\mathcal C}}$ is computed from the FENE-P evolution equation and used to change the shear modes of the LBM~\cite{SbragagliaGupta,Dunweg07,DHumieres02}. The feedback of the polymers is modulated~\cite{Yue04} in space with the functions $f_{\pm}(\phi)$ 
\be
\begin{split}
\rho & \left[ \partial_t \bm u  + ({\bm u} \cdot {\bm \nabla}) \bm u \right] = -{\bm \nabla} {\bm P} \\
&+ {\bm \nabla} \left[(\eta_d f_+(\phi)+\eta_c f_{-}(\phi)) ({\bm \nabla} {\bm u}+({\bm \nabla} {\bm u})^{T} ) \right]\\ 
&+\frac{\eta_P}{\tau_P}{\bm \nabla} [f(r_P){\bm {\mathcal C}} f_{\pm}(\phi) ].
\end{split}
\ee
By using $f_{-}(\phi)$, we recover a case where the viscoelastic properties are confined in the continuous (c) phase, while the use of the function $f_{+}(\phi)$ allows to recover a case where the viscoelastic properties are confined in the dispersed (d) phase.

%%%%%%%%%%%%%%%%%%%%%%%%%%%%%%%%%%%%%%%%%%%%%%%%
% BibTeX users please use one of
%\bibliographystyle{spbasic}      % basic style, author-year citations
%\bibliographystyle{spmpsci}      % mathematics and physical sciences
\bibliographystyle{spphys}       % APS-like style for physics
%-- Rhodes Style --
%\bibliography{}   % name your BibTeX data base
%\bibliographystyle{aipproc}   % if natbib is available
%\bibliographystyle{aipprocl} % if natbib is missing
%-- ------ ----- --

%%%%%%%%%%%%%%%%%%%%%%%%%%%%%%%%%%%%%%%%%%%
\bibliography{sample}

\hyphenation{Post-Script Sprin-ger}
\begin{thebibliography}{10}
\providecommand{\url}[1]{{#1}}
\providecommand{\urlprefix}{URL }
\expandafter\ifx\csname urlstyle\endcsname\relax
  \providecommand{\doi}[1]{DOI \discretionary{}{}{}#1}\else
  \providecommand{\doi}{DOI \discretionary{}{}{}\begingroup
  \urlstyle{rm}\Url}\fi

\bibitem{Christopher07}
G.F. Christopher, S.L. Anna, J. Phys. D Appl. Phys. \textbf{40}, R319 (2007)

\bibitem{Seeman12}
R.~Seemann, M.~Brinkmann, T.~Pfohl, S.~Herminghaus, Rep. Prog. Phys.
  \textbf{75}, 016601 (2012)

\bibitem{Christopher08}
G.F. Christopher, N.N. Noharuddin, J.A. Taylor, S.L. Anna, Phys. Rev. E
  \textbf{78}, 036317 (2008)

\bibitem{Teh08}
S.~Teh, R.~Lin, L.~Hung, A.~Lee, Lab Chip \textbf{8}, 198 (2008)

\bibitem{Baroud10}
C.N. Baroud, F.~Gallaire, R.~Dangla, Lab Chip \textbf{10}, 2032 (2010)

\bibitem{Glawdeletal}
T.~Glawdel, C.~Elbuken, L.~Ren, Phys. Rev. E \textbf{85}, 016322 (2012)

\bibitem{Glawdeletalb}
T.~Glawdel, C.~Elbuken, L.~Ren, Phys. Rev. E \textbf{85}, 016323 (2012)

\bibitem{Glawdeletalbb}
T.~Glawdel, L.~Ren, Phys. Rev. E \textbf{85}, 026308 (2012)

\bibitem{Demenech07}
M.D. Menech, P.~Garstecki, F.~Jousse, H.A. Stone, Jour. Fluid. Mech.
  \textbf{595}, 141 (2008)

\bibitem{Demenech06}
M.D. Menech, Phys. Rev. E \textbf{73}, 031505 (2006)

\bibitem{LiuZhang09}
H.~Liu, Y.~Zhang, J. Appl. Phys. \textbf{106}, 034906 (2009)

\bibitem{LiuZhang11}
H.~Liu, Y.~Zhang, Phys. Fluids \textbf{23}(8), 082101 (2011)

\bibitem{Garstecki13}
L.~Derzsi, M.~Kasprzyk, J.P. Plog, P.~Garstecki, Phys. Fluids \textbf{25},
  092001 (2013)

\bibitem{Garstecki06}
P.~Garstecki, M.J. Fuerstman, H.A. Stone, G.M. Whiteside, Lab Chip \textbf{6},
  437 (2006)

\bibitem{Arratia08}
P.E. Arratia, J.P. Gollub, D.J. Durian, Phys. Rev. E \textbf{77}, 036309 (2008)

\bibitem{Steinhaus}
B.~Steinhaus, A.Q. Shen, R.~Sureshkumar, Phys. Fluids \textbf{19}, 073103
  (2007)

\bibitem{Husny06}
J.~Husny, J.~Cooper-White, J. Non-Newton. Fluid Mech. \textbf{137}, 121 (2006)

\bibitem{Arratia09}
P.E. Arratia, L.A. Cramer, J.P. Gollub, D.J. Durian, New J. Phys. \textbf{11},
  115006 (2009)

\bibitem{Wang11}
W.~Wang, Z.~Liu, Y.~Jin, Y.~Cheng, Chem. Eng. J. \textbf{173}, 828 (2011)

\bibitem{vandersman06}
S.~Van~der Graaf, T.~Nisisako, R.~Schron, C.G.P.H. Van der~Sman, R.~Boom,
  Langmuir \textbf{22}, 4144 (2006)

\bibitem{Legendre}
S.~Arias, D.~Legendre, R.~Gonz\'alez-Cinca, Computers and Fluids \textbf{56},
  49 (2012)

\bibitem{Gupta09}
A.~Gupta, S.M.S. Murshed, R.~Kumar, Appl. Phys. lett. \textbf{94}, 164107
  (2009)

\bibitem{Gupta10}
A.~Gupta, R.~Kumar, Phys. Fluids \textbf{22}, 122001 (2010)

\bibitem{Wagner05}
C.~Wagner, Y.~Amarouchene, D.~Bonn, J.~Eggers, Phys. Rev. Lett.
  \textbf{95}(16), 164504 (2005)

\bibitem{Lindner03}
A.~Lindner, J.~Vermant, D.~Bonn, Physica A \textbf{319}, 125 (2003)

\bibitem{bird}
R.B. Bird, R.C. Armstrong, O.~Hassager, \emph{Dynamics of polymeric liquids}
  (J. Wiley \& Sons, 1987)

\bibitem{Herrchen}
M.~Herrchen, H.~Oettinger, J. Non-Newtonian Fluid Mech. \textbf{68}, 17 (1997)

\bibitem{Zhang11}
J.~Zhang, Microfluid Nanofluid \textbf{10}, 1 (2011)

\bibitem{Aidun10}
C.K. Aidun, J.R. Clausen, Annu. Rev. Fluid Mech. \textbf{42}, 439 (2010)

\bibitem{Xi99}
H.~Xi, C.~Duncan, Phys. Rev. E \textbf{59}(3), 3022 (1999)

\bibitem{vandersman08}
R.G.M.V. der Sman, S.V. der Graaf, Comput. Phys. Commun. \textbf{178}, 492
  (2008)

\bibitem{Komrakova13}
A.E. Komrakovaa, O.~Shardt, D.~Eskinb, J.J. Derksen, Int. J. Multiphas. Flow
  \textbf{59}, 23 (2014)

\bibitem{Liuetal12}
H.~Liu, A.J. Valocchi, Q.~Kang, Phys. Rev. E \textbf{85}, 046309 (2012)

\bibitem{Moradi}
N.~Moradi, F.~Varnik, I.~Steinbach, Europhys. Lett. \textbf{95}, 44003 (2011)

\bibitem{Thampi}
S.~Thampi, R.~Adhikari, R.~Govindarajan, Langmuir \textbf{29}, 3339 (2013)

\bibitem{Yue04}
P.~Yue, J.J. Feng, C.~Liu, J.~Shen, J. Fluid Mech. \textbf{515}, 293 (2004)

\bibitem{Yueetal05}
P.~Yue, J.J. Feng, C.~Liu, J.~Shen, J. Non-Newtonian Fluid Mech.
  \textbf{129}(3), 163 (2005)

\bibitem{Yueetal06a}
P.~Yue, C.~Zhou, J.J. Feng, C.F. Ollivier-Gooch, H.H. Hu, J. Compu. Phys.
  \textbf{219}(1), 47 (2006)

\bibitem{Yueetal06b}
P.~Yue, C.~Zhou, J.J. Feng, Phys. Fluids \textbf{18}(10), 102102 (2006)

\bibitem{Yueetal08}
D.~Zhou, P.~Yue, J.J. Feng, J. Rheol. (1978-present) \textbf{52}(2), 469 (2008)

\bibitem{Yueetal12}
P.~Yue, J.J. Feng, J. Non-Newtonian Fluid Mech. \textbf{189}, 8 (2012)

\bibitem{SbragagliaGuptaScagliarini}
A.~Gupta, M.~Sbragaglia, A.~Scagliarini, J. Compu. Phys. \textbf{291}, 177
  (2015)

\bibitem{SbragagliaGupta}
A.~Gupta, M.~Sbragaglia, Phys. Rev. E \textbf{90}(2), 023305 (2014)

\bibitem{SC93}
X.~Shan, H.~Chen, Phys. Rev. E \textbf{47}, 1815 (1993)

\bibitem{SC94}
X.~Shan, H.~Chen, Phys. Rev. E \textbf{49}, 2941 (1994)

\bibitem{Benzi06}
M.~Sbragaglia, R.~Benzi, L.~Biferale, S.~Succi, F.~Toschi, Phys. Rev. Lett.
  \textbf{97}, 204503 (2006)

\bibitem{sbragaglia08}
M.~Sbragaglia, K.~Sugiyama, L.~Biferale, Jour. Fluid. Mech. \textbf{614}, 471
  (2008)

\bibitem{Greco02}
F.~Greco, J. Non-Newtonian Fluid Mech. \textbf{107}, 111 (2002)

\bibitem{Greco02b}
F.~Greco, Phys. Fluids \textbf{14}(3), 946 (2002)

\bibitem{Minale10}
M.~Minale, S.~Caserta, S.~Guido, Langmuir \textbf{26}, 126 (2010)

\bibitem{Minale04}
M.~Minale, J. Non-Newtonian Fluid Mech. \textbf{123}, 151 (2004)

\bibitem{Minale10b}
M.~Minale, Rheol. Acta \textbf{49}, 789 (2010)

\bibitem{Gladrow00}
D.~Wolf-Gladrow, \emph{Lattice-Gas Cellular Automata and Lattice {B}oltzmann
  Models: An Introduction} (Springer Verlag, 2001)

\bibitem{BowerLee11}
L.~Amaya-Bower, T.~Lee, Philos. T. Roy. Soc. A \textbf{369}(1945), 2405 (2011)

\bibitem{Shonibare}
O.~Shonibare, K.~Feigl, F.X. Tanner,  (2015), pp. 1--12.
\newblock \doi{10.13140/RG.2.1.1198.4806}

\bibitem{Succi01}
S.~Succi, \emph{The Lattice {B}oltzmann Equation for Fluid Dynamics and Beyond}
  (Oxford University Press, 2001)

\bibitem{Dunweg07}
B.~D{\"u}nweg, U.D. Schiller, A.J.C. Ladd, Phys. Rev. E \textbf{76}, 036704
  (2007)

\bibitem{DHumieres02}
D.~d'Humi{\`e}res, I.~Ginzburg, M.~Krafczyk, P.~Lallemand, L.S. Luo, Phil.
  Trans. Roy. Soc. London \textbf{360}(1792), 437 (2002)

\bibitem{Premnath}
K.~Premnath, J.~Abraham, J. Compu. Phys. \textbf{224}, 539 (2007)

\bibitem{SegaSbragaglia13}
M.~Sega, M.S.S.S. Kantorovich, A.O. Ivanovd, Soft Matter \textbf{9}, 10092
  (2013)

\bibitem{Guo}
Z.~Guo, C.~Zheng, B.~Shi, Phys. Rev. E \textbf{65}, 046308 (2002)

\bibitem{CHEM09}
R.~Benzi, M.~Sbragaglia, S.~Succi, M.~Bernaschi, S.~Chibbaro, Jour. Chem. Phys.
  \textbf{131}, 104903 (2009)

\bibitem{sbragaglia12}
M.~Sbragaglia, R.~Benzi, M.~Bernaschi, S.~Succi, Soft Matter \textbf{8}, 10773
  (2012)

\bibitem{SbragagliaBelardinelli}
M.~Sbragaglia, D.~Belardinelli, Phys. Rev. E \textbf{88}, 013306 (2013)

\bibitem{perlekar06}
P.~Perlekar, D.~Mitra, R.~Pandit, Phys. Rev. Lett. \textbf{97}, 264501 (2006)

\bibitem{vaithianathan2003numerical}
T.~Vaithianathan, L.R. Collins, J. Comp. Phys. \textbf{187}, 1 (2003)

\end{thebibliography}
%%%%%%%%%%%%%%%%%%%%%%%%%%%%%%%%%%%%%%%%%%%

\end{document}